\documentclass[a4paper,11pt]{article}
\pdfoutput=1
\usepackage[latin1]{inputenc}
\usepackage{amssymb}
\usepackage{amsmath}
\usepackage{amsthm}
\usepackage{amsfonts}
\usepackage[pdftex]{graphicx}
\usepackage{geometry}
\usepackage{braket}
\usepackage{enumerate}
\usepackage{comment}

\usepackage{subcaption}

\allowdisplaybreaks

\usepackage[table,usenames,dvipsnames]{xcolor}

\usepackage{setspace}
\usepackage[in]{fullpage}
\usepackage{hyperref}
\usepackage{array}
\usepackage{cancel}
\usepackage{wrapfig}
\usepackage[font=small,labelfont=bf]{caption}
\usepackage{pifont}

\usepackage[dvipsnames]{xcolor}
\usepackage{mathtools}

\usepackage{color}

\usepackage{tikz}
\usetikzlibrary{arrows,shapes}
\usetikzlibrary{trees}
\usetikzlibrary{matrix,arrows} % For commutative diagram
\usetikzlibrary{positioning}% For "above of=" commands
\usetikzlibrary{calc,through}% For coordinates
\usetikzlibrary{decorations.pathreplacing}% For curly braces
\usepackage{pgffor}	% For repeating patterns

\numberwithin{equation}{section}

\usepackage[english]{babel}

\usepackage{slashed}
\usepackage{caption}

\usepackage[nottoc,notlot,notlof]{tocbibind}
\usepackage[nosort]{cite}   
\usepackage{color}

\usepackage{parskip}

\usepackage[T1]{fontenc}

\usepackage{lmodern}
\usepackage{sectsty}
\usepackage{yfonts}
\allsectionsfont{\boldmath}

\usepackage{physics}
\usepackage{slashed}
\usepackage{makeidx}
\usepackage[vcentermath]{youngtab}
\usepackage{amsthm}
\usepackage{mathtools}

\hypersetup{
	colorlinks=true,
	linktoc=page,
	citecolor=blue,
	linkcolor=blue,
	urlcolor=blue} 

\newcommand{\agl}[2]{\langle#1 \, #2 \rangle}
\newcommand{\sqr}[2]{\lbrack #1 \, #2 \rbrack}
\newcommand{\sabr}[4]{[#1_{\dot{#2}}\, #3_{#4}\rangle}
\newcommand{\asbr}[4]{\langle#1_{#2} \, #3_{\dot{#4}}]}

\newcommand{\CC}{\mathbb{C}}
\newcommand{\cA}{\mathcal{A}}
\newcommand{\cN}{\mathcal{N}}
\newcommand{\cL}{\mathcal{L}}
\newcommand{\cO}{\mathcal{O}}

\usepackage{float}
\usepackage{marvosym}
\usepackage{empheq}
\usepackage{bbold}

\usepackage{tikz}
\usetikzlibrary{shapes}
\usetikzlibrary{arrows}
\usetikzlibrary{positioning}
\usetikzlibrary{decorations.pathmorphing}
\usetikzlibrary{decorations.pathreplacing}
\usetikzlibrary{decorations.markings}

\begin{document}

%%%% Title page %%%%

\begin{flushright}
	QMUL-PH-19-23\\
	SAGEX-19-21-E\\
	%HU-EP-18/25
\end{flushright}

\vspace{20pt} 

\begin{center}

	{\Large \bf  Complete Form Factors in Yang-Mills from   }  \\
	\vspace{0.3 cm} {\Large \bf  Unitarity and Spinor Helicity in Six Dimensions }

	\vspace{25pt}

	{\mbox {\sf  \!\!\!\!Manuel~Accettulli~Huber, Andreas~Brandhuber, Stefano~De~Angelis and 				Gabriele~Travaglini{\includegraphics[scale=0.05]{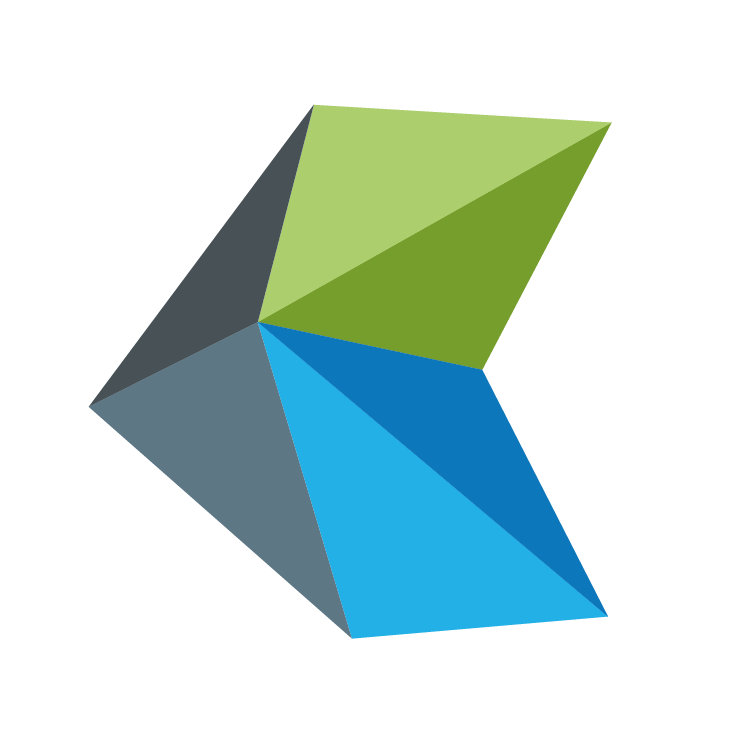}}
	}}
	\vspace{0.5cm}

	\begin{center}
		{\small \em
			Centre for Research in String Theory\\
			School of Physics and Astronomy\\
			Queen Mary University of London\\
			Mile End Road, London E1 4NS, United Kingdom
		}
	\end{center}

	%\vspace{-8pt}

	\vspace{40pt}  %was 40 

	{\bf Abstract}
\end{center}

\vspace{0.3cm}

\noindent

\noindent

We present a systematic procedure to compute complete, analytic form factors of gauge-invariant operators at loop level in pure Yang-Mills. We consider applications to operators of the form $\Tr F^n$ where $F$ is the gluon field strength. Our approach is based on an extension to form factors of the dimensional reconstruction technique, in conjunction with  the six-dimensional spinor-helicity formalism and generalised unitarity.
For form factors this technique  requires the introduction of additional scalar operators, for which we provide a systematic prescription.
 We also discuss a generalisation of  dimensional reconstruction  to any number of loops, both for amplitudes and  form factors.
Several novel results for one-loop minimal and non-minimal form factors of $\Tr F^n$ with $n\!>\!2$  are presented. Finally, we  describe the  \texttt{Mathematica}  package \texttt{SpinorHelicity6D}, which is tailored to handle six-dimensional quantities written in the spinor-helicity formalism.

\vfill
\hrulefill
\newline
\vspace{-1cm}
${\includegraphics[scale=0.05]{Sagex.png}}$~\!\!{\tt\footnotesize\{m.accettullihuber, a.brandhuber, s.deangelis, g.travaglini\}@qmul.ac.uk}

\setcounter{page}{0}
\thispagestyle{empty}
\newpage

%%%%%%%%%%%%%%%%%% TABLE OF CONTENTS %%%%%%%%%%%%%%%%%%%%%%%%%%%%%%%%%

\setcounter{tocdepth}{4}
\hrule height 0.75pt
\tableofcontents
\vspace{0.8cm}
\hrule height 0.75pt
\vspace{1cm}

\setcounter{tocdepth}{2}

\newpage
%%%%%%%%%%%%%%%%%%%%%%%%%%%%%%%%%%%%%%%%%%%%%%%%%%%%%%%%%%%

\section{Introduction}

The aim  of this paper is to construct complete, analytic form factors of gauge-invariant operators at one loop. 
In supersymmetric theories, four-dimensional unitarity \cite{Bern:1994zx,Bern:1994cg}  is sufficient to obtain complete answers for amplitudes at one loop. Without supersymmetry or for   form factors of non-protected operators this is no longer the case because of the appearance of rational contributions.
In the amplitude context this problem  has been addressed  in different ways. In one approach,  one makes use of factorisation to establish a recursion relation that allows to reconstruct   rational terms \cite{Bern:2005hs, Bern:2005ji} (see \cite{Dunbar:2016aux,Dunbar:2016gjb} for  recent elegant applications to two-loop amplitudes in pure Yang-Mills).
Another approach is to shift the dimensionality of internal states in the loop away from four dimensions \cite{vanNeerven:1985xr,Bern:1995db} where  rational terms acquire a singularity which can then be detected using  unitarity cuts. Multiple cuts have also been used efficiently in this context \cite{Brandhuber:2005jw,Anastasiou:2006jv,Anastasiou:2006gt}.
This method requires that the internal lines, corresponding to virtual particles, are kept in $d$ dimensions, while   momenta and polarisation vectors of  external particles live in four dimensions.

Having the internal particles in arbitrary,  non-integer  dimensions introduces complications, since  tree amplitudes are  no longer simple and the power of the  spinor-helicity formalism is lost. A solution to this problem  is offered by  ``dimensional reconstruction'' \cite{Giele:2008ve,Ellis:2008ir,Bern:2010qa,Boughezal:2011br,Davies:2011vt}.
In this approach, one investigates  the dependence of the loop amplitudes on the dimensionality of spacetime, which turns out to be polynomial in pure Yang-Mills theory. Then one computes the amplitudes with virtual particles kept in integer dimension  $d > 4$  to fix the coefficients in the polynomial by interpolation, which leads by analytic continuation to an expression valid for any non-integer dimension $d$. The dimensional reconstruction approach  can  also be effectively  combined with the spinor-helicity formalism in six dimensions of \cite{Cheung:2009dc},  which allows for compact expressions of the on-shell building blocks. 
At higher loops, these techniques were used in   \cite{Badger:2013gxa} to derive the five-point all-plus gluon amplitude integrand in pure Yang-Mills, while a generalisation  to incorporate fermions was  carried out in \cite{Anger:2018ove}. Recent numerical as well as analytical results for arbitrary helicity configurations of five partons   were derived in \cite{Abreu:2018jgq,Badger:2018enw,Abreu:2018zmy}.
In this framework, a systematic prescription for complete  form factors, including rational terms, is still missing, and bridging this gap is one of the main goals of this paper.

A  form factor $F_{\cO}(1,\ldots , n; q)$ is defined as the overlap of an $n$-particle state and the state produced by an
operator $\cO(x)$ acting on the vacuum:
\begin{equation}
	\label{eq::FormFactorDefinition}
	\int \dd[4]{x} e^{-i q \cdot x} \langle 1,  \ldots , n | \cO (x) | 0 \rangle = (2\pi)^4 \delta^{(4)} \Big(q-\sum_{i=1}^{n} p_{i} \Big) F_{\cO} (1, \ldots , n; q)\ .
\end{equation}
Notable examples of   form factors include the form factor of the hadronic electromagnetic current with an external hadronic state, which feature in the   $e^{+} e^{-} \rightarrow $ hadrons and deep inelastic scattering matrix elements, and  as the form factor of the electromagnetic current, which computes the (electron) $g-2$.
The  form factors  which will be  considered in this paper
are related to scattering processes of the Higgs boson and many gluons.
In the large top-quark mass approximation, these  can be described by an effective theory obtained by integrating out  the top quark in QCD. This  generates an infinite series of higher-dimensional  interactions of the   Higgs  with the gluon field strength and its derivatives, in addition to couplings to light quarks.
More precisely, this effective Lagrangian reads
\begin{align}
	\label{one}
	\cL_{\rm eff} \, = \, \hat{C}_0 \mathcal{O}_0 \, + \, \frac{1}{m_{t}^2} \sum_{i=1}^4 \hat{C}_i \mathcal{O}_i \, + \, \mathcal{O}
	\left(\frac{1}{m_{t}^4} \right)\ ,
\end{align}
where the leading-order term in the expansion is $\mathcal{O}_0 \coloneqq H \, \Tr F^2$ \cite{Wilczek:1977zn,Shifman:1979eb,Buchmuller:1985jz,Dawson:1990zj},
$\mathcal{O}_i$, $i=1, \ldots , 4$ are dimension-7 operators made of gluon field strengths and covariant derivatives
\cite{Neill:2009tn, Neill:2009mz, Harlander:2013oja, Dawson:2014ora},  $m_t$ is the mass of the top quark, and 
$\hat{C}_0$, $\hat{C}_i$ are   Wilson  coefficients.%
\footnote{The Wilson coefficients are proportional to $1/v$, where $v$ is the Higgs field  vacuum expectation value. Their precise form will be of no relevance  for this paper.}
After Wick-contracting the Higgs field, what is left to compute is precisely a form factor of partons in the theory of interest, which we will take to be pure Yang-Mills.
It is also interesting to note that at zero momentum transfer, \textit{i.e.}~$q = 0$ in \eqref{eq::FormFactorDefinition}, the form factor of an operator $\cO(x)$ represents the correction to the scattering amplitude due to the inclusion  of a new local interaction proportional to $\cO(x)$. On the other hand, in the study of Higgs + gluon processes one  replaces $q^2$ with the squared mass of the Higgs $m_{\rm H}^2$ to obtain the  amplitudes relevant for this process.

In this paper we will apply the approach discussed so far  to form factors of operators of the form $\Tr F^n$, for $n=2,3,4$, both for minimal and non-minimal form factors up to four external gluons. 
Modern amplitude techniques were applied to form factors  of $\Tr F^2$, which  compute the leading contribution to  Higgs + multi-gluon amplitudes in the  effective Lagrangian approach, including MHV diagrams \cite{Cachazo:2004kj,Brandhuber:2004yw}
at tree level \cite{Dixon:2004za,Badger:2004ty} and one loop \cite{Badger:2007si}, and a combination of one-loop MHV diagrams and recursion relations  \cite{Badger:2009hw}.
Recent work \cite{Brandhuber:2017bkg,Jin:2018fak,Brandhuber:2018xzk,Brandhuber:2018kqb,Jin:2019ile} addressed the computation of  the four-dimensional cut-constructible part of Higgs+multi-gluon scattering from operators of mass dimension seven using generalised unitarity
\cite{Bern:1997sc, Britto:2004nc} applied to form factors 
\cite{Brandhuber:2010ad, Brandhuber:2011tv, Gehrmann:2011xn,Brandhuber:2012vm,Penante:2014sza,Brandhuber:2014ica,Nandan:2014oga,Loebbert:2015ova,Brandhuber:2016fni,Loebbert:2016xkw,  Brandhuber:2018xzk,Brandhuber:2018kqb, 
Bianchi:2018peu,Bianchi:2018rrj}.
The key point of this work is that we extend dimensional reconstruction to any form factor of operators involving vector fields, which requires the subtraction of form factors of an  appropriate class of scalar operators that we identify. Along the way we have also found a generalisation of this procedure to any loop order,  for amplitudes and form factors.

Loop-level calculations are inherently difficult, and no matter how effective the method or how simple the formalism are, sooner or later the complexity of the problems one wishes to tackle will require the use of computer software to deal with the algebra. For this reason we have written a \texttt{Mathematica} package that can handle quantities and perform computations in the six-dimensional spinor-helicity formalism. This package has been inspired by implementations of the analogous four-dimensional formalism in  \cite{Maitre:2007jq,Dixon:2010ik,Panerai2019git} and features analytic as well as numerical~tools.

The rest of the paper is organised as follows.
In Section \ref{sec::GenUni6D} we review the dimensional reconstruction technique at one loop and generalise it to form factors involving vector fields.  We also discuss its generalisation  to any number of loops, which for one and two loops 
 is in agreement  with known results. 
In Section~\ref{sec::tree} we study tree-level form factors  for a wide class of operators involving  field strengths  in four and six dimensions. These quantities are needed in the one-loop unitarity-based calculations  of
Section~\ref{sec::OneLoop}.
There, we begin by  reproducing the well-know one-loop form factors for $\Tr F^2$ with two and three external gluons. Then we prove that the minimal form factor for $\Tr F^3$ has no rational terms, as argued in the literature. Finally, we calculate for the first time the non-minimal one-loop form factor for $\Tr F^3$ and the minimal form factor for $\Tr F^4$ with different helicity configurations. We also generalise some of these results for a class of form factors of the $\Tr F^n$ operators.
A few appendices complete the paper. In the first three we review the spinor-helicity formalism in four and six dimensions, as well as  the structure of six-dimensional tree amplitudes. Appendix \ref{sec:nonminimal} contains detailed calculations of non-minimal tree-level form factors
used as building blocks of loop amplitudes, while Appendix \ref{sec:integrals} lists some useful results on integral functions. Finally,
Appendix \ref{Mathematica} contains a short description of the  \texttt{SpinorHelicity6D} \texttt{MATHEMATICA} package we have used in our numerical calculations, focusing on the  functions required to replicate our results.

\section{The Dimensional Reconstruction Technique}
\label{sec::GenUni6D}

In the first part of the section, we look at the one-loop case from a different perspective which lends itself to a systematic generalisation to form factors. The new viewpoint we adopt presents also a much desirable advantage: it disentangles the number of dimensions in which amplitudes need to be computed from the loop order. This feature allows for a natural generalisation to any loop order, for both amplitudes and form factors, which will be discussed in the second part of the section.

\subsection{One-Loop Dimensional Reconstruction}

The first step in our study is to identify the dependence of the loop amplitude on the dimensionality of the spacetime. In the literature, a common procedure is to distinguish the two sources of this dependence:
\begin{itemize}
	\item the first is the number of spin-eigenstates, which is a function of the dimension of the spacetime $d_s$ (for example, gluons have $d_s-2$ spin degrees of freedom);
	\item the second is the integration over the loop momentum, which lives in a $d$-dimensional space.
\end{itemize}
Specifically, in the following we consider   pure Yang-Mills theory
\begin{equation}
	\cL_{d_s} = -\frac{1}{4} \big(F^{a}_{\mu \nu}F^{a \mu \nu}\big) (x)\ ,
\end{equation}
where $A^{a \mu}$ is a vector in a spacetime of dimension $d_s$ and $x$ is defined on a spacetime of dimension $d>4$.

As we mentioned earlier, we are interested in calculating amplitudes (and form factors) involving four-dimensional external gluons.
At one loop it is possible to write a general amplitude as
\begin{equation}
	\cA_{(d_s,d)}^{(1)}\,(\{p_i,h_i\})=\int\frac{\dd[d]{l}}{(2\pi)^{d}} \frac{\cN^{d_s}(\{p_i,h_i\})}{\prod_{i} d_i}\ ,
\end{equation}
where $\cN^{d_s}(\{p_i,h_i\})$ depends on $d_s$ through the number of spin eigenstates and on $d$ through the loop momentum. However, since all external momenta are four-dimensional, the additional components of the loop momentum enter the amplitude only through its square, which can always be written as
\begin{equation}
	\label{mu2}
	l^2 =l_0^2-l_1^2-l_2^2-l_3^2-\sum_{i=4}^{d-1}l_i^2 \coloneqq (l^{(4)})^2-\mu^2 \>.
\end{equation}

The dependence of the amplitude on $\mu^2$ manifests itself in a number of additional integrals with non-trivial numerators, which have to be added to the usual master integral basis. These integrals have the form:
\begin{equation}\label{eq:muintegrals}
	\int\frac{\dd[d]{l}}{(2 \pi)^{d}}\frac{\mu^{2 p}}{d_1\cdots d_n} \coloneqq I^d_{n}[\mu^{2 p}]\ ,
\end{equation}
which can be evaluated as ordinary integrals, but in higher dimensions \cite{Bern:1995db}. The presence of these integrals cannot be probed using  four-dimensional unitarity cuts.

Consider now the explicit dependence of the amplitude on $d_s$. One-loop amplitudes involving only bosons are {\it linear} in $d_s$, because it appears only in a closed loop of contracted metric tensors coming from vertices and propagators. Consequently, in order to determine the dependence of the amplitude on $d_s$, only two constants need to be fixed and these can be obtained by interpolation. Thus it is sufficient to compute the amplitude in two integer dimensions, for example $d_{0}$ and $d_{1}=d_{0}+1$, and then write the analytic continuation to four dimensions in the Four Dimensional Helicity (FDH) scheme \cite{Bern:1991aq,Bern:2002zk}. The result of the interpolation is given by~\cite{Giele:2008ve}:

\begin{equation}
	\label{eq:1loopdimreconstruction}
	\cA^{(1)}_{(4,d)}=(d_1-4) \cA^{(1)}_{(d_0,d)} -(d_0-4) \cA^{(1)}_{(d_1,d)}\ . %- 2\epsilon \left( \cA_{(5,d)} - \cA_{(6,d)} \right) \>.
\end{equation}

By definition $d$ are the dynamical dimensions of the theory and we can always choose $d_0 \geq d$. By doing so we can consider the extra dimension in the $d_1$-dimensional space as non-dynamical. Then a $d_1$-dimensional gluon behaves as a $d_0$-dimensional one plus a real scalar $A^{a \mu} = \big(A^{a \hat{\mu}}, \phi^a \big)$, and the Lagrangian of the system reads\footnote{The fields are non-dynamical in the $d_1$-dimensional direction of the space-time, thus we can set $\partial_{d_0}\!\!~A^{a \mu} =~0$ and $\partial_{d_0} \phi^{a}= 0$ ($\partial_{d_0} = \partial_{d_1 -1}$).}
\begin{equation}
	\label{KaluzaKlein}
	\cL_{d_i} = -\frac{1}{4} F^{a}_{\mu \nu} F^{a \mu \nu} = -\frac{1}{4} F^{a}_{\hat{\mu} \hat{\nu}} F^{a \hat{\mu} \hat{\nu}} + \frac{1}{2} D_{\hat{\mu}} \phi^a D^{\hat{\mu}} \phi^a\> ,
\end{equation}
where the hatted quantities refer to $d_0$-dimensional Lorentz indices. From the Lagrangian \eqref{KaluzaKlein} we arrive at the conclusion that the one-loop $d_i$-dimensional amplitude $\cA_{(d_i,d)}$ can be expressed as the sum of two contributions: the first contribution is given by the equivalent one-loop gluon amplitude with internal particles living in $d_0$ dimensions $\cA_{(d_0,d)}$; the second one, denoted in the following as $\cA_{(d)}^S$, takes into account also scalar interactions coming from the second term on the right-hand side of \eqref{KaluzaKlein}.
It is also important to stress that $\cA_{(d)}^S$ is a gauge-invariant quantity in its own right. As a result of these observations, \eqref{eq:1loopdimreconstruction} can be written as:
\begin{equation}
	\label{eq::dimensional_reconstruction}
	\cA^{(1)}_{(4,d)}= \cA^{(1)}_{(d_0,d)}- (d_0-4) \cA_{(d)}^S
	\ .
	%+ 2\epsilon \cA_{(d)}^S\ .
\end{equation}
Since we are considering only the one-loop order, it is easy to see that $\cA_{(d)}^S$ can be obtained by trading the gluon loop for a scalar loop.

Up to some additional considerations, the above discussion holds true for form factors as well, and so does  \eqref{eq::dimensional_reconstruction}. In particular, the scalar quantity that we have to subtract from the form factor with $d_0$-dimensional internal gluons is obtained by trading the gluon loop with a scalar one.
However, in contradistinction with the amplitude case, there are two sources for scalars when we are dealing with form factors. Inside the loop, one can have scalars coupled to gluon lines coming from terms of the form $\frac{1}{2} D_{\mu}\phi^a D^{\mu}\phi^a $ in the dimensionally-reduced Lagrangian (as in the case of amplitudes), but also scalars coming from the operator inserted in the form factor. This procedure will be clear in the calculation of the non-minimal $\Tr F^2$ form factor, described in Section~\ref{sec:TrF2nonmin}, where we will emphasise the role of these two distinct contributions (see also \cite{Davies:2011vt}).

Finally, what we need is to identify the scalar operator. The procedure we follow is reminiscent of dimensional reduction, which for the operator $\Tr F^2$ was performed in \eqref{KaluzaKlein}. From this new point of view, the generalization of the Dimensional Reconstruction technique to other operators is straightforward. In particular, for the only two operators with mass-dimension six involving solely gluons, namely $\Tr (D F)^2$ and $\Tr F^3$, the scalar contribution comes from
\begin{equation}
	\label{scalarDF2}
	D_{\mu} F^a_{\nu \rho} D^{\mu} F^{a \nu \rho} \mapsto D_{\mu} D_{\nu} \phi^a D^{\mu} D^{\nu} \phi^a\ ,
\end{equation}
and
\begin{equation}
	\label{scalarF3}
	f^{a b c} F^{a\, \mu}\,_{\nu} F^{b\, \nu}\,_{\rho} F^{c\, \rho}\,_{\mu} \mapsto f^{a b c} D_{\mu} \phi^a D_{\nu} \phi^b F^{c\, \mu \nu}\ ,
\end{equation}
where scalar operators associated to each operator come from the dimensional reduction from $d_1$ to $d_0$. On the other hand for the $\Tr F^4$ operator, which we will consider later in this paper, at one-loop we get
\begin{equation}
	\label{scalarF4}
	\Tr F^{\mu}\,_{\nu} F^{\nu}\,_{\rho} F^{\rho}\,_{\sigma} F^{\sigma}\,_{\mu} \mapsto \Tr D_{\mu} \phi\, D_{\nu} \phi\, F^{\nu}\,_{\rho}\, F^{\rho \mu}\ ,
\end{equation}
where in the last equation the trace is in colour space\footnote{We emphasise that \eqref{scalarF4} is exactly the scalar operator up to an overall factor, which has still to be fixed. In particular, the two scalars in the previous operator have to be adjacent, because the gluon operator involves only contractions between adjacent field strength.}. The proportionality  coefficients are still to be fixed and we will give the right prescription for them within the full tree-level calculation in Section \ref{sec::tree}.

\subsection{An \texorpdfstring{$L$}{L}-loop Generalisation}
\label{sec::Lloop}

The arguments leading to   \eqref{eq:1loopdimreconstruction} can be extended to arbitrary loop order. Considering pure Yang-Mills theory, any $L$-loop amplitude can be written as a degree $L$ polynomial in the dimension $d_s$\footnote{As already mentioned, the $d_s$ dependence comes from traces of $\eta$ tensors, and there can be at most $L$ closed loops leading to such a trace.},
\begin{equation}
	\label{eq::Lloopdimreconstruction}
	\mathcal{A}^{(L)}_{(d_s,d)}=\sum_{i=0}^{L} (d_s - 4)^i \mathcal{K}_{i}\ ,
\end{equation}
where $\mathcal{K}_i$ are quantities to be determined. In particular, note that the four-dimensional amplitude in the FDH scheme \cite{Bern:1991aq,Bern:2002zk} coincides with the zero-degree coefficient: $\mathcal{K}_{0} = \mathcal{A}^{(L)}_{(4,d)}$. In order to find the coefficients $\mathcal{K}_i$, we can interpolate the polynomial in $L+1$ distinct integer dimensions $d_i >4$. Writing the problem in matrix form, one has
\begin{equation}
	\def\arraystretch{1.5}
	\begin{pmatrix}
		\mathcal{A}^{(L)}_{(d_0,d)} \\
		\mathcal{A}^{(L)}_{(d_1,d)} \\
		\> \vdots                   \\
		\mathcal{A}^{(L)}_{(d_L,d)} \\
	\end{pmatrix}
	=
	\begin{pmatrix}
		1      & (d_0-4)   & (d_0-4)^2   & \cdots & (d_0-4)^{L}   \\
		1      & (d_1-4)   & (d_1-4)^2   & \cdots & (d_1-4)^{L}   \\
		\vdots &           &             &        & \vdots        \\
		1      & (d_{L}-4) & (d_{L}-4)^2 & \cdots & (d_{L}-4)^{L} \\
	\end{pmatrix}
	\begin{pmatrix}
		\mathcal{K}_{0} \\
		\mathcal{K}_{1} \\
		\vdots          \\
		\mathcal{K}_{L} \\
	\end{pmatrix}\, ,
\end{equation}
where we recognise the Vandermonde matrix.  Inverting  this  matrix, it is possible to express the $\mathcal{K}_{i}$ as functions of the higher-dimensional amplitudes $\mathcal{A}_{(d_i,d)}^{(L)}$ for $i=0,\ldots,L$. In particular $\mathcal{K}_0$, which is the four-dimensional amplitude we are interested in, can be written as
\begin{equation}
	\label{eq::Lloop_4Damplitude}
	\cA_{(4,d)}^{(L)}=\mathcal{K}_0 = \prod_{j=0}^{L} (d_j -4) \sum_{i=0}^{L} \frac{1}{(d_{i} -4)} \prod_{\substack{k=0 \\ k\neq i}}^{L} \frac{1}{(d_k - d_i)}\cA_{(d_i ,d)}^{(L)}\ .
\end{equation}

We can always choose $d_0 > 4$ to be the smallest dimension among the $d_i$'s, and we also know that at most $d$ dimensions are dynamical, with $4 < d \leq d_0$. Then, we can write the Lagrangian of pure Yang-Mills theory in $d_i > d_0$ dimensions as:
\begin{equation}
	\label{eq::Ds_Lagrangian}
	\cL_{d_i} = -\frac{1}{4} F^{a}_{\mu \nu} F^{a \mu \nu} + \frac{1}{2}\sum_{i=1}^{d_i-d_0} D_{\mu} \phi_i^a D^{\mu} \phi_i^a -\frac{\lambda}{2} f^{a b c} f^{a d e} \sum_{\substack{i,j=1 \\ j>i}}^{d_i-d_0} \phi_{i}^{b} \phi_{j}^{c} \phi_{i}^{d} \phi_{j}^{e}\ ,
\end{equation}
where $\mu, \nu$ are $d_0$-dimensional Lorentz indices, $a, b, c$ are colour indices and $f^{a b c}$ are the structure constants of the gauge group. The vector field in $d_i$ dimensions is decomposed in a ($d_0$-dimensional) vector $A^{a}_{\mu}$ and $d_i - d_0$ scalars $\phi^{a}_{i}$. The coupling of the $\phi^4$ interaction is given by
\begin{equation}
	\lambda = g^2\ ,
\end{equation}
and we call it $\lambda$ for reasons that will be clear in a moment.

From \eqref{eq::Ds_Lagrangian}, we can compute the amplitude with only external gluons%
\footnote{In this section, for the sake of clarity, we reserve the word  \textit{vector} only for the $d_0$-dimensional vector, whereas we refer to the four-dimensional equivalents as gluons.}
\begin{equation}
	\label{eq::amplitude_di_dimensions}
	\cA_{(d_i,d)}^{(L)} = \cA_{(d_0,d)}^{(L)} + \sum_{m=0}^{L-1} (d_i-d_0-1)^m \sum_{n=1}^{L-m} (d_i -d_0)^{n} \cA^{S\, (L)}_{(d_0,d,n,m)}\ ,
\end{equation}
where $A_{(d_0,d)}^{(L)}$ is the complete $L$-loop amplitude where all the internal legs are vectors and $\cA^{S\, (L)}_{(d_0,d,n,m)}$ are specific combinations of diagrams with at least one scalar loop. Specifically, the diagrams  contributing to $\cA^{S\, (L)}_{(d_0,d,n,m)}$ are of order $\lambda^m$, \textit{i.e.}~they contain $m$ four-scalar interactions, and in addition have $n$ distinct purely scalar subdiagrams.

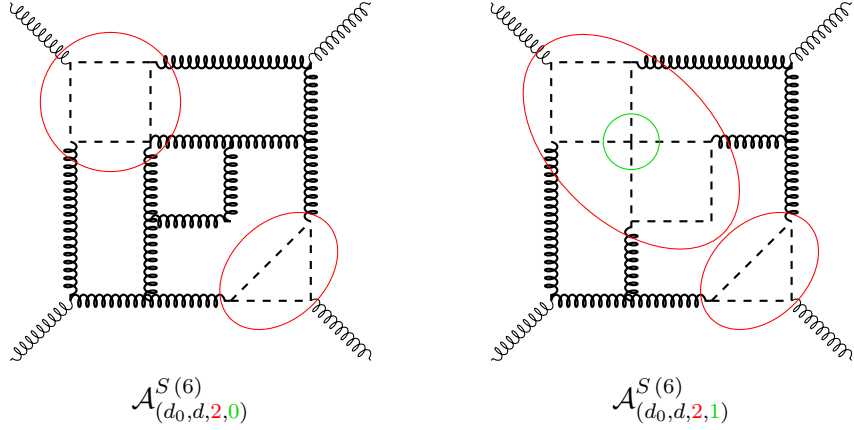
\begin{figure}
	\centering
	\begin{tikzpicture}[scale=15]
		\def\x{0}
		\def\y{0}

		%large square
		\draw [decorate, decoration={coil, amplitude=2.3pt, segment length=3pt},thick](0+\x,0+\y) -- (4pt+\x,0+\y);
		\draw [dashed,thick](4pt+\x,0pt+\y) -- (6pt+\x,0pt+\y);
		\draw [dashed,thick](6pt+\x,0+\y) -- (6pt+\x,2pt+\y);
		\draw [decorate, decoration={coil, amplitude=2.3pt, segment length=3pt},thick](6pt+\x,2pt+\y) -- (6pt+\x,6pt+\y);
		\draw [decorate, decoration={coil, amplitude=2.3pt, segment length=3pt},thick](6pt+\x,6pt+\y) -- (2pt+\x,6pt+\y);
		\draw [dashed,thick](2pt+\x,6pt+\y) -- (0+\x,6pt+\y);
		\draw [dashed,thick](0+\x,6pt+\y) -- (0+\x,4pt+\y);
		\draw [decorate, decoration={coil, amplitude=2.3pt, segment length=3pt},thick](0+\x,4pt+\y) -- (0+\x,0+\y);
		%external lines
		\draw [decorate, decoration={coil, amplitude=2.3pt, segment length=3pt}](0pt+\x,0pt+\y) -- +(-1.5pt,-1.5pt);
		\draw [decorate, decoration={coil, amplitude=2.3pt, segment length=3pt}](6pt+\x,0+\y) -- +(1.5pt,-1.5pt);
		\draw [decorate, decoration={coil, amplitude=2.3pt, segment length=3pt}](6pt+\x,6pt+\y) -- +(1.5pt,1.5pt);
		\draw [decorate, decoration={coil, amplitude=2.3pt, segment length=3pt}](0pt+\x,6pt+\y) -- +(-1.5pt,1.5pt);
		%small square
		\draw [decorate, decoration={coil, amplitude=2.3pt, segment length=3pt},thick](2pt+\x,2pt+\y) -- (4pt+\x,2pt+\y);
		\draw [decorate, decoration={coil, amplitude=2.3pt, segment length=3pt},thick](4pt+\x,2pt+\y) -- (4pt+\x,4pt+\y);
		%connecting lines
		\draw [dashed,thick](4pt+\x,0pt+\y) -- (6pt+\x,2pt+\y);
		\draw [decorate, decoration={coil, amplitude=2.3pt, segment length=3pt},thick](2pt+\x,0pt+\y) -- (2pt+\x,4pt+\y);
		\draw [decorate, decoration={coil, amplitude=2.3pt, segment length=3pt},thick](2pt+\x,4pt+\y) -- (6pt+\x,4pt+\y);
		\draw [dashed,thick](2pt+\x,4pt+\y) -- (2pt+\x,6pt+\y);
		\draw [dashed,thick](2pt+\x,4pt+\y) -- (0pt+\x,4pt+\y);

		%Circles
		\draw [red] (1pt+\x,5pt+\y) circle (1.75pt);
		\draw [red, rotate around={45:(5.2pt+\x,0.75pt+\y)}] (5.2pt+\x,0.75pt+\y) ellipse (1.7pt and 1.2pt);

		\node at (3pt+\x,-2.5pt+\y){$\mathcal{A}^{S\, (6)}_{(d_0,d,\textcolor{red}{2},\textcolor{green!85!black}{0})}$};

		\def\x{12pt}
		\def\y{0}

		%large square
		\draw [decorate, decoration={coil, amplitude=2.3pt, segment length=3pt},thick](0+\x,0+\y) -- (4pt+\x,0+\y);
		\draw [dashed,thick](4pt+\x,0pt+\y) -- (6pt+\x,0pt+\y);
		\draw [dashed,thick](6pt+\x,0+\y) -- (6pt+\x,2pt+\y);
		\draw [decorate, decoration={coil, amplitude=2.3pt, segment length=3pt},thick](6pt+\x,2pt+\y) -- (6pt+\x,6pt+\y);
		\draw [decorate, decoration={coil, amplitude=2.3pt, segment length=3pt},thick](6pt+\x,6pt+\y) -- (2pt+\x,6pt+\y);
		\draw [dashed,thick](2pt+\x,6pt+\y) -- (0+\x,6pt+\y);
		\draw [dashed,thick](0+\x,6pt+\y) -- (0+\x,4pt+\y);
		\draw [decorate, decoration={coil, amplitude=2.3pt, segment length=3pt},thick](0+\x,4pt+\y) -- (0+\x,0+\y);
		%external lines
		\draw [decorate, decoration={coil, amplitude=2.3pt, segment length=3pt}](0pt+\x,0pt+\y) -- +(-1.5pt,-1.5pt);
		\draw [decorate, decoration={coil, amplitude=2.3pt, segment length=3pt}](6pt+\x,0+\y) -- +(1.5pt,-1.5pt);
		\draw [decorate, decoration={coil, amplitude=2.3pt, segment length=3pt}](6pt+\x,6pt+\y) -- +(1.5pt,1.5pt);
		\draw [decorate, decoration={coil, amplitude=2.3pt, segment length=3pt}](0pt+\x,6pt+\y) -- +(-1.5pt,1.5pt);
		%small square
		\draw [dashed,thick](2pt+\x,2pt+\y) -- (4pt+\x,2pt+\y);
		\draw [dashed,thick](4pt+\x,2pt+\y) -- (4pt+\x,4pt+\y);
		\draw [dashed,thick](4pt+\x,4pt+\y) -- (2pt+\x,4pt+\y);
		\draw [dashed,thick](2pt+\x,4pt+\y) -- (2pt+\x,2pt+\y);
		%connecting lines
		\draw [dashed,thick](4pt+\x,0pt+\y) -- (6pt+\x,2pt+\y);
		\draw [decorate, decoration={coil, amplitude=2.3pt, segment length=3pt},thick](2pt+\x,0pt+\y) -- (2pt+\x,2pt+\y);
		\draw [decorate, decoration={coil, amplitude=2.3pt, segment length=3pt},thick](4pt+\x,4pt+\y) -- (6pt+\x,4pt+\y);
		\draw [dashed,thick](2pt+\x,4pt+\y) -- (2pt+\x,6pt+\y);
		\draw [dashed,thick](2pt+\x,4pt+\y) -- (0pt+\x,4pt+\y);

		%Circles
		\draw [red, rotate around={-45:(2pt+\x,4pt+\y)}] (2pt+\x,4pt+\y) ellipse (3.25pt and 2pt);
		\draw [red, rotate around={45:(5.2pt+\x,0.75pt+\y)}] (5.2pt+\x,0.75pt+\y) ellipse (1.7pt and 1.2pt);
		\draw [green!85!black] (2pt+\x,4pt+\y) circle (0.7pt);

		\node at (3pt+\x,-2.5pt+\y){$\mathcal{A}^{S\, (6)}_{(d_0,d,\textcolor{red}{2},\textcolor{green!85!black}{1})}$};

	\end{tikzpicture}
	\caption{Two of the many possible diagrams contributing to the scalar amplitudes at six loops. On the left-hand side an example contribution to $\mathcal{A}^{S\, (6)}_{(d_0,d,2,0)}$ is shown. On the right-hand side the same diagram but with one of the gluon loops involving a four-point interaction replaced by a scalar. The latter diagram contributes to $\mathcal{A}^{S\, (6)}_{(d_0,d,2,1)}$.}
\end{figure}

The coefficients for the scalar contributions in \eqref{eq::amplitude_di_dimensions} can be understood as follows.
\begin{enumerate}
	\item The number of distinct flavours of scalars is $d_i - d_0$ and they all give the same contribution.
	\item Given a set of contiguous scalar propagators inside a diagram, when we draw the first scalar propagator, we need to multiply the 	diagram by a $d_i - d_0$ factor, corresponding to the distinct possible flavours.
	\item Inside the same set of contiguous scalar propagators, each vertex with two scalars and one vector must preserve the scalar flavour, while the four-scalar vertex changes it. Thus each power of $\lambda$ brings a $d_i - d_0 - 1$ factor.
	\item Every distinct set of scalar propagators leads to an additional $d_i - d_0$ factor.
	\item Since there are no external scalars, the number of distinct sets of scalar propagators plus the number of scalar quartic interactions coincides with the number of scalar loops:
	      \begin{equation}
		      n+m= \ \# \ \text{scalar loops}
	      \end{equation}
	\item Clearly the number of scalar loops can be at most $L$.
\end{enumerate}

\begin{figure}
	\centering
	\begin{tikzpicture}[scale=15]

		\def\x{0}
		\def\y{0}

		\node at (0+\x,0+\y) [draw, circle, dashed,thick, inner sep=14pt] (circ1) {};
		\node at (0+\x,0+\y) {\small $d_i-d_0$};
		\node at (0+\x,4.7pt+\y) [draw, circle,thick, dashed, inner sep=14pt] (circ2) {};
		\node at (0+\x,4.7pt+\y) {\small $d_i-d_0$};
		\draw [decorate, decoration={coil, amplitude=2.3pt, segment length=3pt},thick] (circ1) -- (circ2);
		\draw [decorate, decoration={coil, amplitude=2.3pt, segment length=3pt}] (circ1.east) -- +(2pt,0);
		\draw [decorate, decoration={coil, amplitude=2.3pt, segment length=3pt}] (circ2.east) -- +(2pt,0);
		\draw [decorate, decoration={coil, amplitude=2.3pt, segment length=3pt}] (circ1.west) -- +(-2pt,0);
		\draw [decorate, decoration={coil, amplitude=2.3pt, segment length=3pt}] (circ2.west) -- +(-2pt,0);

		%\node at (6pt+\x,2.35pt+\y) {$\simeq (d_i-d_0)^2$};

		\def\x{9pt}
		\def\y{0}

		\node at (0+\x,0+\y) (uno){};
		\node at (3pt+\x,0+\y) (due) {};
		\node at (3pt+\x,4.7pt+\y) (tre){};
		\node at (0+\x,4.7pt+\y) (quattro){};

		\draw [dashed,thick] (uno.center) -- (due.center) -- (quattro.center) -- (tre.center) -- (uno.center);

		\draw [decorate, decoration={coil, amplitude=2.3pt, segment length=3pt}] (due.center) -- +(1.5pt,0);
		\draw [decorate, decoration={coil, amplitude=2.3pt, segment length=3pt}] (tre.center) -- +(1.5pt,0);
		\draw [decorate, decoration={coil, amplitude=2.3pt, segment length=3pt}] (uno.center) -- +(-1.5pt,0);
		\draw [decorate, decoration={coil, amplitude=2.3pt, segment length=3pt}] (quattro.center) -- +(-1.5pt,0);

		\node at (1.5pt+\x,2.35pt+\y) [draw, circle, green!85!black, inner sep=5pt] (point){};
		%\node at (8pt+\x,4pt+\y)[draw, ellipse, green!85!black](comment){\tiny \textcolor{black}{flavour changing}};
		%\draw [->, green!85!black] (comment.west) -- (point);
		%\node at (0+\x,0+\y) [draw, circle, blue,inner sep=5pt] {};
		%\node at (0+\x,4.7pt+\y) [draw, circle, blue,inner sep=5pt] {};
		%\node at (3pt+\x,0+\y) [draw, circle, blue,inner sep=5pt] {};
		%\node at (3pt+\x,4.7pt+\y) [draw, circle, blue,inner sep=5pt] {};
		% \node at (8.5pt+\x,2.35pt+\y) {$\simeq (d_i-d_0)(d_i-d_0-1)$};
		\node at (1.5pt+\x,-0.8pt+\y) {\small $d_i-d_0-1$};
		\node at (1.5pt+\x,5.2pt+\y) {\small $d_i-d_0$};

	\end{tikzpicture}
	\caption{Two two-loop diagrams for comparison. In the first case there are two disconnected scalar loops, and every loop admits $d_i-d_0$ different flavours leading to an overall factor $(d_i-d_0)^2$. The second diagram represents two scalar loops connected by a flavour-changing four-scalar vertex (highlighted in green). In this case there are $d_i-d_0$ allowed flavours in one loop but only $d_i-d_0-1$ in the second loop, which leads to an overall factor $(d_i-d_0)(d_i-d_0-1)$.}
	\label{fig:fourverttwoloop}
\end{figure}
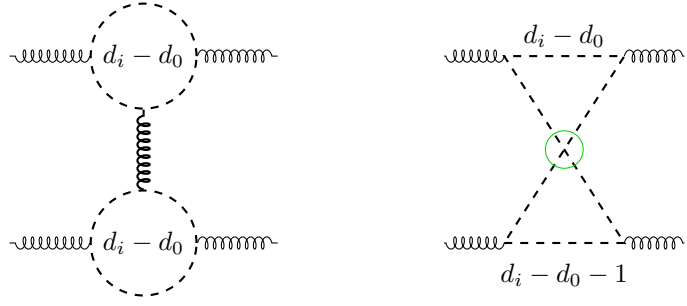

We can substitute \eqref{eq::amplitude_di_dimensions} in \eqref{eq::Lloop_4Damplitude} and, for simplicity, we choose
\begin{equation}
	\label{eq::chosen_dimensions}
	d_i = d_0 + i
\end{equation}
with $i=0,\ldots , L$. The final result should not depend on this choice, because the coefficient of a polynomial cannot depend on which point we choose for the fitting. After some manipulations, we find a closed expression which relates complete $L$-loop four-dimensional amplitudes to the same amplitudes in a higher integer dimension $d_0$ up to subtractions of scalar contribution:
\begin{equation}
	\label{eq::Lloopgen}
	\cA^{(L)}_{(4,d)}=\cA^{(L)}_{(d_0,d)} + \sum_{m=0}^{L-1} \sum_{n=1}^{L-m} (4-d_0)^{n} (3-d_0)^{m} \cA^{S\, (L)}_{(d_0,d,n,m)}\ ,
\end{equation}
where, in order to prove this formula, we have used the identity
\begin{equation}
	\sum_{i=0}^{L} \frac{1}{(d_{i} -4)} \prod_{\substack{k=0 \\ k\neq i}}^{L} \frac{1}{(d_k - d_i)} = \prod_{j=0}^{L} \frac{1}{(d_j -4)}\ .
\end{equation}

The whole reasoning can be applied to a more generic scheme where $d_s=4$ (FHV scheme) is replaced by a generic $d_s$ ({\it e.g.} the HV scheme \cite{tHooft:1972tcz} with $d_s=4-2\epsilon$). As long as we keep $d_i > d_s$ and $d_i \geq d$, all the previous steps are still applicable, and we arrive at
\begin{equation}
	\label{eq::Lloopgen_ds}
	\cA^{(L)}_{(d_s,d)}=\cA^{(L)}_{(d_0,d)} + \sum_{m=0}^{L-1} \sum_{n=1}^{L-m} (d_s-d_0)^{n} (d_s-d_0-1)^{m} \cA^{S\, (L)}_{(d_0,d,n,m)}\ ,
\end{equation}
which at first sight is identical to \eqref{eq::amplitude_di_dimensions}. The non-trivial difference between the two expressions is that $d_s < d_0$, while we need $d_i > d_0$ ($i=1,\ldots , L$) in order to write \eqref{eq::amplitude_di_dimensions}. Moreover, as we stressed before, the quantity $(d_i - d_0)$ has a precise physical meaning: it is the number of distinct flavours of scalars in the dimensional reduced theory \eqref{eq::Ds_Lagrangian}. On the other hand, $(d_s - d_0)$ takes into account the number of extra spin degrees of freedom in dimensional regularisation\footnote{There is no dynamics in the dimensions $d_i - d_0$, while this could be not true for the dimensions $d_0 -d_s$.}.

A posteriori, the fact that the two expressions are exactly the same is a consequence of our previous considerations. Indeed, one could have recognised the polynomial dependence of the amplitude on the dimensionality $d_s$ already from \eqref{eq::amplitude_di_dimensions}, and further identified \eqref{eq::Lloopgen_ds} as its analytic continuation for $d_s - d_0 < 0$. Thus, starting from the dimensionally reduced Lagrangian~\eqref{eq::Ds_Lagrangian}, the dependence on the dimensionality $d_s$ emerges naturally, and the preceding considerations relating the $d_i$ to $d_s$ through the Vandermonde matrix may appear redundant. However, starting from the analysis of the dimensional dependence of the amplitudes provides a clear physical picture of the relation between $d_s$, $d$ and $d_i$.

Our expression reproduces the known results at one loop \cite{Giele:2008ve}
\begin{equation}
	\cA^{(1)}_{(d_s,d)} = \cA^{(1)}_{(d_0,d)} + (d_s-d_0) \cA^{S\, (1)}_{(d_0,d,0,1)}\ ,
\end{equation}
and two loops \cite{Badger:2013gxa}
\begin{equation}
	\cA^{(2)}_{(d_s,d)} = \cA^{(2)}_{(d_0,d)} + (d_s-d_0) \Delta^{S}_{(d_0,d)} + (d_s-d_0)^2 \Delta^{2S}_{(d_0,d)}\ ,
\end{equation}
where
\begin{equation}
	\Delta^{S}_{(d_0,d)} = \cA^{S\, (2)}_{(d_0,d,0,1)}-\cA^{S\, (2)}_{(d_0,d,1,1)}\ , \hspace{1cm} \Delta^{2S}_{(d_0,d)}=\cA^{S\, (2)}_{(d_0,d,0,2)} + \cA^{S\, (2)}_{(d_0,d,1,1)}\ .
\end{equation}
Considering the two-loop expression in more detail, one sees that in \cite{Badger:2013gxa} the four-scalar vertex is interpreted in terms of three fictitious flavour contributions:
\begin{figure}[h]
	\centering
	\begin{tikzpicture}[scale=20,cross/.style={path picture={
							\draw[black]
							(path picture bounding box.south east) -- (path picture bounding box.north west) (path picture bounding box.south west) -- (path picture bounding box.north east);
						}}]

		\def\x{0}
		\def\y{0}

		\draw [dashed, thick] (-1pt+\x,-1pt+\y) -- (1pt+\x,1pt+\y);
		\draw [dashed, thick] (1pt+\x,-1pt+\y) -- (-1pt+\x,1pt+\y);

		% \node at (-1.3pt+\x,-1.3pt+\y) {$j$};
		% \node at (1.3pt+\x,-1.3pt+\y) {$j$};
		% \node at (-1.3pt+\x,1.3pt+\y) {$i$};
		% \node at (1.3pt+\x,1.3pt+\y) {$i$};

		% \node at (0+\x,-2.3pt+\y) {\footnotesize $\sim (d_s-d_0)(d_s-d_0-1)$};

		\node at (3pt+\x,0+\y) {\Large $\longrightarrow$};

		\def\x{6pt}
		\def\y{0}

		\draw [dashed, thick] (-1pt+\x,-1pt+\y) -- (1pt+\x,1pt+\y);
		\draw [dashed, thick] (1pt+\x,-1pt+\y) -- (-1pt+\x,1pt+\y);

		\node [draw, circle, fill, gray!65!white, inner sep=8.5pt] at (0+\x,0+\y) {};
		\draw [thick](0.5pt+\x,-0.5pt+\y) arc (45:135:0.7pt);
		\draw [thick](-0.5pt+\x,0.5pt+\y) arc (225:315:0.7pt);

		% \node at (-1.3pt+\x,-1.3pt+\y) {$j$};
		% \node at (1.3pt+\x,-1.3pt+\y) {$j$};
		% \node at (-1.3pt+\x,1.3pt+\y) {$i$};
		% \node at (1.3pt+\x,1.3pt+\y) {$i$};

		% \node at (0+\x,-2.3pt+\y) {\footnotesize $\sim (d_s-d_0)^2$};

		\node at (2.5pt+\x,0+\y) {\Large +};

		\def\x{11pt}
		\def\y{0}

		\draw [dashed, thick] (-1pt+\x,-1pt+\y) -- (1pt+\x,1pt+\y);
		\draw [dashed, thick] (1pt+\x,-1pt+\y) -- (-1pt+\x,1pt+\y);

		\node [draw, circle, fill, gray!65!white, inner sep=8.5pt] at (0+\x,0+\y) {};
		\draw [thick](0.5pt+\x,-0.5pt+\y) arc (225:135:0.7pt);
		\draw [thick](-0.5pt+\x,-0.5pt+\y) arc (-45:45:0.7pt);

		% \node at (-1.3pt+\x,-1.3pt+\y) {$i$};
		% \node at (1.3pt+\x,-1.3pt+\y) {$j$};
		% \node at (-1.3pt+\x,1.3pt+\y) {$i$};
		% \node at (1.3pt+\x,1.3pt+\y) {$j$};

		%\node at (0+\x,-2.3pt+\y) {\footnotesize $\sim (d_s-d_0)$};

		\node at (2.5pt+\x,0+\y) {\Large +};

		\def\x{16pt}
		\def\y{0}

		\draw [dashed, thick] (-1pt+\x,-1pt+\y) -- (1pt+\x,1pt+\y);
		\draw [dashed, thick] (1pt+\x,-1pt+\y) -- (-1pt+\x,1pt+\y);

		%\node at (0+\x,-2.3pt+\y) {\footnotesize $\sim (d_s-d_0)$};

		\node [draw, circle, fill, gray!65!white, inner sep=8.5pt] at (0+\x,0+\y) {};
		\draw [thick](0.5pt+\x,-0.5pt+\y) -- (-0.5pt+\x,0.5pt+\y);
		\draw [thick](0.5pt+\x,0.5pt+\y) -- (0.08pt+\x,0.08pt+\y);
		\draw [thick](-0.5pt+\x,-0.5pt+\y) -- (-0.08pt+\x,-0.08pt+\y);

		% \node at (-1.3pt+\x,-1.3pt+\y) {$j$};
		% \node at (1.3pt+\x,-1.3pt+\y) {$i$};
		% \node at (-1.3pt+\x,1.3pt+\y) {$i$};
		% \node at (1.3pt+\x,1.3pt+\y) {$j$};

	\end{tikzpicture}
	%\caption{Interpretation of the kinematic four-point vertex in terms of colour four-point vertices of \cite{Badger:2013gxa}.}
\end{figure}

The two continuous lines in the grey blob represent the colour flow inside the vertex. Considering Figure \ref{fig:fourverttwoloop} we see that, in our interpretation, the only diagram which at two loops involves this vertex contributes with a factor $(d_s-d_0)(d_s-d_0-1)$. However, splitting the vertex according to colour flow as above,  the contribution of the same  diagram can be attributed to terms containing   a factor $(d_s-d_0)^2$ as well as $(d_s-d_0)$.
Taking into account this different interpretation of the four-scalar vertex, the two methods perfectly match.

We emphasise that individually each $\cA^{S\, (L)}_{(d_0,d,n,m)}$ is a gauge-invariant quantity: indeed, we know that $\cA^{(L)}_{(d_0,d)}$ is gauge invariant and the same is true for $\cA^{(L)}_{(4,d)}$, regardless of the choice of $d_0$. Since the coefficients of the scalar contributions depend on $d_0$, the single $\cA^{S\, (L)}_{(d_0,d,n,m)}$ must be gauge invariant by themselves.

As in the case of the one-loop procedure, \eqref{eq::Lloopgen} can be applied also to form factors, as far as we bear in mind that more scalar operators are involved in higher-loop calculations, in addition to those entering already at one loop. These additional terms emerge clearly from \eqref{eq::Ds_Lagrangian}. Indeed, for the operator $\Tr F^2$, beyond one-loop calculations we also need to subtract the scalar contribution from the $\phi^4$ operator:
\begin{equation}
	F^{a}_{\mu \nu} F^{a \mu \nu} \mapsto g^2 f^{a b d} f^{a c d} \phi^{b} \widetilde{\phi}^{c} \phi^{d} \widetilde{\phi}^{e}\ ,
\end{equation}
where $\phi$ and $\widetilde{\phi}$ have to be scalars with different flavour. Its contribution has to be carefully taken into account in the subtraction with the right $d_s$-dependence. In particular, in the form factor equivalent of \eqref{eq::amplitude_di_dimensions}, its insertion brings a $(d_i-d_0)(d_i-d_0-1)$ coefficient, because of the flavour changing.

An equivalent reasoning is also valid for higher-dimensional operators. For example, from the dimensional reduction procedure of the $\Tr F^3$ operator, we find that the additional scalar operators entering higher-loop calculations are
\begin{equation}
	f^{a b c} F^{a\, \mu}\,_{\nu} F^{b\, \nu}\,_{\rho} F^{c\, \rho}\,_{\mu} \mapsto
	\begin{cases}
		 & g f^{a b c} f^{a d e} D_{\mu} \phi^{b} D^{\mu} \widetilde{\phi}^{c} \phi^{d} \widetilde{\phi}^{e}                                           \\
		 & g^3 f^{a b c} f^{a d e} f^{b f g} f^{c h i} \phi^d \widetilde{\phi}^{e} \phi^{f} \widehat{\phi}^{g} \widetilde{\phi}^{h} \widehat{\phi}^{i}
	\end{cases}\ ,
\end{equation}
where the former enters the calculation at two-loop level, while the latter from three loops. We stress that $\phi$, $\widetilde{\phi}$ and $\widehat{\phi}$ represent three different scalar flavours. Then, in the generalisation of \eqref{eq::amplitude_di_dimensions} to form factors, the insertion of the scalar operators bring a factor of $(d_i-d_0)(d_i-d_0-1)$ and $(d_i-d_0)(d_i-d_0-1)(d_i-d_0-2)$ respectively. Following the same procedure one can recover the scalar operators for $\Tr F^4$, which we do not write explicitly. 

In the following we are going to apply this technique to one-loop calculations for form factors. We will always choose $d_0 = 6$, due to the existence of a powerful Spinor Helicity Formalism in six dimensions \cite{Cheung:2009dc,Bern:2010qa}. 

A technical comment is in order here.
In performing loop calculations, initially we  treat the loop momenta as living in $d_0\!=\!6$ dimensions, instead of $d$. This procedure is well defined at the integrand level. Indeed, we know the functional dependence of the integrand on the $d-4$ components of the loop momenta, which appear only through rational combinations of $l_i^{(-2 \epsilon)} \cdot l_j^{(-2 \epsilon)}$ and $\mu_{i}^2$. Then, once we identify these combinations, we can treat the loop momenta as being $d$-dimensional and integrate over them\footnote{It is worth mentioning that in terms of the six-dimensional spinor components the quantity mentioned above reads as follows: $l_i^{(-2 \epsilon)} \cdot l_j^{(-2 \epsilon)} = \frac{1}{2} (m_{i} \widetilde{m}_{j} + m_{j} \widetilde{m}_{i})$ and $\mu_{i}^2 = m_{i} \widetilde{m}_{i}$.}.

\section{Tree-Level Form Factors}
\label{sec::tree}

In this section we will provide all the analytic expressions of the tree-level colour-ordered form factors required for loop calculations.

The tensorial structure of the field strength in four dimensions is given by the antisymmetric product of two vector representations
\begin{equation}
	\left(\frac{1}{2},\frac{1}{2}\right)\wedge \left(\frac{1}{2},\frac{1}{2}\right) = \left(1,0\right)\oplus\left(0,1\right)\ ,
\end{equation}
where we can choose each component to correspond to the helicity configurations $\pm 1$. We then define the {\it self-dual} component of the free field strength as\footnote{To clarify the abuse of nomenclature, the quantity it is the field strength in momentum space corresponding to a polarisation vector of given helicity.}
\begin{equation}
	\label{eq::selfdualpart}
	F_{{\rm SD}, \alpha \dot{\alpha} \beta \dot{\beta}} \coloneqq \lambda_{\alpha} \lambda_{\beta} \epsilon_{\dot{\alpha} \dot{\beta}}\ ,
\end{equation}
which has helicity $-1$ and transforms in the $(1,0)$ representation of the Lorentz group\footnote{We could have used as definition the following: $F_{{\rm SD}, \alpha \dot{\alpha} \beta \dot{\beta}} \coloneqq p_{\alpha \dot{\alpha}} \varepsilon^-_{\beta \dot{\beta}}-p_{\beta \dot{\beta}} \varepsilon^-_{\alpha \dot{\alpha}}= -\sqrt{2} \lambda_{\alpha} \lambda_{\beta} \epsilon_{\dot{\alpha} \dot{\beta}}$. As we can see the only difference is an overall $-\sqrt{2}$ factor.}. Then, the {\it anti-self-dual} component, transforming in the $(0,1)$ representation is
\begin{equation}
	\label{eq::antiselfdualpart}
	F_{{\rm ASD}, \alpha \dot{\alpha} \beta \dot{\beta}} = \epsilon_{\alpha \beta} \widetilde{\lambda}_{\dot{\alpha}} \widetilde{\lambda}_{\dot{\beta}}\ .
\end{equation}

In terms of SU$^*$(4) representations, the six-dimensional free field strength transforms in the $\mathbf{6}\wedge \mathbf{6} = \mathbf{15}$, which is the traceless part of $\mathbf{4} \otimes \bar{\mathbf{4}}$. Thus it can be written as
\cite{Dennen:2009vk}
\begin{equation}
	\label{eq::FieldStrength6D}
	F_{a \dot{a}}^{A B}\,_{C D} = \alpha \, \delta^{[A}_{[C} F_{a \dot{a}}\,^{B]}\,_{D]}\ ,
\end{equation}
where $\alpha$ is a numerical coefficient to be fixed and $F_{a \dot{a}}\,^{A}\,_{B}$ is such that $F_{a \dot{a}}\,^{A}\,_{A}=0$\footnote{$A,B,\ldots = 1,\ldots ,4$ are indices in the (anti)fundamental representation of SU$^*$(4) and $a,\dot{a}$ are indices of the six-dimensional little group (for a detailed discussion see Appendix \ref{sec:spinorhel6D}).}. In spinor helicity variables  this quantity is \cite{Dennen:2009vk}
\begin{equation}
	\label{fssix}
	F_{a \dot{a}}\,^{A}\,_{B} = \lambda_{a}^{A} \widetilde{\lambda}_{\dot{a} B}\ ,
\end{equation}
which is indeed traceless thanks to the six-dimensional Dirac equation \eqref{6DDirac}. Upon dimensionally reducing  \eqref{fssix}
down to four dimensions we match it with \eqref{eq::selfdualpart} and \eqref{eq::antiselfdualpart}, which  fixes the proportionality coefficient to be $\alpha = 2$.

\subsection{\texorpdfstring{$\Tr F^2$}{TrF2} Form Factors}

In this section we consider the operator
\begin{equation}
	\cO_2 \coloneqq F^{a}_{\mu \nu} F^{a \mu \nu}\ .
\end{equation}
In four dimensions $\cO_2$ splits naturally into the sum of the traces of the self-dual and the anti-self-dual components of the field strength:
\begin{equation}
	\Tr F^2 = \Tr F^2_{\rm SD} +\Tr F^2_{\rm ASD}\ .
\end{equation}
It is trivial to identify these two four-dimensional components of the colour-ordered form factor:
\begin{equation}
	\setlength{\jot}{7pt}
	\begin{split}
		\label{eq::four_dim_F2}
		&F^{(0)}_{\cO_2} (1^{+}, 2^{+}; q) = 2 \sqr{1}{2} \sqr{2}{1}\ ,\\
		&F^{(0)}_{\cO_2} (1^{-}, 2^{-}; q) = 2 \agl{1}{2} \agl{2}{1}\ .
	\end{split}
\end{equation}
On the other hand, the six-dimensional form factor is
\begin{equation}
	\label{eq::six_dim_F2}
	F^{(0)}_{\cO_2}(1_{a\dot{a}},2_{b \dot{b}};q) = 2 \asbr{1}{a}{2}{b} \asbr{2}{b}{1}{a}\ .
\end{equation}
Using the particular embedding of the four-dimensional  into the six-dimensional space introduced in Appendix~\ref{sec::from_six_to_four}  we find that%
\footnote{Four-dimensional limit here means choosing appropriate little-group indices corresponding to the desired helicity configuration in four dimensions, and the taking $m_{i},\tilde{m}_{i} \to 0$ for any particle $i$.}
\begin{equation}
	F^{(0)}_{\cO_2}(1_{1\dot{1}},2_{1 \dot{1}};q)\Big|_{\text{4D}}=F^{(0)}_{\cO_2}(1^+,2^+;q) \>, \hspace{1cm} F^{(0)}_{\cO_2}(1_{2\dot{2}},2_{2 \dot{2}};q)\Big|_{\text{4D}}=F^{(0)}_{\cO_2}(1^-,2^-;q) \>.
\end{equation}
An analogous statement is true also for amplitudes, where all the four-dimensional helicity configurations can be recovered from the six-dimensional amplitude.%
\footnote{Further details on the relation between four and six-dimensional tree-level quantities can be found in Appendix~\ref{sec:SixDAmp}.}

The scalar form factor is obtained from~\eqref{KaluzaKlein}, and we find
\begin{equation}\label{eq:TrF2scalartree}
	F^{(0)}_{\cO_{2,s}} (1,2; q) = - \langle 1_{a}, 2_{\dot{b}} \rbrack \langle 1^{a}, 2^{\dot{b}} \rbrack  = 2 s_{1 2}\ ,
\end{equation}
where
\begin{equation}
	\cO_{2,s} \propto (D\phi)^2 \coloneqq D_{\mu} \phi^a D^{\mu} \phi^a\ .
\end{equation}

The normalisation of~\eqref{eq:TrF2scalartree} has been fixed by matching the four-dimensional limit of this operator with that of the scalar components of~\eqref{eq::six_dim_F2}, which must yield the same result. Of course, if one starts from the Lagrangian~\eqref{KaluzaKlein} and computes the  minimal form factors of the two operators on the right-hand side, the resulting relative normalisation would be the same. The four-dimensional matching prescription is much faster for more complex operators. Let us stress that it would not be possible to implement the scalar subtraction just by excluding the little group components that in four dimensions behave like scalars. Indeed, this would bring us to a result which is not invariant under a little group transformation of the internal six-dimensional legs. In particular, for the subtraction we need a quantity that behaves as a scalar in six dimensions and matches the scalar components of the dimensional-reduced gluon in four dimensions, as shown in Appendix~\ref{sec:SixDAmp}.
%Notice however that in six-dimensions the interpretation of the two objects is different, and thus one is not allowed to perform a ``scalar subtraction'' by subtracting out the components of~\eqref{eq::six_dim_F2} which in four dimensions correspond to scalars.

%We now briefly discuss  three-point form factors.
Using BFCW recursion relation \cite{Britto:2004nc,Britto:2005fq} in six dimensions \cite{Cheung:2009dc} we have derived the six-dimensional non-minimal form factors with three external legs at tree level, both for the gluon and the scalar operators. The results for $\cO_2$ with three gluons reads
\begin{equation}
	\label{eq:TrF2nonminimal}
	\setlength{\jot}{7pt}
	\begin{split}
		F^{(0)}_{\cO_2} (1_{a \dot{a}}, 2_{b \dot{b}}, 3_{c \dot{c}}; q)  &= \frac{2}{s_{2 3} s_{3 1}} \asbr{1}{a}{2}{b}\asbr{2}{b}{1}{a} \langle 3_{c} | \slashed{p}_1 \slashed{p}_2 | 3_{\dot{c}} \rbrack + \mathrm{cyclic} \\
		&+ 2\left(\frac{1}{s_{1 2}} + \frac{1}{s_{2 3}} + \frac{1}{s_{3 1}}\right) \left(\asbr{1}{a}{2}{b} \asbr{2}{b}{3}{c} \asbr{3}{c}{1}{a} - \sabr{1}{a}{2}{b} \sabr{2}{b}{3}{c} \sabr{3}{c}{1}{a}\right)\ ,
	\end{split}
\end{equation}
which agrees with the analogous result computed from Feynman diagrams in~\cite{Davies:2011vt}, upon some algebraic manipulation. As a further consistency check we verified that in the four-dimensional limit the different helicity components match the results of~\cite{Dixon:2004za}.

Furthermore, in the scalar subtraction we need to take into account an additional contribution, namely  the form factor of the operator $\cO_2$ with two external scalars and one gluon, which is different from zero. Indeed, this is given by:
\begin{equation}
	\label{eq::TrF2nonmininaml_scalars}
	F^{(0)}_{\cO_2} (1, 2, 3_{c \dot{c}}; q) = - \frac{2}{s_{1 2}}  \langle 3_c | \slashed{p}_{1} \slashed{p}_{2} | 3_{\dot{c}}\rbrack\ .
\end{equation}

Finally, the non-minimal scalar form factor of $\cO_{2,s}$ can be shown to be
\begin{equation}
	\label{eq::Dphi2nonminimal}
	F^{(0)}_{\cO_{2,s}} (1, 2, 3_{c \dot{c}}; q) = - \frac{2 q^2}{s_{2 3}s_{3 1}} \langle 3_c | \slashed{p}_{1} \slashed{p}_{2} | 3_{\dot{c}}\rbrack\ .
\end{equation}
For a detailed derivation of \eqref{eq:TrF2nonminimal}-\eqref{eq::Dphi2nonminimal} see Appendix~\ref{sec:nonminimal}.
The sum of \eqref{eq::TrF2nonmininaml_scalars} and \eqref{eq::Dphi2nonminimal} agrees with the result of \cite{Davies:2011vt}.

\subsection{\texorpdfstring{$\Tr F^3$}{TrF3} Form Factors}

Consider now the operator
\begin{equation}
	\cO_3 \coloneqq \Tr F^{\mu}\,_{\nu} F^{\nu}\,_{\rho} F^{\rho}\,_{\mu} \ .
\end{equation}
Similarly to the case of $\Tr F^2$, this operator splits,  in four dimensions, into a self-dual and anti-self dual part
\begin{equation}
	\cO_3 \coloneqq \Tr F^3 = \Tr F^3_{\rm SD} +\Tr F^3_{\rm ASD}\ .
\end{equation}
Consequently, the only possible helicity configurations of the minimal tree-level form factors are  all-plus and all-minus:
\begin{equation}
	\setlength{\jot}{7pt}
	\begin{split}
		\label{eq::four_dim_F3}
		&F^{(0)}_{\cO_{3}} (1^{+}, 2^{+}, 3^{+}; q) = -2\sqr{1}{2}\sqr{2}{3}\sqr{3}{1}\ ,\\
		&F^{(0)}_{\cO_{3}} (1^{-}, 2^{-}, 3^{-}; q) = 2\agl{1}{2}\agl{2}{3}\agl{3}{1}\ .
	\end{split}
\end{equation}

In six dimensions the minimal form factor is given by
\begin{equation}
	\label{minimalF3_6D}
	\setlength{\jot}{7pt}
	\begin{split}
		F^{(0)}_{\cO_{3}} (1_{a \dot{a}},2_{b \dot{b}},3_{c \dot{c}})&= F_{1\, a \dot{a}}^{A B}\,_{C D} F_{2\, b \dot{b}}^{C D}\,_{E F} F_{3\, c \dot{c}}^{E F}\,_{A B}
		\\
		&=-\langle 1_{a} 2_{\dot{b}}\rbrack \langle 2_{b} 3_{\dot{c}}\rbrack \langle 3_c 1_{\dot{a}}\rbrack+\lbrack 1_{\dot{a}} 2_{b}\rangle \lbrack 2_{\dot{b}} 3_{c} \rangle \lbrack 3_{\dot{c}} 1_{a}\rangle\ ,
	\end{split}
\end{equation}
where $F_{a \dot{a}}^{A B}\,_{C D}$ is defined in \eqref{eq::FieldStrength6D}.
We can obtain  the corresponding scalar operator from~\eqref{scalarF3}, which states that
\begin{equation}
	\cO_{3,s}\propto \Tr (D \phi)^2 F \coloneqq \Tr D_{\mu} \phi D_{\nu} \phi F^{\mu \nu} \ .
\end{equation}
Thus
\begin{equation}
	F^{(0)}_{\cO_{3,s}} (1,2, 3_{c \dot{c}}) \coloneqq \frac{1}{2} p_{1}^{A B} p_{2 C D} F_{3\, c \dot{c}}^{C D}\,_{A B} = \langle 3_{c} | \slashed{p}_{1} \slashed{p}_{2} | 3_{\dot{c}}\rbrack\ ,
\end{equation}
where, once again, the normalisation is fixed  by matching the four-dimensional limits of this quantity with the scalar configuration of \eqref{minimalF3_6D}.

As a final remark, we point out that $\cO_3$ is not the only mass-dimension six operator which appears in the Yang-Mills theories (also with matter). One also has a contribution from
\begin{equation}
	\widetilde{\mathcal{O}}_3 \coloneqq D^{\alpha}F^{a\mu \nu}D_{\alpha}F^{a}_{\mu\nu}\ .
\end{equation}
However, it is easy to see that the minimal form factor for $\widetilde{\cO}_3$ can be related to the one of $\cO_2$ as
\begin{equation}
	F_{\widetilde{\cO}_3}^{(0)} (1_{a \dot{a}},2_{b \dot{b}}; q) = s_{12} F_{\cO_2}^{(0)} (1_{a \dot{a}},2_{b \dot{b}}; q) \ .
\end{equation}
Further Lorentz contractions of two covariant derivatives and two field strengths, such as $D^{\mu}F^{a\nu}_{\mu}D^{\rho}F^a_{\rho\nu}$, are ruled out or expressed in terms of the operators previously mentioned thanks to the equations of motion. In particular, in the case of pure Yang-Mills theory $\widetilde{\cO}_3$ can be expressed as a linear combination of $\cO_2$ and $\cO_3$ through the equations of motion. For a detailed discussion, see~\cite{Dawson:2015gka}.

Finally, we provide the tree-level expressions needed for the one-loop computation of the non-minimal form factor of $\mathcal{O}_3$ which are:
\begin{itemize}
	\item the non-minimal tree-level form factor of $\mathcal{O}_3$ with four gluons
	      \begin{equation}\label{eq:F3nmtree}
		      F^{(0)}_{\mathcal{O}_3}(1_{a\dot{a}},2_{b\dot{b}},3_{c\dot{c}},4_{d\dot{d}};q)= \mathcal{B}_{a\dot{a}b\dot{b}c\dot{c}d\dot{d}}+\mathcal{C}_{a\dot{a}b\dot{b}c\dot{c}d\dot{d}}+\mathcal{D}_{a\dot{a}b\dot{b}c\dot{c}d\dot{d}} \> ,
	      \end{equation}
	      with
	      \begin{equation}
		      \setlength{\jot}{7pt}
		      \begin{split}
			      \mathcal{B}_{a\dot{a}b\dot{b}c\dot{c}d\dot{d}}&=\left( -\asbr{1}{a}{2}{b}\asbr{2}{b}{3}{c}\asbr{3}{c}{1}{a}+ \sabr{1}{a}{2}{b}\sabr{2}{b}{3}{c}\sabr{3}{c}{1}{a} \right)\dfrac{\langle 4_d | \slashed{p}_1 \slashed{p}_3 | 4_{\dot{d}}]}{s_{34}s_{41}} + \text{ cyclic} \> ,\\
			      %\end{equation}
			      %\begin{equation}
			      \mathcal{C}_{a\dot{a}b\dot{b}c\dot{c}d\dot{d}}&=\dfrac{\asbr{1}{a}{2}{b}\asbr{2}{b}{4}{d}\asbr{4}{d}{3}{c}\asbr{3}{c}{1}{a}+\sabr{1}{a}{2}{b}\sabr{2}{b}{4}{d}\sabr{4}{d}{3}{c}\sabr{3}{c}{1}{a}}{s_{34}}+ \text{ cyclic} \> ,\\
			      %\end{equation}
			      %\begin{equation}
			      %\mathcal{D}_{a\dot{a}b\dot{b}c\dot{c}d\dot{d}}=-\left( \dfrac{1}{s_{12}}+\dfrac{1}{s_{23}}+\dfrac{1}{s_{34}}+\dfrac{1}{s_{41}} \right)\left( \asbr{1}{a}{2}{b}\asbr{2}{b}{3}{c}\asbr{3}{c}{4}{d}\asbr{4}{d}{1}{a}+ \sabr{1}{a}{2}{b}\sabr{2}{b}{3}{c}\sabr{3}{c}{4}{d}\sabr{4}{d}{1}{a} \right) \> .
			      %\end{equation}
			      %\begin{equation}
			      \mathcal{D}_{a\dot{a}b\dot{b}c\dot{c}d\dot{d}}&=-\left( \sum_{i=1}^4 \dfrac{1}{s_{i \, i+1}}\right)\left( \asbr{1}{a}{2}{b}\asbr{2}{b}{3}{c}\asbr{3}{c}{4}{d}\asbr{4}{d}{1}{a}+ \sabr{1}{a}{2}{b}\sabr{2}{b}{3}{c}\sabr{3}{c}{4}{d}\sabr{4}{d}{1}{a} \right) \> 
		      \end{split}
	      \end{equation}
	      %where in the sum in the last expression $i+1=5\equiv 1$.

	\item the non-minimal tree-level form factor of $\mathcal{O}_3$ with two external scalars

	      \begin{equation}
		      \begin{array}{lr}
			      F^{(0)}_{\mathcal{O}_3}(1,2,3_{c\dot{c}},4_{d\dot{d}};q)= & \dfrac{1}{s_{12}}\left( \asbr{3}{c}{4}{d} \langle 4_d |\slashed{p}_1 \slashed{p}_2 | 3_{\dot{c}}]-\asbr{4}{d}{3}{c}\langle 3_c |\slashed{p}_1 \slashed{p}_2 | 4_{\dot{d}}] \right)
		      \end{array}
	      \end{equation}

	\item the non-minimal tree-level form factor of $\mathcal{O}_{3,s}$ with two external scalars

	      \begin{equation}
		      \def\arraystretch{2}
		      \begin{array}{rl}
			      F^{(0)}_{\mathcal{O}_{3,s}}(1,2,3_{c\dot{c}},4_{d\dot{d}};q)= & \dfrac{\langle 3_c | \slashed{p}_4 \slashed{p}_2 | 3_{\dot{c}}]\langle 4_d | \slashed{p}_1 \slashed{p}_2|4_{\dot{d}}]}{s_{23}s_{34}} + \dfrac{\langle 4_d | \slashed{p}_1 \slashed{p}_3|4_{\dot{d}}]\langle 3_c | \slashed{p}_1 \slashed{p}_2 | 3_{\dot{c}}]}{s_{34}s_{41}} \\
			      +                                                             & \langle 3_c |\slashed{p}_2 \slashed{p}_1 |4_{\dot{d}}]\asbr{4}{d}{3}{c}\left( \dfrac{1}{s_{34}}+\dfrac{1}{s_{23}}+\dfrac{1}{s_{41}}\right)                                                                                                                                  \\
			      -                                                             & \langle 4_d |\slashed{p}_2 \slashed{p}_1 |3_{\dot{c}}]\asbr{3}{c}{4}{d}\dfrac{1}{s_{34}}  + \langle 3_{c} | \slashed{p}_{2} | 4_{d} \rangle \lbrack 3_{\dot{c}} | \slashed{p}_{1} | 4_{\dot{d}} \rbrack \left(\dfrac{1}{s_{23}}+\dfrac{1}{s_{41}}\right)\> .
		      \end{array}
	      \end{equation}
\end{itemize}

These formulas have been obtained by requiring the six-dimensional form factor to match, upon taking the four-dimensional limit, the known four-dimensional expressions in different helicity configurations~\cite{Brandhuber:2017bkg,Neill:2009mz,Dixon:1993xd,Broedel:2012rc}. The resulting ansatz was then numerically compared with the results from Feynman diagrams and a complete match was found.

\subsection{\texorpdfstring{$\Tr F^4$}{TrF4} and Higher Dimensional Form Factors}

The fourth power in the field strength can be considered as a turning point in the general behaviour of the operators, for reasons which will become clear in a moment.
%This is because up to $\Tr F^3$ the operators were constrained to split into a self-dual and anti-self-dual part, which is no longer true when four or more field strengths are contracted.
It turns out that we can have four possible independent operators involving different contractions of four field strengths \cite{Tseytlin:1986ti, Gracey:2002he}:
\begin{equation}
	\setlength{\jot}{7pt}
	\begin{split}
		&\Tr F^{\mu}\,_{\nu} F^{\nu}\,_{\rho} F^{\rho}\,_{\sigma} F^{\sigma}\,_{\mu}\ , \hspace{1cm} \Tr F^{\mu \nu} F_{\mu \nu} F^{\rho \sigma} F_{\rho \sigma}\ ,\\
		&\Tr F^{\mu}\,_{\nu} F^{\rho}\,_{\sigma} F^{\nu}\,_{\rho} F^{\sigma}\,_{\mu}\ , \hspace{1cm} \Tr F^{\mu \nu} F^{\rho \sigma} F_{\mu \nu} F_{\rho \sigma}\ .
	\end{split}
\end{equation}
In pure gauge theories, which we are considering in this work, all these operators can appear with independent coefficients, while they are no more independent in the low energy effective action from the superstring theory \cite{Green:1981xx,Schwarz:1982jn,Tseytlin:1986ti}. In this section we will focus only on the first operator, which we will refer to as $\Tr F^4$:
\begin{equation}
	\cO_4 \coloneqq \Tr F^4 \coloneqq \Tr F^{\mu}\,_{\nu} F^{\nu}\,_{\rho} F^{\rho}\,_{\sigma} F^{\sigma}\,_{\mu}\ .
\end{equation}
This encloses all the main features of the operators with higher powers in the field strength, and at the end of this section we will be able to generalise some results to a peculiar operator involving a consecutive chain of $n$ field strengths.

In four dimensions the main difference between $\Tr F^4$ and the lower-power cases is that the structure of this operator allows the mixing of the self- and anti-self-dual components, {\it i.e.} schematically
\begin{equation}
	\Tr F^4 \simeq \Tr F^{4}_{\rm SD}  + \Tr (F^{2}_{\rm SD} F^{2}_{\rm ASD}) + \Tr F^{4}_{\rm ASD}\ .
\end{equation}
Thus the usual all-plus (all-minus) minimal form factors appear along with MHV-like quantities:
\begin{equation}
	\setlength{\jot}{7pt}
	\begin{split}
		\label{eq::four_dim_F4}
		&F^{(0)}_{\cO_{4}} (1^{+}, 2^{+}, 3^{+}, 4^{+}; q) = 2\sqr{1}{2}\sqr{2}{3}\sqr{3}{4} \sqr{4}{1}\ ,\\
		&F^{(0)}_{\cO_{4}} (1^{+}, 2^{+}, 3^{-}, 4^{-}; q) = \sqr{1}{2}^2 \agl{3}{4}^2\ ,\\
		&F^{(0)}_{\cO_{4}} (1^{+}, 2^{-}, 3^{+}, 4^{-}; q) = \sqr{1}{3}^2 \agl{2}{4}^2\ ,
	\end{split}
\end{equation}
and all the other configurations can be obtained by symmetry and parity arguments.

In six dimensions the minimal form factor is
\begin{equation}
	\label{eq::six_dim_F4}
	\setlength{\jot}{7pt}
	\begin{split}
		F^{(0)}_{\cO_{4}} (1_{a \dot{a}}, 2_{b \dot{b}}, 3_{c \dot{c}}, 4_{d \dot{d}}; q) = &F_{1\, a \dot{a}}^{A B}\,_{C D} F_{2\, b \dot{b}}^{C D}\,_{E F} F_{3\, c \dot{c}}^{E F}\,_{G H} F_{4\, d \dot{d}}^{G H}\,_{A B}\\
		\overset{\eqref{4id1}}{=}& \langle 1_{a} 2_{\dot{b}}\rbrack \langle 2_{b} 3_{\dot{c}}\rbrack \langle 3_c 4_{\dot{d}}\rbrack\langle 4_{d} 1_{\dot{a}}\rbrack+\lbrack 1_{\dot{a}} 2_{b}\rangle \lbrack 2_{\dot{b}} 3_{c} \rangle \lbrack 3_{\dot{c}} 4_{d}\rangle \lbrack 4_{\dot{d}} 1_{a}\rangle\\
		&+ \langle 1_{a} 2_{b} 3_{c} 4_{d} \rangle\lbrack 1_{\dot{a}} 2_{\dot{b}} 3_{\dot{c}} 4_{\dot{d}}\rbrack\ ,
	\end{split}
\end{equation}
where we notice that at this power of the field strength the new structure $\langle \cdot\,\cdot\,\cdot\,\cdot \rangle\lbrack \cdot\,\cdot\,\cdot\,\cdot\rbrack$ involving four-spinor invariants appears, which is very reminiscent of the four-point amplitude. This new structure gives us the MHV-like components in \eqref{eq::four_dim_F4} when  we consider the appropriate little-group configurations in the four-dimensional limit (see Appendix~\ref{sec:SixDAmp}).

We have already identified the scalar operator associated to $\Tr F^4$ in \eqref{scalarF4} and we define
\begin{equation}
	\cO_{4,s} \propto \Tr D_{\mu} \phi D_{\nu} \phi F^{\nu}\,_{\rho} F^{\rho \mu}
\end{equation}
such that its minimal form factor is
\begin{equation}
	\setlength{\jot}{7pt}
	\begin{split}
		F_{\cO_{4,s}} (1,2, 3_{c \dot{c}}, 4_{d \dot{d}}; q)  &= \frac{1}{2} p_{1}^{A B} p_{2 C D} F_{3\, c \dot{c}}^{C D}\,_{E F} F_{4\, d \dot{d}}^{E F}\,_{A B} \\
		&= - \langle 3_{c} | \slashed{p}_2 \slashed{p}_1 | 4_{\dot{d}}\rbrack \asbr{4}{d}{3}{c} +\frac{1}{4}\langle 2^{a} 2_{a} 3_{c} 4_{d} \rangle\lbrack 1_{\dot{a}} 1^{\dot{a}} 3_{\dot{c}} 4_{\dot{d}}\rbrack\ .
	\end{split}
\end{equation}

The expression of $\Tr F^4$ gives us some insight about the operators involving the $n^{\rm th}$ power of the field strength, where the Lorentz indices are contracted between adjacent field strengths, which we will refer to as $\Tr F^{n}$:
\begin{equation}\label{eq:TrFndef}
	\cO_n \coloneqq \Tr F^{n} = \Tr F_{\mu_1}\,^{\mu_2} F_{\mu_2}\,^{\mu_3}\cdots F_{\mu_{n-1}}\,^{\mu_n} F_{\mu_n}\,^{\mu_1}\ .
\end{equation}

It is easy to show that this operator can be decomposed in a sum of double traces (in the Lorentz indices) on the self-dual and anti-self-dual parts, schematically%
\footnote{We stress that this general structure was hidden by lower power-operators because the field strength is traceless: $\Tr F_{SD}^{n-1} F_{ASD} = \Tr F_{SD} F_{ASD}^{n-1} = 0$.
	%: $\Tr F = \Tr F_{\rm SD} = \Tr F_{\rm ASD} = 0$.
}:
\begin{equation}
	\Tr F^n \simeq \sum_{i = 0}^{n} \Tr (F^{n - i}_{\rm SD}  F^{i}_{\rm ASD})\ .
\end{equation}

Take two disjoint and ordered subsets of labels $S_{+}=\{p_k\}_{k=1\ldots i}$ and $S_{-}=\{q_k\}_{k=1\ldots n-i}$, with $S_{+} \cup S_{-} = \{1,\ldots, n\}$. Then  all tree level form factors, for any helicity configuration, can be written in a very compact way:
\begin{equation}
	\label{eq::four_dim_Fn}
	F_{\cO_{n}}^{(0)} (1^{h_1}, \ldots, n^{h_n}; q) = c_{n,i} \prod_{k = 1}^{i} \sqr{p_{k}}{p_{k+1}} \prod_{k = i+1}^{n} \agl{q_{k}}{q_{k+1}}\ ,
\end{equation}
where the overall coefficient is
\begin{equation}
	c_{n, i} =
	\begin{cases}
		2          & i = 0           \\
		(-1)^{n-i} & i\neq 0,\, n\ . \\
		(-)^n 2    & i = n
	\end{cases}
\end{equation}
An explicit example of this general formula is given by
\begin{equation}
	F_{\cO_{5}}^{(0)} (1^-, 2^+, 3^-, 4^-, 5^+; q) = - \agl{1}{3} \agl{3}{4} \agl{4}{1} \sqr{2}{5} \sqr{5}{2}\ .
\end{equation}

The structure of $\Tr F^n$ form factors in six dimensions is much more complicated than the four-dimensional one, the number of terms grows very fast, but nonetheless some general pattern can be observed. In particular if we restrict to a kinematic configuration for which only some of the legs are truly six dimensional and the others are defined on the embedded four-dimensional subspace, the formulae are much easier and compact. In principle, this is all we need in order to calculate rational terms with the dimensional reconstruction technique, since we need to consider only the limited number of internal loop legs as six dimensional. As an example, consider the minimal form factor of $\Tr F^n$ with two six-dimensional legs and $n-2$ four-dimensional legs in the all-plus helicity configuration. The general expression is given by
\begin{equation}
	\label{trFn}
	\setlength{\jot}{7pt}
	\begin{split}
		\Tr F^n (1_{a \dot{a}}, 2_{b \dot{b}}, 3^+, \ldots , n^+) = & \big(\langle 1_{a} 2_{\dot{b}}\rbrack \langle 2_{b} 3_{\dot{1}}\rbrack \sqr{3}{4} \langle n_{1} 1_{\dot{a}}\rbrack+\lbrack 1_{\dot{a}} 2_{b}\rangle \lbrack 2_{\dot{b}} 3_{1} \rangle  \sqr{3}{4} \lbrack n_{\dot{1}} 1_{a}\rangle + \\
		&+ \langle 1_{a} 2_{b} n_{1} 3_{1} \rangle\lbrack 1_{\dot{a}} 2_{\dot{b}} 3_{\dot{1}} 4_{\dot{1}}\rbrack \big) \prod_{i=4}^{n-1} \sqr{i}{i+1}\ .
	\end{split}
\end{equation}
This result can be found by observing that the combination $\lambda_{i \, a}^{A} \widetilde{\lambda}_{i \, B \dot{a}}$ appears only once for each six-dimensional leg, which allows to write an ansatz comprising every possible combination with arbitrary coefficients to be fixed. The coefficients can then be determined by taking the four-dimensional limit of the six-dimensional gluons and requiring the form factor to match \eqref{eq::four_dim_Fn}. For the sake of comparison, if we take $n=6$ the three terms of~\eqref{trFn} come from a fully six-dimensional expression of 39 terms which has already been reduced from initial 52 terms using Schouten identity.

\section{One-Loop Form Factors}
\label{sec::OneLoop}

In this section we will consider a number of  one-loop applications of the dimensional reconstruction procedure  discussed so far. The results obtained for the minimal form factors of $\Tr F^2$ and $\Tr F^3$ were already known in the literature. We prove that the latter has no rational terms, as it has also been argued by \cite{Neill:2009mz}. These calculations will be useful to set the stage and give an example of the procedure before dealing with more involved operators and kinematic configurations. In particular, we reproduce the known non-minimal form factor  of $\Tr F^2$ with three positive-helicity external gluons. Finally, we compute the complete minimal form factor of $\Tr F^n$ with $n=4$ at one loop and generalise some of the results to arbitrary~$n$.

\subsection{The Minimal \texorpdfstring{$\Tr F^2$}{TrF2} Form Factors}

As a first proof of concept of the method we will confirm the well known statement that the minimal form factor of the operator $\Tr F^2$ in pure Yang-Mills does not have any rational terms.
In particular, we will consider the all-plus helicity configuration. %Of course the all-minus can be obtained by applying a parity transformation to the this result whereas the mixed helicity configurations are trivially zero.

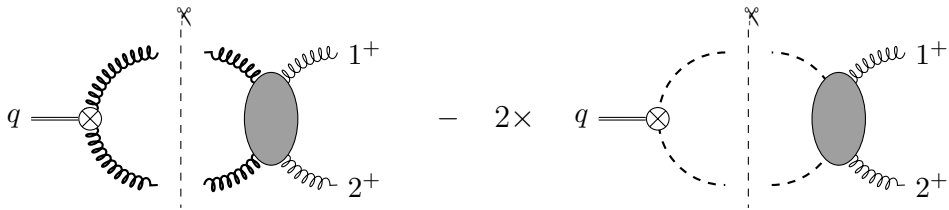
\begin{figure}[H]
	\centering
	\begin{tikzpicture}[scale=25,auto,cross/.style={path picture={
							\draw[black]
							(path picture bounding box.south east) -- (path picture bounding box.north west) (path picture bounding box.south west) -- (path picture bounding box.north east);
						}}]

		%One loop the form factors

		\def\y{-7pt}
		\def\x1{11pt}

		\node at (0.5pt+\x1,2pt+\y) (one){};
		\draw [thick,decorate, decoration={coil, amplitude=2.3pt, segment length=3pt}](one.east) arc (90:270:1pt);
		\node [draw,circle,cross, inner sep=3pt,fill=white] at (-0.36pt+\x1,1pt+\y)(cross){};
		\node at (-1.5pt+\x1,1pt+\y)(Q){$q$};

		\draw [double](Q) -- (cross);

		\node at (3.75pt+\x1,2pt+\y) (three2){$1^+$};
		\node at (3.75pt+\x1,0pt+\y) (four2){$2^+$};
		\node at (1.35pt+\x1,0pt+\y)(six){};

		\draw [thick,decorate, decoration={coil, amplitude=2.3pt, segment length=3pt}](six) arc (-90:90:1pt);
		\draw [decorate, decoration={coil, amplitude=2.3pt, segment length=3pt}](three2.west) arc (90:270:1pt);
		\draw [fill=gray!75] (2.35pt+\x1,1pt+\y) ellipse (0.4pt and 0.7pt);

		\node at (1pt+\x1,2.5pt+\y) (cut1){};
		\node at (1pt+\x1,-0.5pt+\y) (cut2){};
		\draw [dashed] (cut1) -- (cut2);
		\node at (1.1pt+\x1,2.4pt+\y)[rotate around={-90:(1pt,2.5pt)}]{\Cutright};

		\node at (5pt+\x1,1pt+\y){$-$};
		\node at (6pt+\x1,1pt+\y){\large $2 \times$};

		\def\x1{19.5}
		\def\y{-7pt}

		\node at (0.5pt+\x1,2pt+\y) (one){};
		\draw [thick,dashed](one.east) arc (90:270:1pt);
		\node [draw,circle,cross, inner sep=3pt,fill=white] at (-0.36pt+\x1,1pt+\y)(cross){};
		\node at (-1.5pt+\x1,1pt+\y)(Q){$q$};

		\draw [double](Q) -- (cross);

		\node at (3.75pt+\x1,2pt+\y) (three2){$1^+$};
		\node at (3.75pt+\x1,0pt+\y) (four2){$2^+$};
		\node at (1.35pt+\x1,0pt+\y)(six){};

		\draw [thick,dashed](six) arc (-90:90:1pt);
		\draw [decorate, decoration={coil, amplitude=2.3pt, segment length=3pt}](three2.west) arc (90:270:1pt);
		\draw [fill=gray!75] (2.35pt+\x1,1pt+\y) ellipse (0.4pt and 0.7pt);

		\node at (1pt+\x1,2.5pt+\y) (cut1){};
		\node at (1pt+\x1,-0.5pt+\y) (cut2){};
		\draw [dashed] (cut1) -- (cut2);
		\node at (1.1pt+\x1,2.4pt+\y)[rotate around={-90:(1pt,2.5pt)}]{\Cutright};

	\end{tikzpicture}
	\caption{Two-particle cut of the one-loop form factor $\Tr F^2$ in six dimensions.}\label{fig:doublecutf2}
\end{figure}
\noindent
The quantity we want to compute can be written as
\begin{equation}
	\def\arraystretch{1.5}
	\begin{array}{rl}
		F^{(1)}_{\cO_{2}} (1^+,2^+;q) \coloneqq & F^{(0)}_{\cO_{2}} (1^+,2^+;q)\cdot f^{(2)}\left(s_{1 2}\right) \\
		=                                       & 2\sqr{1}{2} \sqr{2}{1}\cdot f^{(2)}\left(s_{1 2}\right)\ ,
	\end{array}
\end{equation}
where we factored out all the helicity dependence in the tree-level prefactor, and $f^{(2)}(s_{1 2})$ is a function only of the Mandelstam variable $s_{12}$. As explained in Section \ref{sec::GenUni6D} this quantity can be computed using \eqref{eq::dimensional_reconstruction}:
\begin{equation}
	f^{(2)}\left(s_{1 2}\right) = f_{\mathrm{6D}}^{(2)} \left(s_{1 2}\right) - 2 f_{\phi}^{(2)} \left(s_{1 2}\right)\ ,
\end{equation}
where $f_{\mathrm{6D}}^2 \left(s_{1 2}\right)$ and $f_{\phi}^2 \left(s_{1 2}\right)$ are the form factors with six-dimensional internal gluons or scalars respectively, normalised by the corresponding  tree-level quantity.

At one loop, the  two-particle cut represented in Figure \ref{fig:doublecutf2} is%
\footnote{The explicit expression of $\mathcal{A}_g$ can be found in Appendix \ref{sec:SixDAmp}.}

%\begin{equation}
%	\label{eq::F2doublecut}
%	\left. f_{\mathrm{6D}}^2\left(s_{1 2}\right)\right|_{\rm 2-cut}^{\rm 1-loop} = 1+ \int \dd{\rm LIPS} \frac{1}{\sqr{1}{2}^2 \, s_{1 2}\ s_{2 l_2}}\, \langle l_{1 a}\, l_{2 \dot{b}} \rbrack \langle l_{2 b}\, l_{1 \dot{a}}\rbrack \langle 1_{1}\, 2_{1}\, {l_{2}}^{b} \, {l_{1}}^{a} \rangle\lbrack 1_{\dot{1}}\, 2_{\dot{1}}\, {l_{2}}^{\dot{b}}\, {l_{1}}^{\dot{a}}\rbrack\ ,
%\end{equation}
\begin{equation}
	\label{eq::F2doublecut}
	\left. f_{\mathrm{6D}}^{(2)}\left(s_{1 2}\right)\right|_{\rm 2-cut} = \frac{1}{2 \sqr{1}{2} \sqr{2}{1}} \int \dd{\rm LIPS} \, F^{(0)}_{\cO_{2}}(-l_1^{a \dot{a}},-l_2^{b \dot{b}}) \; \mathcal{A}^{(0)}_g (l_{2 \, b \dot{b}},l_{1 \, a \dot{a}},1_{1 \dot{1}},2_{2 \dot{2}})\ .
\end{equation}
In order to simplify this expression we decompose the six-dimensional quantities in terms of four-dimensional ones, as explained in detail in Section \ref{sec::from_six_to_four}. These  calculations are rather lengthy and we have devised a  \texttt{MATHEMATICA} package  to deal with them (see Appendix \ref{Mathematica} for a detailed presentation). In general, we write six-dimensional expressions in terms of $\{\lambda_{i \alpha},\widetilde{\lambda}_{i \dot{\alpha}},\mu_{i \alpha},\widetilde{\mu}_{i \dot{\alpha}},m_i,\widetilde{m}_i \}$ with $i=1,2,l_1,l_2$, as explained in Appendix \ref{sec::from_six_to_four}. Imposing that the external legs are defined in four dimensions is equivalent to setting $m_j=0$ and $\widetilde{m}_j=0$ for $j=1,2$, which automatically removes any dependence of $f^{(2)}$ on $\mu_{j \alpha}$ and $\widetilde{\mu}_{j \dot{\alpha}}$. From \eqref{eq:momentum6D}, momentum conservation implies
\begin{equation}
	\sum_{i} m_i=0\ , \hspace{1.5cm} \sum_{i} \widetilde{m}_i =0\ .
\end{equation}
Only the two internal legs $l_1$ and $l_2$ have to be kept in  six dimensions, in other words $p_i^5,p_i^6 \neq 0$  for $i=l_1,l_2$, which implies
\begin{equation}\label{eq:extramass}
	m_{l_2} = -m_{l_1} \coloneqq -m\ , \hspace{1.5cm} \widetilde{m}_{l_2}= -\widetilde{m}_{l_1} \coloneqq -\widetilde{m}\ ,
\end{equation}
where
\begin{equation}
	\mu^2 = m \widetilde{m}\ ,
\end{equation}
with $\mu^2$ defined in \eqref{mu2}.
The  result for the complete integrand in \eqref{eq::F2doublecut} is, schematically,
\begin{equation}\label{eq:TrF2ugly}
	\mathcal{I} = \dfrac{i \agl{l_1}{l_2}^2 \sqr{l_2}{l_1}^2}{s_{12}s_{2l_2}}+ \mu^2 \; (\text{4 terms}) + \mu^4 \; (\text{17 terms}) + \mu^6 \;(\text{5 terms}) + \mu^8 \; (\text{1 term}),
\end{equation}
where the Mandelstam invariants are defined in terms of six-dimensional momenta.
It is important to note that the dependence on  $\mu_{i}$ and $\widetilde{\mu}_i$ is spurious and we can choose these ``reference momenta'' in order to cancel as many terms as possible from our result. After doing so one has to be careful in identifying the loop momenta and Mandelstam invariants consistently with this choice. A particularly convenient choice is
\begin{equation}
	\mu_{l_1}\rightarrow \lambda_{l_2}\ , \hspace{0.5cm}	\mu_{l_2}\rightarrow \lambda_{l_1}\ , \hspace{0.5cm}
	\widetilde{\mu}_{l_2} \rightarrow \widetilde{\mu}_{l_1}\ .
\end{equation}
Doing so, we immediately arrive at
%\begin{equation}\label{eq:TrF2gluon}
%	\left. f_{\mathrm{6D}}^2\left(s_{1 2}\right)\right|_{\rm 2-cut}^{\rm 1-loop} = 1+ \int \dd{\rm LIPS} \left(- i \frac{s_{1 2}}{s_{2 l_2}} +2 i \frac{\mu^2}{s_{2 l_2}}\right)\ .
%\end{equation}
\begin{equation}\label{eq:TrF2gluon}
	\left. f_{\mathrm{6D}}^{(2)}\left(s_{1 2}\right)\right|_{\rm 2-cut} = \int \dd{\rm LIPS} \left(- i \frac{s_{1 2}}{s_{2 l_2}} +2 i \frac{\mu^2}{s_{2 l_2}}\right)\ .
\end{equation}
%Before doing any uplift to the full integral with propagators off-shell, we need to subtract two times the contribution coming from the two scalars running in the loop. The double-cut of the scalar form factor, normalized is

Next we repeat a similar computation for the two-particle cut with  internal gluons replaced by scalars:
%\begin{equation}\label{eq:TrF2scalar}
%	\left. f^{2}_{\phi}\left(s_{1 2}\right)\right|_{\rm 2-cut}^{\rm 1-loop}=\int \dd{\rm LIPS} \frac{s_{l_1 l_2}}{\sqr{1}{2}^2} \frac{- i}{4\, s_{1 2}\, s_{2 l_2}}  \langle 1_{1} \, 2_{1}\, l_{2 c}\, {l_{2}}^{c} \rangle\lbrack 1_{\dot{1}}\, 2_{\dot{1}}\, l_{1 \dot{c}}\, {l_{1}}^{\dot{c}} \rbrack =\int \dd{\rm LIPS} i\frac{\mu^2}{s_{2 l_2}}\ .
%\end{equation}

\begin{equation}
	\label{eq:TrF2scalar}
	\setlength{\jot}{7pt}
	\begin{split}
		\left. f^{(2)}_{\phi}\left(s_{1 2}\right)\right|_{\rm 2-cut}&=\frac{1}{2 \sqr{1}{2} \sqr{2}{1}} \int\dd{\rm LIPS} F^{(0)}_{\cO_{2,s}}(-l_1,-l_2) \; \mathcal{A}^{(0)}(l_2, l_1,1_{a \dot{a}},2_{b \dot{b}})\\
		&=\int \dd{\rm LIPS} i\frac{\mu^2}{s_{2 l_2}}\ .
	\end{split}
\end{equation}

Taking the difference between  \eqref{eq:TrF2gluon} and twice \eqref{eq:TrF2scalar} leads to the desired
four-dimen\-sional result
\begin{equation}\label{eq:TrF2sub}
	f^{(2)}\left(s_{1 2}\right)\big|_{\rm 2-cut}=-is_{1 2}\int \dd{\rm LIPS} \frac{1}{s_{2 l_2}} \>.
\end{equation}
It is important to stress that in order to perform the scalar subtraction consistently, one needs first to write both $f_{6D}^{(2)}$ and $f_{\phi}^{(2)}$ as functions of the the full $d$-dimensional momenta and Mandelstam invariants, in order to eliminate any dependence on the choice of the arbitrary helicity spinors $\mu_{i}$ and $\widetilde{\mu}_i$. We can directly read off the one-loop result from~\eqref{eq:TrF2sub}:
\begin{equation}\label{eq:TrF2final}
	f^{(2)}\left(s_{1 2}\right)=-is_{1 2} \cdot \>
	\begin{tikzpicture}[baseline={([yshift=-1mm]uno.base)},scale=15]
		\def\x{0}
		\def\y{0}

		\clip (-1.4pt,-2pt) rectangle (3.5pt,2pt);

		\node at (0+\x,0+\y)(uno){};
		\node at (1.5pt+\x,1pt+\y)(due){};
		\node at (1.5pt+\x,-1pt+\y)(tre){};
		%\node at (-0.75pt+\x,0.5pt+\y)(unoB){};
		%\node at (-0.75pt+\x,0.5pt+\y)(unoB){};
		\node at (-0.75pt+\x,0+\y)(unoS){};
		\node at (2.25pt+\x,1.5pt+\y)(dueB){};
		\node [right=0.02pt of dueB, yshift=0.4pt]{\footnotesize$p_1$};
		\node at (2.25pt+\x,-1.5pt+\y)(treB){};
		\node [right=-0.02pt of treB, yshift=-0.4pt]{\footnotesize$p_2$};
		\node at (0.75pt+\x,-1pt+\y) (end){};
		\node at (-4pt+\x,-2pt+\y)(start){};
		\node at (-1.2pt+\x,0+\y){\footnotesize$q$};
		\node at (0.9pt+\x,0+\y)[thick,scale=1.5]{};

		\draw [thick] (uno.center) -- (due.center);
		\draw [thick] (due.center) -- (tre.center);
		\draw [thick] (uno.center) -- node [yshift=-0.2cm] {\footnotesize$l$} (tre.center);
		%\draw [thick] (uno.center) -- (unoB.center);
		%\draw [thick] (uno.center) -- (unoC.center);
		\draw [thick,double] (uno.center) -- (unoS.center);
		\draw [thick] (due.center) -- (dueB.center);
		\draw [thick] (tre.center) -- (treB.center);

		%\draw [dotted,thick](-0.6pt+\x,0.3pt+\y) arc (140:240:0.4pt);
	\end{tikzpicture}
\end{equation}
%\begin{equation}
%	\left. f^{(2)}\left(s_{1 2}\right)\right|_{\rm 2-cut}^{\rm 1-loop} = 1-is_{1 2}\int \dd{\rm LIPS} \frac{1}{s_{2 l_2}} \,\overset{\rm Uplift}{ \xrightarrow{\hspace*{1.2cm}}}\, 1-is_{1 2} I_{3}(p_1,p_2,q)\ ,
%\end{equation}
where the triangle integral with outgoing momenta $(p_1,p_2,q)$ is defined in Appendix \ref{sec:integrals}.
%\begin{equation}\label{eq:TrF2result}
%	\begin{tikzpicture}[baseline={([yshift=-1mm]uno.base)},scale=15]
%		\def\x{0}
%		\def\y{0}
%
%		\clip (-1.4pt,-2pt) rectangle (3.5pt,2pt);
%
%		\node at (0+\x,0+\y)(uno){};
%		\node at (1.5pt+\x,1pt+\y)(due){};
%		\node at (1.5pt+\x,-1pt+\y)(tre){};
%\node at (-0.75pt+\x,0.5pt+\y)(unoB){};
%\node at (-0.75pt+\x,0.5pt+\y)(unoB){};
%		\node at (-0.75pt+\x,0+\y)(unoS){};
%		\node at (2.25pt+\x,1.5pt+\y)(dueB){};
%		\node [right=0.02pt of dueB, yshift=0.4pt]{\footnotesize$p_1$};
%		\node at (2.25pt+\x,-1.5pt+\y)(treB){};
%		\node [right=-0.02pt of treB, yshift=-0.4pt]{\footnotesize$p_2$};
%		\node at (0.75pt+\x,-1pt+\y) (end){};
%		\node at (-4pt+\x,-2pt+\y)(start){};
%		\node at (-1.2pt+\x,0+\y){\footnotesize$q$};
%		\node at (0.9pt+\x,0+\y)[thick,scale=1.5]{};

%		\draw [thick] (uno.center) -- (due.center);
%		\draw [thick] (due.center) -- (tre.center);
%		\draw [thick] (uno.center) -- node [yshift=-0.2cm] {\footnotesize$l$} (tre.center);
%\draw [thick] (uno.center) -- (unoB.center);
%\draw [thick] (uno.center) -- (unoC.center);
%		\draw [thick,double] (uno.center) -- (unoS.center);
%		\draw [thick] (due.center) -- (dueB.center);
%		\draw [thick] (tre.center) -- (treB.center);

%\draw [dotted,thick](-0.6pt+\x,0.3pt+\y) arc (140:240:0.4pt);
%	\end{tikzpicture}
%\coloneqq I_{3}(p_1,p_2,q)
%	= \int \frac{\dd^d l}{(2 \pi)^{d}} \;\dfrac{1}{l^2\, (l+p_2)^2\, (l+p_1+p_2)^2}
%\end{equation}

As anticipated, our result \eqref{eq:TrF2final} does not contain any $\mu^2$ term {\it i.e.} any rational term, and is thus in agreement with the very well known result. An equivalent result holds for  the all-minus helicity configuration. As expected, there are no bubbles in the result, because both $\Tr F_{\rm SD}^2$ and $\Tr F_{\rm ASD}^2$ are protected operators.
%The result is a triangle integral and in complete agreement with the known result (ref needed).

\subsection{The Non-Minimal \texorpdfstring{$\Tr F^2$}{TrF2} Form Factor}\label{sec:TrF2nonmin}

In this section we address the computation of the one-loop non-minimal form factor of the operator $\Tr F^2$. As usual we begin by defining the normalised quantity $f^{(2;3)}$ as
\begin{equation}
	F^{(1)}_{\cO_{2}} (1^+,2^+,3^+;q) \coloneqq 2 \sqr{1}{2}\sqr{2}{3}\sqr{3}{1}\cdot f^{(2;3)}\left(s_{1 2}, s_{23},s_{13}\right)\ ,
\end{equation}
with
\begin{equation}
	f^{(2;3)}\left(s_{1 2}, s_{23},s_{13}\right) = f_{\mathrm{6D}}^{(2;3)} \left(s_{1 2}, s_{23},s_{13}\right) - 2 f_{\phi}^{(2;3)} \left(s_{1 2}, s_{23},s_{13}\right)\ ,
\end{equation}
Notice that we decided not to normalise by the corresponding tree-level form-factor, which carries additional non-trivial dependence on the Mandelstam variables, but simply by a factor $\sqr{1}{2}\sqr{2}{3}\sqr{3}{1}$ which only captures the complete helicity dependence of the operator. Computing the discontinuity in the $s_{12}$-channel we have
\begin{equation}
	\label{eq::F2nmdoublecut}
	\left. f_{\mathrm{6D}}^{(3)}\left(\{ s_{i j}\}\right)\right|_{s_{1 2}-\textrm{cut}} =\frac{1}{2\sqr{1}{2}\sqr{2}{3}\sqr{3}{1}} \int \dd{\rm LIPS}\, F^{(0)}_{\cO_{2}}(l_1^{a \dot{a}},l_2^{b \dot{b}},3_{1 \dot{1}})\; \mathcal{A}^{(0)} (-l_{2 \, a \dot{a}},-l_{1 \, b \dot{b}},1_{1 \dot{1}},2_{2 \dot{2}}) \> ,
\end{equation}
which, upon making use of momentum conservation in the form of~\eqref{eq:extramass}, is a 356-term expression. We make use of the redundant degrees of freedom to simply the expression by choosing
\begin{equation}\label{eq:muTrF2nm}
	\mu_{l_1} \mapsto \lambda_{l_2} \ , \hspace{0.5cm} \tilde{\mu}_{l_1} \mapsto \tilde{\lambda }_{3}\ , \hspace{0.5cm} \mu_{l_2} \mapsto \lambda_{l_1} \ , \hspace{0.5cm}\tilde{\mu }_{l_2}\mapsto \tilde{\lambda }_{3} \>,
\end{equation}
which leads to
\begin{equation}\label{eq:F2nmpartial}
	\left. f_{\mathrm{6D}}^{(3)}\left(\{ s_{i j}\}\right)\right|_{s_{1 2}-\textrm{cut}} \propto \int \dd{\rm LIPS}\, \left[3\medspace l_1\right] \langle l_1 \; l_2 \rangle \lbrack l_2 \; 3 \rbrack  %\left(
	%\frac{-2 i m^2}{s_{12} s_{3l_1} s_{2- l_2}} \> + \> 
	% \ldots \right)
	\>.
\end{equation}
Note that, after using \eqref{eq:muTrF2nm}, the last expression apparently is no longer  invariant with respect to little-group transformations of $l_1$ and $l_2$, since these transformations mix the $\lambda$ and $\mu$. In other words, looking at the numerator of~\eqref{eq:F2nmpartial}, $l_1$ and $l_2$ appear as four-dimensional massless momenta, whereas they should really be massive. Hence in order to further manipulate the expression in a consistent manner we have to restore the masses, \textit{i.e.} restore explicit little-group invariance. This is achieved by the replacement
\begin{equation}
	\lambda_i^{\alpha}\tilde{\lambda}_i^{\dot{\alpha}} \ \mapsto \underbrace{\left(\lambda_i^{\alpha}\tilde{\lambda}_i^{\dot{\alpha}} + \frac{\mu^2}{\agl{\lambda_{i} }{\mu_{i}}\sqr{\widetilde{\mu}_{i}}{\widetilde{\lambda}_{i}}} \mu_i^{\alpha}\tilde{\mu}_i^{\dot{\alpha}} \right)}_{\displaystyle p_i^{(4)\, \alpha\dot{\alpha}}} - \frac{\mu^2}{\agl{\lambda_{i} }{\mu_{i}}\sqr{\widetilde{\mu}_{i}}{\widetilde{\lambda}_{i}}} \mu_i^{\alpha}\tilde{\mu}_i^{\dot{\alpha}} \ ,
\end{equation}
which in the particular case of~\eqref{eq:F2nmpartial} becomes
\begin{equation}
	|l_1 \rangle \lbrack l_1 | \mapsto \slashed{l}^{(4)}_1 -\frac{\mu^2}{\agl{l_1 }{l_2}\sqr{3}{l_1}}  |l_2 \rangle \lbrack 3 | \ , \hspace{0.5cm}  |l_2 \rangle \lbrack l_2 | \mapsto \slashed{l}^{(4)}_2 -\frac{\mu^2}{\agl{l_2 }{l_1}\sqr{3}{l_2}}  |l_1 \rangle \lbrack 3 | \ ,
\end{equation}
where the replacements~\eqref{eq:muTrF2nm} have already been applied. After  this substitution and some further manipulation, \eqref{eq:F2nmpartial} becomes
\begin{equation}\label{eq:f23gluon}
	f^{(2;3)}_{\rm 6D}(s_{12},s_{23},s_{13})\big|_{s_{12}\rm{-cut}}= i\frac{\sqr{1}{2}}{\sqr{2}{3}\sqr{3}{1}}\int \dd{\rm LIPS} \lbrack 3|\, \slashed{l}_1^{(4)}\, \slashed{l}_2^{(4)} | 3\rbrack\; \mathcal{I}_{\rm 6D}^{(2;3)}\ ,
\end{equation}
where
\begin{equation}\label{eq:F2nmgluon}
	\mathcal{I}_{\rm 6D}^{(2;3)}=\frac{ q^4 s_{12} - 2 \mu^2 q^2 s_{12} -4 \mu^2 s_{3 l_1} s_{3 l_2} }{s_{12}^2 s_{2, -l_2} s_{3 l_1} s_{3 l_2}} \> .
\end{equation}

\begin{figure}
	\centering
	\begin{tikzpicture}[scale=25,auto,cross/.style={path picture={
							\draw[black]
							(path picture bounding box.south east) -- (path picture bounding box.north west) (path picture bounding box.south west) -- (path picture bounding box.north east);
						}}]

		\def\x1{15.5}
		\def\y{-7pt}

		\node at (-5pt+\x1,1pt+\y) {$f_{\phi}^{(2;3)}\big|_{s_{12}\rm{-cut}}=$};

		\draw (-1.9pt+\x1,3.3pt+\y) -- (-2.2pt+\x1,3.3pt+\y) -- (-2.2pt+\x1,-1.2pt+\y) -- (-1.9pt+\x1,-1.2pt+\y);

		\node at (0.5pt+\x1,2pt+\y) (one){};
		\draw [thick,dashed](one.east) arc (90:270:1pt);
		\node [draw,circle,cross, inner sep=3pt,fill=white] at (-0.36pt+\x1,1pt+\y)(cross){};
		\node at (-1.5pt+\x1,1pt+\y)(Q){$q$};
		\node at (-1.5pt+\x1,-0.5pt+\y) (five){$3^+$};
		\draw [decorate, decoration={coil, amplitude=2.3pt, segment length=3pt}] (five) -- (cross);
		\node [draw, rectangle, red, inner sep=10pt] at (-0.36pt+\x1,1pt+\y)(square){};
		\node at (-1pt+\x1,2.7pt+\y) (operator){$\mathcal{O}_2$};
		\draw [->,red] (square) -- (operator);

		\draw [double](Q) -- (cross);

		\node at (1.5pt+\x1,1pt+\y) {\large $+$};

		\def\x1{19.5}
		\def\y{-7pt}

		\node at (0.5pt+\x1,2pt+\y) (one){};
		\draw [thick,dashed](one.east) arc (90:270:1pt);
		\node [draw,circle,cross, inner sep=3pt,fill=white] at (-0.36pt+\x1,1pt+\y)(cross){};
		\node at (-1.5pt+\x1,1pt+\y)(Q){$q$};
		\node [draw, rectangle, red, inner sep=10pt] at (-0.36pt+\x1,1pt+\y)(square){};
		\node at (-1pt+\x1,2.7pt+\y) (operator){$\mathcal{O}_{2,s}$};
		\draw [->,red] (square) -- (operator);
		\draw [double](Q) -- (cross);
		\node at (-1.5pt+\x1,-0.5pt+\y) (five){$3^+$};

		\draw (1pt+\x1,3.3pt+\y) -- (1.3pt+\x1,3.3pt+\y) -- (1.3pt+\x1,-1.2pt+\y) -- (1pt+\x1,-1.2pt+\y);

		\def\x1{20.5}

		\node at (3.75pt+\x1,2pt+\y) (three2){$1^+$};
		\node at (3.75pt+\x1,0pt+\y) (four2){$2^+$};
		\node at (1.35pt+\x1,0pt+\y)(six){};

		\draw [thick,dashed](six) arc (-90:90:1pt);
		\draw [decorate, decoration={coil, amplitude=2.3pt, segment length=3pt}] (five) -- (cross);
		\draw [decorate, decoration={coil, amplitude=2.3pt, segment length=3pt}](three2.west) arc (90:270:1pt);
		\draw [fill=gray!75] (2.35pt+\x1,1pt+\y) ellipse (0.4pt and 0.7pt);

		\node at (1pt+\x1,2.5pt+\y) (cut1){};
		\node at (1pt+\x1,-0.5pt+\y) (cut2){};
		\draw [dashed] (cut1) -- (cut2);
		\node at (1.1pt+\x1,2.4pt+\y)[rotate around={-90:(1pt,2.5pt)}]{\Cutright};
	\end{tikzpicture}
	\caption{A double cut of the scalar contribution to $\Tr F^2$ non-minimal. The red boxes highlight the two different operator insertions.}
	\label{fig:doublecutF2nm}
\end{figure}
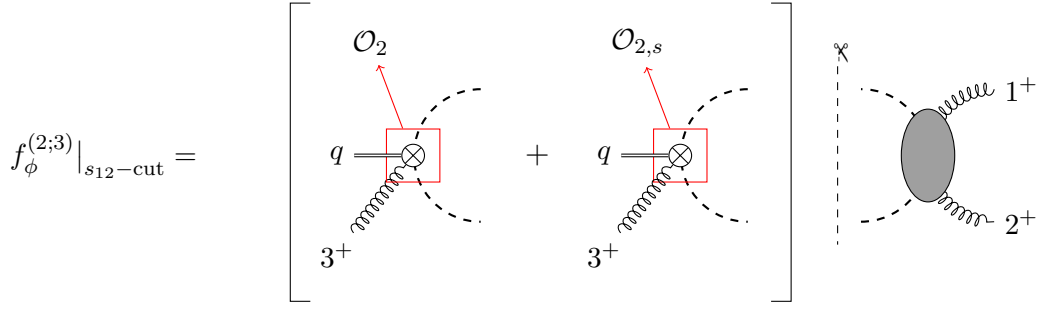

Performing the appropriate scalar subtraction for the non-minimal configuration of the operator $\Tr F^2$ is more subtle than in the minimal case. The double cut one needs to compute is represented in Figure~\ref{fig:doublecutF2nm}. There are two different tree-level form factors to be inserted into the cut: the non-minimal form factors with two external scalars and one gluon of the operators $\Tr F^2$ and $(D \phi)^2$.
%, both contributing to the double-cut of the scalar loop, these are
%\begin{equation}
%\def\arraystretch{1.5}
%\begin{array}{c}
%    F_{\mathcal{O}_2}^{(0)}(l_1,l_2,3_{c\dot{c}};q)\coloneqq (2 \pi)^4 \; \delta^4\left(q-\sum_i p_i\right)\; \langle l_1,l_2,3_{c \dot{c}} | \Tr F^2 |0\rangle \\
%        F_{\mathcal{O}_{2,s}}^{(0)}(l_1,l_2,3_{c\dot{c}};q)\coloneqq (2 \pi)^4 \; \delta^4\left(q-\sum_i p_i\right)\; \langle l_1,l_2,3_{c \dot{c}} | \Tr \phi^2 |0\rangle
%\end{array}
%\end{equation}
The tree-level expression for these form factors are given in~\eqref{eq::TrF2nonmininaml_scalars} and~\eqref{eq::Dphi2nonminimal} respectively.
Computing the complete result for the double-cut of the scalar contribution leads to
\begin{equation}\label{eq:f23scal}
	f^{(2;3)}_{\phi}(s_{12},s_{23},s_{13})\big|_{s_{12}\rm{-cut}}= i\frac{\sqr{1}{2}}{\sqr{2}{3}\sqr{3}{1}}\int \dd{\rm LIPS} \lbrack 3|\, \slashed{l}_1^{(4)}\, \slashed{l}_2^{(4)} | 3\rbrack\; \mathcal{I}_{\phi}^{(2;3)}\ ,
\end{equation}
with
\begin{equation}
	\mathcal{I}_{\phi}^{(2;3)}=- \mu^2 \frac{q^2 s_{12} + s_{3 l_1} s_{3 l_2} }{s_{12}^2 s_{2, -l_2} s_{3 l_1} s_{3 l_2}} \> .
\end{equation}
Upon subtracting twice~\eqref{eq:f23scal} from~\eqref{eq:f23gluon}, uplifting the cut and performing some algebraic manipulations on the numerator, one ends up with the final expression:

\begin{equation}
	\label{eq:TrF2nonminimalfinal}
	\def\arraystretch{4}
	\begin{array}{rl}
		f^{(2,3)}(s_{12},s_{23},s_{13})\big|_{\rm s_{12}-disc}= - & \frac{i q^4}{2s_{31}}
		\begin{tikzpicture}[baseline={([yshift=-1mm]Base.base)},scale=15]

			\def\x{0}
			\def\y{0}

			\clip (-1.4pt,-1.4pt) rectangle (3.5pt,3.5pt);

			\node at (0pt+\x,1pt+\y) (Base) {};

			\node at (0+\x,0+\y) (A) {};
			\node at (2pt+\x,0+\y) (B) {};
			\node at (2pt+\x,2pt+\y) (C) {};
			\node at (0+\x,2pt+\y) (D) {};

			\node at (-1pt+\x, -1pt+\y) (A1) {\footnotesize$p_3$};
			\node at (3pt+\x,-1pt+\y) (B1) {\footnotesize$p_2$};
			\node at (3pt+\x,3pt+\y) (C1) {\footnotesize$p_1$};
			\node at (-1pt+\x,3pt+\y) (D1) {\footnotesize$q$};

			\draw [thick] (A.center) -- (B.center);
			\draw [thick] (B.center) -- (C.center);
			\draw [thick] (C.center) -- (D.center);
			\draw [thick] (D.center) -- (A.center);
			\draw [thick] (A1) -- (A.center);
			\draw [thick] (B1) -- (B.center);
			\draw [thick] (C1) -- (C.center);
			\draw [thick,double] (D1) -- (D.center);

		\end{tikzpicture}
		-\frac{i q^4}{2s_{23}}
		\begin{tikzpicture}[baseline={([yshift=-1mm]Base.base)},scale=15]

			\def\x{0}
			\def\y{0}

			\clip (-1.4pt,-1.4pt) rectangle (3.5pt,3.5pt);

			\node at (0pt+\x,1pt+\y) (Base) {};

			\node at (0+\x,0+\y) (A) {};
			\node at (2pt+\x,0+\y) (B) {};
			\node at (2pt+\x,2pt+\y) (C) {};
			\node at (0+\x,2pt+\y) (D) {};

			\node at (-1pt+\x, -1pt+\y) (A1) {\footnotesize$p_2$};
			\node at (3pt+\x,-1pt+\y) (B1) {\footnotesize$p_1$};
			\node at (3pt+\x,3pt+\y) (C1) {\footnotesize$p_3$};
			\node at (-1pt+\x,3pt+\y) (D1) {\footnotesize$q$};

			\draw [thick] (A.center) -- (B.center);
			\draw [thick] (B.center) -- (C.center);
			\draw [thick] (C.center) -- (D.center);
			\draw [thick] (D.center) -- (A.center);
			\draw [thick] (A1) -- (A.center);
			\draw [thick] (B1) -- (B.center);
			\draw [thick] (C1) -- (C.center);
			\draw [thick,double] (D1) -- (D.center);

		\end{tikzpicture}
		\\
		                                                         - & \frac{i q^4 (s_{31}+s_{23})}{ s_{12} s_{23} s_{31}} \cdot\left( \begin{tikzpicture}[baseline={([yshift=-1mm]uno.base)},scale=15]
				\def\x{0}
				\def\y{0}

				\clip (-1.4pt,-2pt) rectangle (3.5pt,2pt);

				\node at (0+\x,0+\y)(uno){};
				\node at (1.5pt+\x,1pt+\y)(due){};
				\node at (1.5pt+\x,-1pt+\y)(tre){};
				%\node at (-0.75pt+\x,0.5pt+\y)(unoB){};
				%\node at (-0.75pt+\x,0.5pt+\y)(unoB){};
				\node at (-0.75pt+\x,0+\y)(unoS){};
				\node at (-1.2pt+\x,0pt+\y){\footnotesize$q$};
				\node at (2.25pt+\x,1.5pt+\y)(dueB){};
				\node [right=0.02pt of dueB, yshift=0.4pt]{\footnotesize$p_1$};
				\node at (2.25pt+\x,0.5pt+\y)(dueC){};
				\node [right=0.02pt of dueC, yshift=-0.4pt]{\footnotesize$p_2$};
				\node at (2.25pt+\x,-1.5pt+\y)(treB){};
				\node [right=-0.02pt of treB, yshift=-0.4pt]{\footnotesize$p_3$};
				\node at (0.75pt+\x,-1pt+\y) (end){};
				\node at (-4pt+\x,-2pt+\y)(start){};
				\node at (0.9pt+\x,0+\y)[thick,scale=1.5]{};

				\draw [thick] (uno.center) -- (due.center);
				\draw [thick] (due.center) -- (tre.center);
				\draw [thick] (uno.center) -- (tre.center);
				%\draw [thick] (uno.center) -- (unoB.center);
				%\draw [thick] (uno.center) -- (unoC.center);
				\draw [thick,double] (uno.center) -- (unoS.center);
				\draw [thick] (due.center) -- (dueB.center);
				\draw [thick] (due.center) -- (dueC.center);
				\draw [thick] (tre.center) -- (treB.center);

				%\draw [dotted,thick](-0.6pt+\x,0.3pt+\y) arc (140:240:0.4pt);
			\end{tikzpicture} +
		\begin{tikzpicture}[baseline={([yshift=-1mm]uno.base)},scale=15]
				\def\x{0}
				\def\y{0}

				\clip (-1.4pt,-2pt) rectangle (3.5pt,2pt);

				\node at (0+\x,0+\y)(uno){};
				\node at (1.5pt+\x,1pt+\y)(due){};
				\node at (1.5pt+\x,-1pt+\y)(tre){};
				%\node at (-0.75pt+\x,0.5pt+\y)(unoB){};
				%\node at (-0.75pt+\x,0.5pt+\y)(unoB){};
				\node at (-0.75pt+\x,0+\y)(unoS){};
				\node at (-1.2pt+\x,0pt+\y){\footnotesize$q$};
				\node at (2.25pt+\x,1.5pt+\y)(dueB){};
				\node [right=0.02pt of dueB, yshift=0.4pt]{\footnotesize$p_3$};
				\node at (2.25pt+\x,-0.5pt+\y)(treC){};
				\node [right=0.02pt of treC, yshift=+0.4pt]{\footnotesize$p_1$};
				\node at (2.25pt+\x,-1.5pt+\y)(treB){};
				\node [right=-0.02pt of treB, yshift=-0.4pt]{\footnotesize$p_2$};
				\node at (0.75pt+\x,-1pt+\y) (end){};
				\node at (-4pt+\x,-2pt+\y)(start){};
				\node at (0.9pt+\x,0+\y)[thick,scale=1.5]{};

				\draw [thick] (uno.center) -- (due.center);
				\draw [thick] (due.center) -- (tre.center);
				\draw [thick] (uno.center) -- (tre.center);
				%\draw [thick] (uno.center) -- (unoB.center);
				%\draw [thick] (uno.center) -- (unoC.center);
				\draw [thick,double] (uno.center) -- (unoS.center);
				\draw [thick] (due.center) -- (dueB.center);
				\draw [thick] (tre.center) -- (treC.center);
				\draw [thick] (tre.center) -- (treB.center);

				%\draw [dotted,thick](-0.6pt+\x,0.3pt+\y) arc (140:240:0.4pt);
			\end{tikzpicture}\right)                                                                                                                       \\
		                                                        + & \frac{4 i}{s_{12}^2} \cdot \begin{tikzpicture}[baseline={([yshift=-1mm]uno.base)},scale=15]
			\def\x{0}
			\def\y{0}

			\clip (-2.5pt,-2pt) rectangle (2.5pt,2pt);

			\def\x{0pt}
			\def\y{0}

			\node at (-0.75pt+\x,0pt+\y)(uno){};
			\node at (-1.5pt+\x,0.75pt+\y)(unoA){};
			\node at (-1.5pt+\x,-0.75pt+\y)(unoB){};
			\node at (-2pt+\x,0.75pt+\y){\footnotesize $q$};
			\node at (-2pt+\x,-0.75pt+\y){\footnotesize $p_3$};
			\node at (-1.5pt+\x,0+\y)(tre){};
			%\node at (-2pt+\x,0+\y){\footnotesize $p_4$};
			\node at (0.75pt+\x,0pt+\y)(due){};
			\node at (1.5pt+\x,0.75pt+\y)(dueA){};
			\node at (1.5pt+\x,-0.75pt+\y)(dueB){};
			\node at (2pt+\x,0.75pt+\y){\footnotesize $p_1$};
			\node at (2pt+\x,-0.75pt+\y){\footnotesize $p_2$};
			\node at (0pt+\x,0+\y)[thick,scale=1.5]{};
			\node at (0+\x,0+\y){\footnotesize $\mu^2$};
			\node at (0+\x,-1.2pt+\y){};

			\draw [thick] (0+\x,0+\y) circle (0.75pt);
			\draw [thick,double] (uno.center) -- (unoA.center);
			\draw [thick] (uno.center) -- (unoB.center);
			%\draw [thick] (uno.center) -- (tre.center);
			\draw [thick] (due.center) -- (dueA.center);
			\draw [thick] (due.center) -- (dueB.center);
		\end{tikzpicture} + \frac{2 i}{s_{12}} \cdot
		\begin{tikzpicture}[baseline={([yshift=-1mm]uno.base)},scale=15]
			\def\x{0}
			\def\y{0}

			\clip (-1.8pt,-2pt) rectangle (3.5pt,2pt);

			\node at (0+\x,0+\y)(uno){};
			\node at (1.5pt+\x,1pt+\y)(due){};
			\node at (1.5pt+\x,-1pt+\y)(tre){};
			\node at (-0.75pt+\x,0.5pt+\y)(unoB){};
			\node at (-0.75pt+\x,-0.5pt+\y)(unoC){};
			\node at (2.25pt+\x,1.5pt+\y)(dueB){};
			\node [right=0.02pt of dueB, yshift=0.4pt]{\footnotesize$p_1$};
			\node at (2.25pt+\x,-1.5pt+\y)(treB){};
			\node [right=-0.02pt of treB, yshift=-0.4pt]{\footnotesize$p_2$};
			\node at (0.75pt+\x,-1pt+\y) (end){};
			\node at (-4pt+\x,-2pt+\y)(start){};
			\node at (-1.25pt+\x,0.75pt+\y){\footnotesize$q$};
			\node at (-1.2pt+\x,-0.75pt+\y){\footnotesize$p_3$};
			%		\node at (0.9pt+\x,0+\y)[thick,scale=1.5]{$\circlearrowright$};
			\node at (1pt+\x,0.1pt+\y){\footnotesize$\mu^2$};
			\node at (-0.75pt+\x,0+\y)(cinque){};
			%\node at (-1.25pt+\x,0+\y){\footnotesize $p_4$};

			\draw [thick] (uno.center) -- (due.center);
			\draw [thick] (due.center) -- (tre.center);
			\draw [thick] (uno.center) -- (tre.center);
			\draw [thick,double] (uno.center) -- (unoB.center);
			\draw [thick] (uno.center) -- (unoC.center);
			\draw [thick] (due.center) -- (dueB.center);
			\draw [thick] (tre.center) -- (treB.center);
		\end{tikzpicture}
	\end{array}
\end{equation}
where all integrals can be found in Appendix \ref{sec:integrals}.

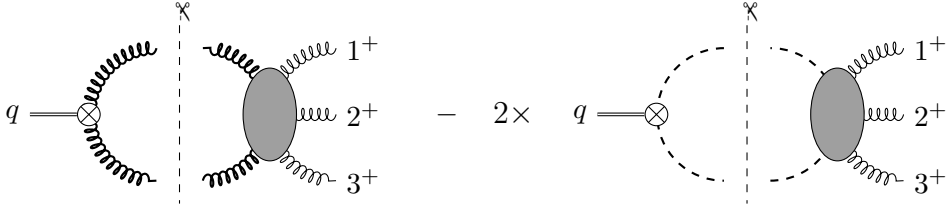
\begin{figure}
	\centering
	\begin{tikzpicture}[scale=25,auto,cross/.style={path picture={
							\draw[black]
							(path picture bounding box.south east) -- (path picture bounding box.north west) (path picture bounding box.south west) -- (path picture bounding box.north east);
						}}]

		%One loop the form factors

		\def\y{-7pt}
		\def\x1{11pt}

		\node at (0.5pt+\x1,2pt+\y) (one){};
		\draw [thick,decorate, decoration={coil, amplitude=2.3pt, segment length=3pt}](one.east) arc (90:270:1pt);
		\node [draw,circle,cross, inner sep=3pt,fill=white] at (-0.36pt+\x1,1pt+\y)(cross){};
		\node at (-1.5pt+\x1,1pt+\y)(Q){$q$};

		\draw [double](Q) -- (cross);

		\node at (3.75pt+\x1,2pt+\y) (three2){$1^+$};
		\node at (3.75pt+\x1,0pt+\y) (four2){$3^+$};
		\node at (3.75pt+\x1,1pt+\y) (five){$2^+$};
		\node at (1.35pt+\x1,0pt+\y)(six){};

		\draw [thick,decorate, decoration={coil, amplitude=2.3pt, segment length=3pt}](six) arc (-90:90:1pt);
		\draw [decorate, decoration={coil, amplitude=2.3pt, segment length=3pt}](three2.west) arc (90:270:1pt);
		\draw [decorate, decoration={coil, amplitude=2.3pt, segment length=3pt}] (five) -- (2.35pt+\x1,1pt+\y);
		\draw [fill=gray!75] (2.35pt+\x1,1pt+\y) ellipse (0.4pt and 0.7pt);

		\node at (1pt+\x1,2.5pt+\y) (cut1){};
		\node at (1pt+\x1,-0.5pt+\y) (cut2){};
		\draw [dashed] (cut1) -- (cut2);
		\node at (1.1pt+\x1,2.4pt+\y)[rotate around={-90:(1pt,2.5pt)}]{\Cutright};

		\node at (5pt+\x1,1pt+\y){$-$};
		\node at (6pt+\x1,1pt+\y){\large $2 \times$};

		\def\x1{19.5}
		\def\y{-7pt}

		\node at (0.5pt+\x1,2pt+\y) (one){};
		\draw [thick,dashed](one.east) arc (90:270:1pt);
		\node [draw,circle,cross, inner sep=3pt,fill=white] at (-0.36pt+\x1,1pt+\y)(cross){};
		\node at (-1.5pt+\x1,1pt+\y)(Q){$q$};

		\draw [double](Q) -- (cross);

		\node at (3.75pt+\x1,2pt+\y) (three2){$1^+$};
		\node at (3.75pt+\x1,0pt+\y) (four2){$3^+$};
		\node at (3.75pt+\x1,1pt+\y) (five){$2^+$};
		\node at (1.35pt+\x1,0pt+\y)(six){};

		\draw [thick,dashed](six) arc (-90:90:1pt);
		\draw [decorate, decoration={coil, amplitude=2.3pt, segment length=3pt}] (five) -- (2.35pt+\x1,1pt+\y);
		\draw [decorate, decoration={coil, amplitude=2.3pt, segment length=3pt}](three2.west) arc (90:270:1pt);
		\draw [fill=gray!75] (2.35pt+\x1,1pt+\y) ellipse (0.4pt and 0.7pt);

		\node at (1pt+\x1,2.5pt+\y) (cut1){};
		\node at (1pt+\x1,-0.5pt+\y) (cut2){};
		\draw [dashed] (cut1) -- (cut2);
		\node at (1.1pt+\x1,2.4pt+\y)[rotate around={-90:(1pt,2.5pt)}]{\Cutright};

		\node at (5pt+\x1,1pt+\y){};

	\end{tikzpicture}
	\caption{Two-particle cut of the one-loop form factor $\Tr F^2$ in the $s_{123}$ channel in six dimensions.}\label{fig:doublecutF2s123}
\end{figure}

%As can be seen, not only bubbles and triangles appear but also boxes, and in particular both the box-topologies probed by the cut.
Clearly the double cuts in the channels $s_{23}$ and $s_{13}$ can be derived from~\eqref{eq:TrF2nonminimalfinal} by symmetry arguments, thus the only invariant channel left to compute would be $s_{123}$, see Figure~\ref{fig:doublecutF2s123}. This double-cut involves the use of the five-point amplitudes in six dimensions with five gluons as well as with three gluons and two scalars\footnote{Their analytic expression is given in Appendix \ref{sec:SixDAmp}.}, combined with the minimal form factor of $\cO_2$ and $\cO_{2,s}$ respectively. The only topology probed by this cut, which is not probed by any of the previous cuts, is the bubble with the form factor in one of the two vertices and all the momenta in the other. Performing the calculation the associated coefficient turns out to be zero. Thus \eqref{eq:TrF2nonminimalfinal} and its permutations give the complete result, which matches the one given in \cite{Schmidt:1997wr,Davies:2011vt}.

\subsection{The Minimal \texorpdfstring{$\Tr F^3$}{TrF3} Form Factors}

We now consider the  $\Tr F^3$ form factor in the all-plus helicity configuration. The procedure we follow is exactly the same as in the $\Tr F^2$ case. First we factor out the helicity dependence as an overall tree-level prefactor:
\begin{equation}
	\def\arraystretch{1.5}
	\begin{array}{rl}
		F^{(1)}_{\cO_{3}} (1^+,2^+,3^+;q) \coloneqq & F^{(0)}_{\cO_{3}} (1^+,2^+,3^+;q)\cdot f^{(3)}\left(s_{1 2}, s_{23},s_{13}\right)   \\
		=                                           & -2\sqr{1}{2}\sqr{2}{3}\sqr{3}{1}\cdot f^{(3)}\left(s_{1 2}, s_{23},s_{13}\right)\ ,
	\end{array}
\end{equation}
then we compute $f^{(3)}$ as the difference $f^{(3)}_{\rm 6D}-2f^{(3)}_{\phi}$. We start with the two-particle cut in the $s_{12}$ channel represented in Figure~\ref{fig:doublecutF3}, which reads
\begin{equation}
	\label{eq::F3doublecut}
	\left. f_{\mathrm{6D}}^{(3)}\left(s_{1 2}\right)\right|_{s_{1 2}-\textrm{cut}} =-\frac{1}{2\sqr{1}{2}\sqr{2}{3}\sqr{3}{1}} \int \dd{\rm LIPS}\, F^{(0)}_{\cO_{3}}(-l_1^{a \dot{a}},-l_2^{b \dot{b}},3_{1 \dot{1}})\; \mathcal{A}^{(0)} (l_{2 \, a \dot{a}},l_{1 \, b \dot{b}},1_{1 \dot{1}},2_{2 \dot{2}})\, .
\end{equation}
Upon expanding the six-dimensional invariants we get a 168-term expression. This can be considerably simplified  using momentum conservation as in \eqref{eq:extramass} and choosing the $\mu$s to be
\begin{equation}\label{eq:muTrF3}
	\mu_{l_1} \mapsto \lambda_{l_2} \ , \hspace{0.5cm} \tilde{\mu}_{l_1} \mapsto \tilde{\lambda }_{3}\ , \hspace{0.5cm} \mu_{l_2} \mapsto \lambda_{l_1} \ , \hspace{0.5cm}\tilde{\mu }_{l_2}\mapsto \tilde{\lambda }_{3} \>.
\end{equation}
Doing so, we arrive at the compact expression
\begin{equation}
	f^{(3)}_{\rm 6D}(s_{12},s_{23},s_{13})\big|_{s_{12}\rm{-cut}}= i\int \dd{\rm LIPS}\left(\frac{\sqr{1}{2} \lbrack 3|\, \slashed{l}_1^{(4)}\, \slashed{l}_2^{(4)} | 3\rbrack }{s_{2 l_2} \sqr{2}{3}\sqr{3}{1}} + \mu^2 \frac{\lbrack3|\, \slashed{l}_1^{(4)}\, \slashed{l}_2^{(4)} | 3\rbrack}{\lbrack3|\, \slashed{p}_1\, \slashed{p}_2 | 3\rbrack} \right)\ ,
\end{equation}
where we have already reconstructed the full $d$-dimensional momenta. Computing the scalar contribution in a similar fashion\footnote{For the case of the scalar contribution it turns out that the most convenient choice for the $\mu$s  is the same as in the  gluon case. Notice that it is for this particular reason that we would have been allowed to perform the subtraction between the two contributions without writing them in terms of full $d$-dimensional quantities first. Indeed, if this were not the case, we would have had to reconstruct the form of the loop momenta in terms of general $\mu$s before doing the subtraction.} leads to
\begin{equation}
	f^{(3)}_{\phi}(s_{12},s_{23},s_{13})\big|_{s_{12}\rm{-cut}} = \frac{i}{2} \int \dd{\rm LIPS} \mu^2 \frac{\lbrack3|\, \slashed{l}_1^{(4)}\, \slashed{l}_2^{(4)} | 3\rbrack}{\lbrack3|\, \slashed{p}_1\, \slashed{p}_2 | 3\rbrack}\ ,
\end{equation}
and finally
\begin{equation}
	f^{(3)}(s_{12},s_{23},s_{13})\big|_{s_{12}\rm{-cut}}= i\frac{\sqr{1}{2}}{\sqr{2}{3}\sqr{3}{1}}\int \dd{\rm LIPS}\frac{ \lbrack 3|\, \slashed{l}_1^{(4)}\, \slashed{l}_2^{(4)} | 3\rbrack }{s_{2 l_2}}\ .
\end{equation}
After using \eqref{eq:muTrF3}, it is possible to write $f^{(3)}$ in terms of Mandelstam invariants:
\begin{equation}
	f^{(3)}(s_{12},s_{23},s_{13})\big|_{s_{12}\rm{-cut}}= - i \int \dd{\rm LIPS} \left( \frac{s_{1 2}}{s_{2 l_2}} + 2\right)
\end{equation}
modulo terms which integrate to zero. Uplifting this result leads to:

\begin{equation}
	\label{eq::TrF3comp}
	f^{(3)}(s_{12},s_{23},s_{13})\big|_{s_{12}-{\rm disc}}= - i s_{1 2}\cdot
	\begin{tikzpicture}[baseline={([yshift=-1mm]uno.base)},scale=15]
		\def\x{0}
		\def\y{0}

		\clip (-1.8pt,-2pt) rectangle (3.5pt,2pt);

		\node at (0+\x,0+\y)(uno){};
		\node at (1.5pt+\x,1pt+\y)(due){};
		\node at (1.5pt+\x,-1pt+\y)(tre){};
		\node at (-0.75pt+\x,0.5pt+\y)(unoB){};
		\node at (-0.75pt+\x,-0.5pt+\y)(unoC){};
		\node at (2.25pt+\x,1.5pt+\y)(dueB){};
		\node [right=0.02pt of dueB, yshift=0.4pt]{\footnotesize$p_1$};
		\node at (2.25pt+\x,-1.5pt+\y)(treB){};
		\node [right=-0.02pt of treB, yshift=-0.4pt]{\footnotesize$p_2$};
		\node at (0.75pt+\x,-1pt+\y) (end){};
		\node at (-4pt+\x,-2pt+\y)(start){};
		\node at (-1.25pt+\x,0.75pt+\y){\footnotesize$q$};
		\node at (-1.2pt+\x,-0.75pt+\y){\footnotesize$p_3$};
		%		\node at (0.9pt+\x,0+\y)[thick,scale=1.5]{$\circlearrowright$};
		\node at (0.9pt+\x,0+\y)[thick,scale=1.5]{};

		\draw [thick] (uno.center) -- (due.center);
		\draw [thick] (due.center) -- (tre.center);
		\draw [thick] (uno.center) -- node [yshift=-0.2cm] {\footnotesize$l$} (tre.center);
		\draw [thick,double] (uno.center) -- (unoB.center);
		\draw [thick] (uno.center) -- (unoC.center);
		\draw [thick] (due.center) -- (dueB.center);
		\draw [thick] (tre.center) -- (treB.center);

		%\draw [dotted,thick](-0.6pt+\x,0.3pt+\y) arc (140:240:0.4pt);
	\end{tikzpicture} -2 i \cdot
	\begin{tikzpicture}[baseline={([yshift=-1mm]uno.base)},scale=15]
		\def\x{0}
		\def\y{0}

		\clip (-2.5pt,-2pt) rectangle (2.5pt,2pt);

		\def\x{0pt}
		\def\y{0}

		\node at (-0.75pt+\x,0pt+\y)(uno){};
		\node at (-1.5pt+\x,0.75pt+\y)(unoA){};
		\node at (-1.5pt+\x,-0.75pt+\y)(unoB){};
		\node at (-2pt+\x,0.75pt+\y){\footnotesize $q$};
		\node at (-2pt+\x,-0.75pt+\y){\footnotesize $p_3$};
		\node at (0.75pt+\x,0pt+\y)(due){};
		\node at (1.5pt+\x,0.75pt+\y)(dueA){};
		\node at (1.5pt+\x,-0.75pt+\y)(dueB){};
		\node at (2pt+\x,0.75pt+\y){\footnotesize $p_1$};
		\node at (2pt+\x,-0.75pt+\y){\footnotesize $p_2$};
		%\node at (0pt+\x,0+\y)[thick,scale=1.5]{$\circlearrowright$};
		\node at (0pt+\x,0+\y)[thick,scale=1.5]{};
		\node at (0+\x,-1.2pt+\y){\footnotesize $l$};

		\draw [thick] (0+\x,0+\y) circle (0.75pt);
		\draw [thick,double] (uno.center) -- (unoA.center);
		\draw [thick] (uno.center) -- (unoB.center);
		\draw [thick] (due.center) -- (dueA.center);
		\draw [thick] (due.center) -- (dueB.center);
	\end{tikzpicture}\ .
\end{equation}
%with
%\begin{equation}
%	\begin{tikzpicture}[baseline={([yshift=-1mm]uno.base)},scale=15]
%		\def\x{0}
%		\def\y{0}
%
%		\clip (-2.5pt,-2pt) rectangle (2.5pt,2pt);
%
%
%		\def\x{0pt}
%		\def\y{0}
%
%		\node at (-0.75pt+\x,0pt+\y)(uno){};
%		\node at (-1.5pt+\x,0.75pt+\y)(unoA){};
%		\node at (-1.5pt+\x,-0.75pt+\y)(unoB){};
%		\node at (-2pt+\x,0.75pt+\y){\footnotesize $q$};
%		\node at (-2pt+\x,-0.75pt+\y){\footnotesize $p_3$};
%		\node at (0.75pt+\x,0pt+\y)(due){};
%		\node at (1.5pt+\x,0.75pt+\y)(dueA){};
%		\node at (1.5pt+\x,-0.75pt+\y)(dueB){};
%		\node at (2pt+\x,0.75pt+\y){\footnotesize $p_1$};
%		\node at (2pt+\x,-0.75pt+\y){\footnotesize $p_2$};
%		\node at (0pt+\x,0+\y)[thick,scale=1.5]{};
%		\node at (0+\x,-1.2pt+\y){\footnotesize $l$};
%
%		\draw [thick] (0+\x,0+\y) circle (0.75pt);
%		\draw [thick,double] (uno.center) -- (unoA.center);
%		\draw [thick] (uno.center) -- (unoB.center);
%		\draw [thick] (due.center) -- (dueA.center);
%		\draw [thick] (due.center) -- (dueB.center);
%	\end{tikzpicture} \> = \int \frac{\dd^d l}{(2 \pi)^{d}} \;\dfrac{1}{l^2\, (l+p_1+p_2)^2} \> .
%\end{equation}

Combining the discontinuities in the three channels $s_{1 2}$, $s_{2 3}$ and $s_{3 1}$ we arrive at the complete one-loop form factor
\begin{equation}
	\label{eq::sum_cuts}
	f^{(3)}(\{s_{ij}\})= \sum_{k=1}^nf^{(3)}(\{s_{ij}\})\big|_{s_{k\, k+1}-{\rm disc}}\ ,
\end{equation}
where every term in the sum can be obtained from \eqref{eq::TrF3comp} by relabelling the external legs.

\begin{figure}
	\centering
	\begin{tikzpicture}[scale=25,auto,cross/.style={path picture={
							\draw[black]
							(path picture bounding box.south east) -- (path picture bounding box.north west) (path picture bounding box.south west) -- (path picture bounding box.north east);
						}}]

		%One loop the form factors

		\def\y{-7pt}
		\def\x1{11pt}

		\node at (0.5pt+\x1,2pt+\y) (one){};
		\draw [thick,decorate, decoration={coil, amplitude=2.3pt, segment length=3pt}](one.east) arc (90:270:1pt);
		\node [draw,circle,cross, inner sep=3pt,fill=white] at (-0.36pt+\x1,1pt+\y)(cross){};
		\node at (-1.5pt+\x1,1pt+\y)(Q){$q$};

		\draw [double](Q) -- (cross);

		\node at (3.75pt+\x1,2pt+\y) (three2){$1^+$};
		\node at (3.75pt+\x1,0pt+\y) (four2){$2^+$};
		\node at (-1.5pt+\x1,-0.5pt+\y) (five){$3^+$};
		\node at (1.35pt+\x1,0pt+\y)(six){};

		\draw [thick,decorate, decoration={coil, amplitude=2.3pt, segment length=3pt}](six) arc (-90:90:1pt);
		\draw [decorate, decoration={coil, amplitude=2.3pt, segment length=3pt}](three2.west) arc (90:270:1pt);
		\draw [fill=gray!75] (2.35pt+\x1,1pt+\y) ellipse (0.4pt and 0.7pt);
		\draw [decorate, decoration={coil, amplitude=2.3pt, segment length=3pt}] (five) -- (cross);

		\node at (1pt+\x1,2.5pt+\y) (cut1){};
		\node at (1pt+\x1,-0.5pt+\y) (cut2){};
		\draw [dashed] (cut1) -- (cut2);
		\node at (1.1pt+\x1,2.4pt+\y)[rotate around={-90:(1pt,2.5pt)}]{\Cutright};

		\node at (5pt+\x1,1pt+\y){$-$};
		\node at (6pt+\x1,1pt+\y){\large $2 \times$};

		\def\x1{19.5}
		\def\y{-7pt}

		\node at (0.5pt+\x1,2pt+\y) (one){};
		\draw [thick,dashed](one.east) arc (90:270:1pt);
		\node [draw,circle,cross, inner sep=3pt,fill=white] at (-0.36pt+\x1,1pt+\y)(cross){};
		\node at (-1.5pt+\x1,1pt+\y)(Q){$q$};

		\draw [double](Q) -- (cross);

		\node at (3.75pt+\x1,2pt+\y) (three2){$1^+$};
		\node at (3.75pt+\x1,0pt+\y) (four2){$2^+$};
		\node at (-1.5pt+\x1,-0.5pt+\y) (five){$3^+$};
		\node at (1.35pt+\x1,0pt+\y)(six){};

		\draw [thick,dashed](six) arc (-90:90:1pt);
		\draw [decorate, decoration={coil, amplitude=2.3pt, segment length=3pt}] (five) -- (cross);
		\draw [decorate, decoration={coil, amplitude=2.3pt, segment length=3pt}](three2.west) arc (90:270:1pt);
		\draw [fill=gray!75] (2.35pt+\x1,1pt+\y) ellipse (0.4pt and 0.7pt);

		\node at (1pt+\x1,2.5pt+\y) (cut1){};
		\node at (1pt+\x1,-0.5pt+\y) (cut2){};
		\draw [dashed] (cut1) -- (cut2);
		\node at (1.1pt+\x1,2.4pt+\y)[rotate around={-90:(1pt,2.5pt)}]{\Cutright};

	\end{tikzpicture}
	\caption{Two-particle cut of the one-loop form factor $\Tr F^3$ in the $s_{12}$ channel in six dimensions.}\label{fig:doublecutF3}
\end{figure}
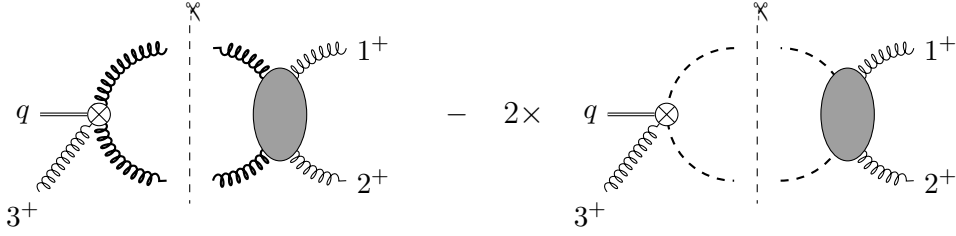

\subsection{The Non-Minimal \texorpdfstring{$\Tr F^3$}{TrF3}  Form Factor}

In the last sections we showed how the dimensional reconstruction can be applied to form factors. In this section we derive for the first time  the complete form factor of the operator $\Tr F^3$ with four gluons in the all-plus helicity configuration.

The procedure we follow has been described in detail earlier, hence we now only sketch the relevant derivations and provide the main results. Up to cyclic permutations there are two independent unitarity cuts to be computed, say in the $s_{12}$-channel and $s_{123}$-channel. 
Starting from the $s_{123}$-cut, one needs to evaluate the following difference to obtain the complete result:

\begin{figure}[h]
	\centering
	\begin{tikzpicture}[scale=25,auto,cross/.style={path picture={
							\draw[black]
							(path picture bounding box.south east) -- (path picture bounding box.north west) (path picture bounding box.south west) -- (path picture bounding box.north east);
						}}]

		%One loop the form factors

		\def\y{-7pt}
		\def\x1{11pt}

		\node at (0.5pt+\x1,2pt+\y) (one){};
		\draw [thick,decorate, decoration={coil, amplitude=2.3pt, segment length=3pt}](one.east) arc (90:270:1pt);
		\node [draw,circle,cross, inner sep=3pt,fill=white] at (-0.36pt+\x1,1pt+\y)(cross){};
		\node at (-1.5pt+\x1,1pt+\y)(Q){$q$};

		\draw [double](Q) -- (cross);

		\node at (3.75pt+\x1,2pt+\y) (three2){$1^+$};
		\node at (3.75pt+\x1,0pt+\y) (four2){$3^+$};
		\node at (-1.5pt+\x1,-0.5pt+\y) (five){$4^+$};
		\node at (3.75pt+\x1,1pt+\y) (sette){$2^+$};
		\node at (1.35pt+\x1,0pt+\y)(six){};

		\draw [thick,decorate, decoration={coil, amplitude=2.3pt, segment length=3pt}](six) arc (-90:90:1pt);
		\draw [decorate, decoration={coil, amplitude=2.3pt, segment length=3pt}](three2.west) arc (90:270:1pt);
		\draw [decorate, decoration={coil, amplitude=2.3pt, segment length=3pt}](2.35pt+\x1,1pt+\y) -- (sette);
		\draw [fill=gray!75] (2.35pt+\x1,1pt+\y) ellipse (0.4pt and 0.7pt);
		\draw [decorate, decoration={coil, amplitude=2.3pt, segment length=3pt}] (five) -- (cross);

		\node at (1pt+\x1,2.5pt+\y) (cut1){};
		\node at (1pt+\x1,-0.5pt+\y) (cut2){};
		\draw [dashed] (cut1) -- (cut2);
		\node at (1.1pt+\x1,2.4pt+\y)[rotate around={-90:(1pt,2.5pt)}]{\Cutright};

		\node at (5pt+\x1,1pt+\y){$-$};
		\node at (6pt+\x1,1pt+\y){\large $2 \times$};

		\def\x1{19.5}
		\def\y{-7pt}

		\node at (0.5pt+\x1,2pt+\y) (one){};
		\draw [thick,dashed](one.east) arc (90:270:1pt);
		\node [draw,circle,cross, inner sep=3pt,fill=white] at (-0.36pt+\x1,1pt+\y)(cross){};
		\node at (-1.5pt+\x1,1pt+\y)(Q){$q$};

		\draw [double](Q) -- (cross);

		\node at (3.75pt+\x1,2pt+\y) (three2){$1^+$};
		\node at (3.75pt+\x1,0pt+\y) (four2){$3^+$};
		\node at (-1.5pt+\x1,-0.5pt+\y) (five){$4^+$};
		\node at (3.75pt+\x1,1pt+\y) (sette){$2^+$};
		\node at (1.35pt+\x1,0pt+\y)(six){};

		\draw [thick,dashed](six) arc (-90:90:1pt);
		\draw [decorate, decoration={coil, amplitude=2.3pt, segment length=3pt}] (five) -- (cross);
		\draw [decorate, decoration={coil, amplitude=2.3pt, segment length=3pt}](three2.west) arc (90:270:1pt);
		\draw [decorate, decoration={coil, amplitude=2.3pt, segment length=3pt}](2.35pt+\x1,1pt+\y) -- (sette);
		\draw [fill=gray!75] (2.35pt+\x1,1pt+\y) ellipse (0.4pt and 0.7pt);

		\node at (1pt+\x1,2.5pt+\y) (cut1){};
		\node at (1pt+\x1,-0.5pt+\y) (cut2){};
		\draw [dashed] (cut1) -- (cut2);
		\node at (1.1pt+\x1,2.4pt+\y)[rotate around={-90:(1pt,2.5pt)}]{\Cutright};

	\end{tikzpicture}
	%\caption{Two-particle cut of the one-loop non-minimal form factor $\Tr F^3$ in the $s_{123}$ channel in six dimensions.}
	\label{fig:doublecutF3_4}
\end{figure}
where the tree-level form factor in the scalar subtraction term (second term in the figure above) is the minimal form factor  of the operator $\Tr \big(D_{\mu}\phi D_{\nu}\phi F^{\mu \nu}\big)$. The Passarino-Veltman reductions of the resulting tensor integrals have been performed using  the \texttt{Mathematica} package \texttt{FeynCalc} \cite{Mertig:1990an,Shtabovenko:2016sxi}. From the two-particle cuts in the $s_{123}$-channel we obtain the following functions: 
%Among the different contributions we obtained the coefficients of the two-mass triangles and the one-mass boxes, which can be %obtained also from the cuts in the $s_{ij}$ channels. This be used as a sanity check of the results. The bubble integral on the other %hand is peculiar to this cut.

\begin{equation}
	\label{eq:TrF3nonminimalfinals123}
	\def\arraystretch{4}
	\begin{array}{rl}
		F^{(1)}_{\cO_{3}} (1^+,2^+,3^+,4^+;q)\big|_{\rm s_{123}-disc} = & D^{[0]}_0
		\begin{tikzpicture}[baseline={([yshift=-1mm]Base.base)},scale=15]

			\def\x{0}
			\def\y{0}

			\clip (-1.55pt,-1.4pt) rectangle (3.5pt,3.5pt);

			\node at (0pt+\x,1pt+\y) (Base) {};

			\node at (0+\x,0+\y) (A) {};
			\node at (2pt+\x,0+\y) (B) {};
			\node at (2pt+\x,2pt+\y) (C) {};
			\node at (0+\x,2pt+\y) (D) {};

			\node at (-1pt+\x, -1pt+\y) (A1) {\footnotesize$p_3$};
			\node at (3pt+\x,-1pt+\y) (B1) {\footnotesize$p_2$};
			\node at (3pt+\x,3pt+\y) (C1) {\footnotesize$p_1$};
			\node at (0pt+\x,3.3pt+\y) (D1) {\footnotesize$q$};
			\node at (-1.3pt+\x,2pt+\y) (D2) {\footnotesize$p_4$};

			\draw [thick] (A.center) -- (B.center);
			\draw [thick] (B.center) -- (C.center);
			\draw [thick] (C.center) -- (D.center);
			\draw [thick] (D.center) -- (A.center);
			\draw [thick] (A1) -- (A.center);
			\draw [thick] (B1) -- (B.center);
			\draw [thick] (C1) -- (C.center);
			\draw [thick,double] (D1) -- (D.center);
			\draw [thick] (D2) -- (D.center);

		\end{tikzpicture}

		+C^{[0]}_0\begin{tikzpicture}[baseline={([yshift=-1mm]uno.base)},scale=15]
			\def\x{0}
			\def\y{0}

			\clip (-1.8pt,-2pt) rectangle (3.5pt,2pt);

			\node at (0+\x,0+\y)(uno){};
			\node at (1.5pt+\x,1pt+\y)(due){};
			\node at (1.5pt+\x,-1pt+\y)(tre){};
			\node at (-0.75pt+\x,0.5pt+\y)(unoB){};
			\node at (-0.75pt+\x,-0.5pt+\y)(unoC){};
			\node at (2.25pt+\x,1.5pt+\y)(dueB){};
			\node [right=0.02pt of dueB, yshift=0.4pt]{\footnotesize$p_1$};
			\node at (2.25pt+\x,-1.5pt+\y)(treB){};
			\node [right=-0.02pt of treB, yshift=-0.4pt]{\footnotesize$p_3$};
			\node at (0.75pt+\x,-1pt+\y) (end){};
			\node at (-4pt+\x,-2pt+\y)(start){};
			\node at (-1.25pt+\x,0.75pt+\y){\footnotesize$q$};
			\node at (-1.2pt+\x,-0.75pt+\y){\footnotesize$p_4$};
			%		\node at (0.9pt+\x,0+\y)[thick,scale=1.5]{$\circlearrowright$};
			%\node at (1pt+\x,0.1pt+\y){\footnotesize$\mu^2$};
			\node at (-0.75pt+\x,0+\y)(cinque){};
			%\node at (-1.25pt+\x,0+\y){\footnotesize $p_4$};
			\node at (2.25pt+\x,-0.5pt+\y)(treC){};
			\node [right=0.02pt of treC, yshift=+0.4pt]{\footnotesize$p_2$};

			\draw [thick] (uno.center) -- (due.center);
			\draw [thick] (due.center) -- (tre.center);
			\draw [thick] (uno.center) -- (tre.center);
			\draw [thick,double] (uno.center) -- (unoB.center);
			\draw [thick] (uno.center) -- (unoC.center);
			\draw [thick] (due.center) -- (dueB.center);
			\draw [thick] (tre.center) -- (treB.center);
			\draw [thick] (tre.center) -- (treC.center);
		\end{tikzpicture}                                     \\

		+ & C^{[0]}_1\begin{tikzpicture}[baseline={([yshift=-1mm]uno.base)},scale=15]
			\def\x{0}
			\def\y{0}

			\clip (-1.8pt,-2pt) rectangle (3.5pt,2pt);

			\node at (0+\x,0+\y)(uno){};
			\node at (1.5pt+\x,1pt+\y)(due){};
			\node at (1.5pt+\x,-1pt+\y)(tre){};
			\node at (-0.75pt+\x,0.5pt+\y)(unoB){};
			\node at (-0.75pt+\x,-0.5pt+\y)(unoC){};
			\node at (2.25pt+\x,1.5pt+\y)(dueB){};
			\node [right=0.02pt of dueB, yshift=0.4pt]{\footnotesize$p_1$};
			\node at (2.25pt+\x,-1.5pt+\y)(treB){};
			\node [right=-0.02pt of treB, yshift=-0.4pt]{\footnotesize$p_3$};
			\node at (0.75pt+\x,-1pt+\y) (end){};
			\node at (-4pt+\x,-2pt+\y)(start){};
			\node at (-1.25pt+\x,0.75pt+\y){\footnotesize$q$};
			\node at (-1.2pt+\x,-0.75pt+\y){\footnotesize$p_4$};
			%\node at (1pt+\x,0.1pt+\y){\footnotesize$\mu^2$};
			\node at (-0.75pt+\x,0+\y)(cinque){};
			\node at (2.25pt+\x,0.5pt+\y)(treC){};
			\node [right=0.02pt of treC, yshift=+0.4pt]{\footnotesize$p_2$};

			\draw [thick] (uno.center) -- (due.center);
			\draw [thick] (due.center) -- (tre.center);
			\draw [thick] (uno.center) -- (tre.center);
			\draw [thick,double] (uno.center) -- (unoB.center);
			\draw [thick] (uno.center) -- (unoC.center);
			\draw [thick] (due.center) -- (dueB.center);
			\draw [thick] (tre.center) -- (treB.center);
			\draw [thick] (due.center) -- (treC.center);
		\end{tikzpicture}

		+B^{[0]}_0 \begin{tikzpicture}[baseline={([yshift=-1mm]uno.base)},scale=15]
			\def\x{0}
			\def\y{0}

			\clip (-2.5pt,-2pt) rectangle (2.5pt,2pt);

			\def\x{0pt}
			\def\y{0}

			\node at (-0.75pt+\x,0pt+\y)(uno){};
			\node at (-1.5pt+\x,0.75pt+\y)(unoA){};
			\node at (-1.5pt+\x,-0.75pt+\y)(unoB){};
			\node at (-2pt+\x,0.75pt+\y){\footnotesize $q$};
			\node at (-2pt+\x,-0.75pt+\y){\footnotesize $p_4$};
			\node at (-1.5pt+\x,0+\y)(tre){};
			%\node at (-2pt+\x,0+\y){\footnotesize $p_4$};
			\node at (0.75pt+\x,0pt+\y)(due){};
			\node at (1.5pt+\x,0.75pt+\y)(dueA){};
			\node at (1.5pt+\x,-0.75pt+\y)(dueB){};
			\node at (2.2pt+\x,0.85pt+\y){\footnotesize $p_1$};
			\node at (2.2pt+\x,-0.85pt+\y){\footnotesize $p_3$};
			\node at (0pt+\x,0+\y)[thick,scale=1.5]{};
			%\node at (0+\x,0+\y){\footnotesize $\mu^2$};
			\node at (0+\x,-1.2pt+\y){};
			\node at (2.2pt+\x,0pt+\y) (dueC) {\footnotesize $p_2$};

			\draw [thick] (0+\x,0+\y) circle (0.75pt);
			\draw [thick,double] (uno.center) -- (unoA.center);
			\draw [thick] (uno.center) -- (unoB.center);
			%\draw [thick] (uno.center) -- (tre.center);
			\draw [thick] (due.center) -- (dueA.center);
			\draw [thick] (due.center) -- (dueB.center);
			\draw [thick] (due.center) -- (dueC);
		\end{tikzpicture}\ ,
	\end{array}
\end{equation}
with the coefficients
\begin{equation}
	\setlength{\jot}{7pt}
	\begin{split}
		D^{[0]}_0 &= -i \, {\rm F}(1,2,3;4)\ ,\\
		C^{[0]}_0 &= - i\frac{s_{12}+s_{31}}{s_{12} s_{23}}\, {\rm F}(1,2,3;4)\ ,\\
		C^{[0]}_1 &= - i\frac{s_{23}+s_{31}}{s_{12} s_{23}}\, {\rm F}(1,2,3;4)\ ,\\
	\end{split}
\end{equation}
where
\begin{equation}
	{\rm F}(1,2,3;4)\coloneqq \frac{s_{123}}{s_{31}^2}\left(s_{12}\sqr{1}{3}\sqr{2}{4}+s_{31}\sqr{1}{2}\sqr{3}{4}\right)\left(s_{23}\sqr{1}{3}\sqr{2}{4}+s_{31} \sqr{2}{3}\sqr{1}{4}\right)\ .
\end{equation}
Finally the coefficient of the bubble can be written as
\begin{equation}
	B^{[0]}_0 = 2i\sqr{1}{2}\sqr{2}{3}\sqr{3}{4}\sqr{4}{1}\, \mathrm{b}^{[0]}_0\ ,
\end{equation}
where the helicity-blind function $\mathrm{b}^{[0]}_0$ is defined as
\begin{equation}
	\mathrm{b}^{[0]}_0 = \frac{s_{123}^2}{s_{12}s_{23}}\left(\frac{1}{s_{12}+s_{31}}+\frac{1}{s_{23}+s_{31}}\right)+\frac{\sqr{1}{3}\sqr{2}{4}}{\sqr{1}{2}\sqr{3}{4}} \frac{s_{123}^2}{s_{23}s_{31}}\cdot\frac{1}{s_{12}+s_{31}}+\frac{\sqr{1}{3}\sqr{2}{4}}{\sqr{1}{4}\sqr{2}{3}} \frac{s_{123}^2}{s_{12}s_{31}}\cdot\frac{1}{s_{23}+s_{31}}\ .
\end{equation}
The result also contains a box  integral with a $\mu^2$ i numerator, which after integration is of $\cO (\epsilon)$. For completeness we quote its coefficient:  
\begin{equation}
	D^{[2]}_0 = -2 i s_{123} \left(\frac{\sqr{1}{2}^2 \sqr{3}{4}^2}{s_{12}}+\frac{\sqr{2}{3}^2\sqr{4}{1}^2}{s_{23}}+\frac{\sqr{1}{3}^2\sqr{2}{4}^2}{s_{31}}\right)\ .
\end{equation}

Next we consider the two-particle cut in the $s_{12}$-channel and, as discussed in earlier sections,  the discontinuity of the complete form factor is determined from the difference 
\begin{equation}
	F^{(1)}_{\rm 6D} (1^+,2^+,3^+,4^+;q)\big|_{s_{12}\rm{-cut}} -2 F^{(1)}_{\phi} (1^+,2^+,3^+,4^+;q)\big|_{s_{12}\rm{-cut}}\ , 
\end{equation}
where the second term is the scalar subtraction. 
As in the case of the non-minimal form factor of $\Tr F^2$, there are two contributions to the scalar quantity $F^{(1)}_{\phi} (1^+,2^+,3^+,4^+;q)$, which are represented in Figure~\ref{fig:doublecutF3nm}. The first contribution comes from the operator $\Tr F^3$ with two scalars and two gluons, whereas the second one comes from the scalar operator $\Tr D_{\mu}\phi D_{\nu}\phi F^{\mu \nu}$. 

\begin{figure}[h]
	\centering
	\begin{tikzpicture}[scale=25,auto,cross/.style={path picture={
							\draw[black]
							(path picture bounding box.south east) -- (path picture bounding box.north west) (path picture bounding box.south west) -- (path picture bounding box.north east);
						}}]

		\def\x1{15.5}
		\def\y{-7pt}

		\node at (-5.6pt+\x1,1pt+\y) {$F^{(1)}_{\phi} (1^+,2^+,3^+,4^+;q)\big|_{s_{12}\rm{-cut}}=$};

		\draw (-1.9pt+\x1,3.3pt+\y) -- (-2.2pt+\x1,3.3pt+\y) -- (-2.2pt+\x1,-1.2pt+\y) -- (-1.9pt+\x1,-1.2pt+\y);

		\node at (0.5pt+\x1,2pt+\y) (one){};
		\draw [thick,dashed](one.east) arc (90:270:1pt);
		\node [draw,circle,cross, inner sep=3pt,fill=white] at (-0.36pt+\x1,1pt+\y)(cross){};
		\node at (-1.5pt+\x1,1pt+\y)(Q){$q$};
		\node at (-1.5pt+\x1,-0.5pt+\y) (five){$4^+$};
		\node at (-0.3pt+\x1,-0.5pt+\y) (sei) {$3^+$};
		\draw [decorate, decoration={coil, amplitude=2.3pt, segment length=3pt}] (five) -- (cross);
		\draw [decorate, decoration={coil, amplitude=2.3pt, segment length=3pt}] (sei) -- (cross);
		\node [draw, rectangle, red, inner sep=10pt] at (-0.36pt+\x1,1pt+\y)(square){};
		\node at (-1pt+\x1,2.7pt+\y) (operator){$\mathcal{O}_3$};
		\draw [->,red] (square) -- (operator);

		\draw [double](Q) -- (cross);

		\node at (1.5pt+\x1,1pt+\y) {\large $+$};

		\def\x1{19.5}
		\def\y{-7pt}

		\node at (0.5pt+\x1,2pt+\y) (one){};
		\draw [thick,dashed](one.east) arc (90:270:1pt);
		\node [draw,circle,cross, inner sep=3pt,fill=white] at (-0.36pt+\x1,1pt+\y)(cross){};
		\node at (-1.5pt+\x1,1pt+\y)(Q){$q$};
		\node [draw, rectangle, red, inner sep=10pt] at (-0.36pt+\x1,1pt+\y)(square){};
		\node at (-1pt+\x1,2.7pt+\y) (operator){$\mathcal{O}_{3,s}$};
		\draw [->,red] (square) -- (operator);
		\draw [double](Q) -- (cross);
		\node at (-1.5pt+\x1,-0.5pt+\y) (five){$4^+$};
		\node at (-0.3pt+\x1,-0.5pt+\y) (sei) {$3^+$};

		\draw (1pt+\x1,3.3pt+\y) -- (1.3pt+\x1,3.3pt+\y) -- (1.3pt+\x1,-1.2pt+\y) -- (1pt+\x1,-1.2pt+\y);

		\def\x1{20.5}

		\node at (3.75pt+\x1,2pt+\y) (three2){$1^+$};
		\node at (3.75pt+\x1,0pt+\y) (four2){$2^+$};
		\node at (1.35pt+\x1,0pt+\y)(six){};

		\draw [thick,dashed](six) arc (-90:90:1pt);
		\draw [decorate, decoration={coil, amplitude=2.3pt, segment length=3pt}] (five) -- (cross);
		\draw [decorate, decoration={coil, amplitude=2.3pt, segment length=3pt}] (sei) -- (cross);
		\draw [decorate, decoration={coil, amplitude=2.3pt, segment length=3pt}](three2.west) arc (90:270:1pt);
		\draw [fill=gray!75] (2.35pt+\x1,1pt+\y) ellipse (0.4pt and 0.7pt);

		\node at (1pt+\x1,2.5pt+\y) (cut1){};
		\node at (1pt+\x1,-0.5pt+\y) (cut2){};
		\draw [dashed] (cut1) -- (cut2);
		\node at (1.1pt+\x1,2.4pt+\y)[rotate around={-90:(1pt,2.5pt)}]{\Cutright};
	\end{tikzpicture}
	\caption{A two-particle cut of the scalar contribution to the non-minimal $\Tr F^3$ form factor. The red boxes highlight the two different operator insertions.}
	\label{fig:doublecutF3nm}
\end{figure}

After  tensor reductions, we find

\begin{equation}
	\label{eq:TrF3nonminimalfinals12}
	\def\arraystretch{4}
	\begin{array}{rl}
		F^{(1)}_{\cO_{3}} (1^+,2^+,3^+,4^+;q)\big|_{\rm s_{12}-disc}= & D^{[0]}_0\;
		\begin{tikzpicture}[baseline={([yshift=-1mm]Base.base)},scale=15]

			\def\x{0}
			\def\y{0}

			\clip (-1.55pt,-1.4pt) rectangle (3.5pt,3.5pt);

			\node at (0pt+\x,1pt+\y) (Base) {};

			\node at (0+\x,0+\y) (A) {};
			\node at (2pt+\x,0+\y) (B) {};
			\node at (2pt+\x,2pt+\y) (C) {};
			\node at (0+\x,2pt+\y) (D) {};

			\node at (-1pt+\x, -1pt+\y) (A1) {\footnotesize$p_3$};
			\node at (3pt+\x,-1pt+\y) (B1) {\footnotesize$p_2$};
			\node at (3pt+\x,3pt+\y) (C1) {\footnotesize$p_1$};
			\node at (0pt+\x,3.3pt+\y) (D1) {\footnotesize$q$};
			\node at (-1.3pt+\x,2pt+\y) (D2) {\footnotesize$p_4$};

			\draw [thick] (A.center) -- (B.center);
			\draw [thick] (B.center) -- (C.center);
			\draw [thick] (C.center) -- (D.center);
			\draw [thick] (D.center) -- (A.center);
			\draw [thick] (A1) -- (A.center);
			\draw [thick] (B1) -- (B.center);
			\draw [thick] (C1) -- (C.center);
			\draw [thick,double] (D1) -- (D.center);
			\draw [thick] (D2) -- (D.center);

		\end{tikzpicture}
		+D^{[0]}_1\;
		\begin{tikzpicture}[baseline={([yshift=-1mm]Base.base)},scale=15]

			\def\x{0}
			\def\y{0}

			\clip (-1.55pt,-1.4pt) rectangle (3.5pt,3.5pt);

			\node at (0pt+\x,1pt+\y) (Base) {};

			\node at (0+\x,0+\y) (A) {};
			\node at (2pt+\x,0+\y) (B) {};
			\node at (2pt+\x,2pt+\y) (C) {};
			\node at (0+\x,2pt+\y) (D) {};

			\node at (-1pt+\x, -1pt+\y) (A1) {\footnotesize$p_2$};
			\node at (3pt+\x,-1pt+\y) (B1) {\footnotesize$p_1$};
			\node at (3pt+\x,3pt+\y) (C1) {\footnotesize$p_4$};
			\node at (0pt+\x,3.3pt+\y) (D1) {\footnotesize$q$};
			\node at (-1.3pt+\x,2pt+\y) (D2) {\footnotesize$p_3$};

			\draw [thick] (A.center) -- (B.center);
			\draw [thick] (B.center) -- (C.center);
			\draw [thick] (C.center) -- (D.center);
			\draw [thick] (D.center) -- (A.center);
			\draw [thick] (A1) -- (A.center);
			\draw [thick] (B1) -- (B.center);
			\draw [thick] (C1) -- (C.center);
			\draw [thick,double] (D1) -- (D.center);
			\draw [thick] (D2) -- (D.center);

		\end{tikzpicture}
		\\

		+& C^{[0]}_1\begin{tikzpicture}[baseline={([yshift=-1mm]uno.base)},scale=15]
			\def\x{0}
			\def\y{0}

			\clip (-1.8pt,-2pt) rectangle (3.5pt,2pt);

			\node at (0+\x,0+\y)(uno){};
			\node at (1.5pt+\x,1pt+\y)(due){};
			\node at (1.5pt+\x,-1pt+\y)(tre){};
			\node at (-0.75pt+\x,0.5pt+\y)(unoB){};
			\node at (-0.75pt+\x,-0.5pt+\y)(unoC){};
			\node at (2.25pt+\x,1.5pt+\y)(dueB){};
			\node [right=0.02pt of dueB, yshift=0.4pt]{\footnotesize$p_1$};
			\node at (2.25pt+\x,-1.5pt+\y)(treB){};
			\node [right=-0.02pt of treB, yshift=-0.4pt]{\footnotesize$p_3$};
			\node at (0.75pt+\x,-1pt+\y) (end){};
			\node at (-4pt+\x,-2pt+\y)(start){};
			\node at (-1.25pt+\x,0.75pt+\y){\footnotesize$q$};
			\node at (-1.2pt+\x,-0.75pt+\y){\footnotesize$p_4$};
			%\node at (1pt+\x,0.1pt+\y){\footnotesize$\mu^2$};
			\node at (-0.75pt+\x,0+\y)(cinque){};
			\node at (2.25pt+\x,0.5pt+\y)(treC){};
			\node [right=0.02pt of treC, yshift=+0.4pt]{\footnotesize$p_2$};

			\draw [thick] (uno.center) -- (due.center);
			\draw [thick] (due.center) -- (tre.center);
			\draw [thick] (uno.center) -- (tre.center);
			\draw [thick,double] (uno.center) -- (unoB.center);
			\draw [thick] (uno.center) -- (unoC.center);
			\draw [thick] (due.center) -- (dueB.center);
			\draw [thick] (tre.center) -- (treB.center);
			\draw [thick] (due.center) -- (treC.center);
		\end{tikzpicture}

		+C^{[0]}_2\begin{tikzpicture}[baseline={([yshift=-1mm]uno.base)},scale=15]
			\def\x{0}
			\def\y{0}

			\clip (-1.8pt,-2pt) rectangle (3.5pt,2pt);

			\node at (0+\x,0+\y)(uno){};
			\node at (1.5pt+\x,1pt+\y)(due){};
			\node at (1.5pt+\x,-1pt+\y)(tre){};
			\node at (-0.75pt+\x,0.5pt+\y)(unoB){};
			\node at (-0.75pt+\x,-0.5pt+\y)(unoC){};
			\node at (2.25pt+\x,1.5pt+\y)(dueB){};
			\node [right=0.02pt of dueB, yshift=0.4pt]{\footnotesize$p_4$};
			\node at (2.25pt+\x,-1.5pt+\y)(treB){};
			\node [right=-0.02pt of treB, yshift=-0.4pt]{\footnotesize$p_2$};
			\node at (0.75pt+\x,-1pt+\y) (end){};
			\node at (-4pt+\x,-2pt+\y)(start){};
			\node at (-1.25pt+\x,0.75pt+\y){\footnotesize$q$};
			\node at (-1.2pt+\x,-0.75pt+\y){\footnotesize$p_3$};
			%		\node at (0.9pt+\x,0+\y)[thick,scale=1.5]{$\circlearrowright$};
			%\node at (1pt+\x,0.1pt+\y){\footnotesize$\mu^2$};
			\node at (-0.75pt+\x,0+\y)(cinque){};
			%\node at (-1.25pt+\x,0+\y){\footnotesize $p_4$};
			\node at (2.25pt+\x,-0.5pt+\y)(treC){};
			\node [right=0.02pt of treC, yshift=+0.4pt]{\footnotesize$p_1$};

			\draw [thick] (uno.center) -- (due.center);
			\draw [thick] (due.center) -- (tre.center);
			\draw [thick] (uno.center) -- (tre.center);
			\draw [thick,double] (uno.center) -- (unoB.center);
			\draw [thick] (uno.center) -- (unoC.center);
			\draw [thick] (due.center) -- (dueB.center);
			\draw [thick] (tre.center) -- (treB.center);
			\draw [thick] (tre.center) -- (treC.center);
		\end{tikzpicture}
		\\

		+& C^{[0]}_3 \begin{tikzpicture}[baseline={([yshift=-1mm]uno.base)},scale=15]
			\def\x{0}
			\def\y{0}

			\clip (-2pt,-2pt) rectangle (3.5pt,2pt);

			\node at (0+\x,0+\y)(uno){};
			\node at (1.5pt+\x,1pt+\y)(due){};
			\node at (1.5pt+\x,-1pt+\y)(tre){};
			\node at (-0.75pt+\x,0.75pt+\y)(unoB){};
			\node at (-0.75pt+\x,-0.75pt+\y)(unoC){};
			\node at (2.25pt+\x,1.5pt+\y)(dueB){};
			\node [right=0.02pt of dueB, yshift=0.4pt]{\footnotesize$p_1$};
			\node at (2.25pt+\x,-1.5pt+\y)(treB){};
			\node [right=-0.02pt of treB, yshift=-0.4pt]{\footnotesize$p_2$};
			\node at (0.75pt+\x,-1pt+\y) (end){};
			\node at (-4pt+\x,-2pt+\y)(start){};
			\node at (-1.25pt+\x,1pt+\y){\footnotesize$q$};
			\node at (-1.2pt+\x,-1pt+\y){\footnotesize$p_3$};
			\node at (-1.4pt+\x,0pt+\y) (quattro){\footnotesize$p_4$};
			%		\node at (0.9pt+\x,0+\y)[thick,scale=1.5]{$\circlearrowright$};
			%\node at (1pt+\x,0.1pt+\y){\footnotesize$\mu^2$};
			\node at (-0.75pt+\x,0+\y)(cinque){};
			%\node at (-1.25pt+\x,0+\y){\footnotesize $p_4$};

			\draw [thick] (uno.center) -- (due.center);
			\draw [thick] (due.center) -- (tre.center);
			\draw [thick] (uno.center) -- (tre.center);
			\draw [thick,double] (uno.center) -- (unoB.center);
			\draw [thick] (uno.center) -- (unoC.center);
			\draw [thick] (due.center) -- (dueB.center);
			\draw [thick] (tre.center) -- (treB.center);
			\draw [thick] (uno.center) -- (quattro);
		\end{tikzpicture} 
		+C^{[2]}_3 \begin{tikzpicture}[baseline={([yshift=-1mm]uno.base)},scale=15]
			\def\x{0}
			\def\y{0}

			\clip (-2pt,-2pt) rectangle (3.5pt,2pt);

			\node at (0+\x,0+\y)(uno){};
			\node at (1.5pt+\x,1pt+\y)(due){};
			\node at (1.5pt+\x,-1pt+\y)(tre){};
			\node at (-0.75pt+\x,0.75pt+\y)(unoB){};
			\node at (-0.75pt+\x,-0.75pt+\y)(unoC){};
			\node at (2.25pt+\x,1.5pt+\y)(dueB){};
			\node [right=0.02pt of dueB, yshift=0.4pt]{\footnotesize$p_1$};
			\node at (2.25pt+\x,-1.5pt+\y)(treB){};
			\node [right=-0.02pt of treB, yshift=-0.4pt]{\footnotesize$p_2$};
			\node at (0.75pt+\x,-1pt+\y) (end){};
			\node at (-4pt+\x,-2pt+\y)(start){};
			\node at (-1.25pt+\x,1pt+\y){\footnotesize$q$};
			\node at (-1.2pt+\x,-1pt+\y){\footnotesize$p_3$};
			\node at (-1.4pt+\x,0pt+\y) (quattro){\footnotesize$p_4$};
			%		\node at (0.9pt+\x,0+\y)[thick,scale=1.5]{$\circlearrowright$};
			\node at (1pt+\x,0.1pt+\y){\footnotesize$\mu^2$};
			\node at (-0.75pt+\x,0+\y)(cinque){};
			%\node at (-1.25pt+\x,0+\y){\footnotesize $p_4$};

			\draw [thick] (uno.center) -- (due.center);
			\draw [thick] (due.center) -- (tre.center);
			\draw [thick] (uno.center) -- (tre.center);
			\draw [thick,double] (uno.center) -- (unoB.center);
			\draw [thick] (uno.center) -- (unoC.center);
			\draw [thick] (due.center) -- (dueB.center);
			\draw [thick] (tre.center) -- (treB.center);
			\draw [thick] (uno.center) -- (quattro);
		\end{tikzpicture}\\

		+& B^{[0]}_1 \begin{tikzpicture}[baseline={([yshift=-1mm]uno.base)},scale=15]
			\def\x{0}
			\def\y{0}

			\clip (-2.9pt,-2pt) rectangle (2.5pt,2pt);

			\def\x{0pt}
			\def\y{0}

			\node at (-0.75pt+\x,0pt+\y)(uno){};
			\node at (-1.5pt+\x,0.75pt+\y)(unoA){};
			\node at (-1.5pt+\x,-0.75pt+\y)(unoB){};
			\node at (-2pt+\x,0.85pt+\y){\footnotesize $q$};
			\node at (-2pt+\x,-0.85pt+\y){\footnotesize $p_3$};
			\node at (-1.5pt+\x,0+\y)(tre){};
			%\node at (-2pt+\x,0+\y){\footnotesize $p_4$};
			\node at (0.75pt+\x,0pt+\y)(due){};
			\node at (1.5pt+\x,0.75pt+\y)(dueA){};
			\node at (1.5pt+\x,-0.75pt+\y)(dueB){};
			\node at (2pt+\x,0.75pt+\y){\footnotesize $p_1$};
			\node at (2pt+\x,-0.75pt+\y){\footnotesize $p_2$};
			\node at (0pt+\x,0+\y)[thick,scale=1.5]{};
			%\node at (0+\x,0+\y){\footnotesize $\mu^2$};
			\node at (0+\x,-1.2pt+\y){};
			\node at (-2.2pt+\x,0pt+\y) (unoC) {\footnotesize $p_4$};

			\draw [thick] (0+\x,0+\y) circle (0.75pt);
			\draw [thick,double] (uno.center) -- (unoA.center);
			\draw [thick] (uno.center) -- (unoB.center);
			%\draw [thick] (uno.center) -- (tre.center);
			\draw [thick] (due.center) -- (dueA.center);
			\draw [thick] (due.center) -- (dueB.center);
			\draw [thick] (uno.center) -- (unoC);
		\end{tikzpicture} 
		+ B^{[2]}_1 \begin{tikzpicture}[baseline={([yshift=-1mm]uno.base)},scale=15]
			\def\x{0}
			\def\y{0}

			\clip (-2.9pt,-2pt) rectangle (2.5pt,2pt);

			\def\x{0pt}
			\def\y{0}

			\node at (-0.75pt+\x,0pt+\y)(uno){};
			\node at (-1.5pt+\x,0.75pt+\y)(unoA){};
			\node at (-1.5pt+\x,-0.75pt+\y)(unoB){};
			\node at (-2pt+\x,0.85pt+\y){\footnotesize $q$};
			\node at (-2pt+\x,-0.85pt+\y){\footnotesize $p_3$};
			\node at (-1.5pt+\x,0+\y)(tre){};
			%\node at (-2pt+\x,0+\y){\footnotesize $p_4$};
			\node at (0.75pt+\x,0pt+\y)(due){};
			\node at (1.5pt+\x,0.75pt+\y)(dueA){};
			\node at (1.5pt+\x,-0.75pt+\y)(dueB){};
			\node at (2pt+\x,0.75pt+\y){\footnotesize $p_1$};
			\node at (2pt+\x,-0.75pt+\y){\footnotesize $p_2$};
			\node at (0pt+\x,0+\y)[thick,scale=1.5]{};
			\node at (0+\x,0+\y){\footnotesize $\mu^2$};
			\node at (0+\x,-1.2pt+\y){};
			\node at (-2.2pt+\x,0pt+\y) (unoC) {\footnotesize $p_4$};

			\draw [thick] (0+\x,0+\y) circle (0.75pt);
			\draw [thick,double] (uno.center) -- (unoA.center);
			\draw [thick] (uno.center) -- (unoB.center);
			%\draw [thick] (uno.center) -- (tre.center);
			\draw [thick] (due.center) -- (dueA.center);
			\draw [thick] (due.center) -- (dueB.center);
			\draw [thick] (uno.center) -- (unoC);
		\end{tikzpicture}\ ,
	\end{array}
\end{equation}
where we checked that the coefficients $D^{[p]}_0$, $D^{[p]}_1$, $C^{[p]}_0$ and $C^{[p]}_2$ match the ones found in the previous calculation, up to relabelling. The other coefficients for the triangles are
\begin{equation}
	\setlength{\jot}{7pt}
	\begin{split}
		C^{[0]}_3 &= i \sqr{1}{2}\sqr{2}{3}\sqr{3}{4}\sqr{4}{1}\, \mathrm{c}^{[0]}_{3}\ ,\\
		C^{[2]}_3 &= \frac{4 i}{s_{12}}\sqr{1}{2}\sqr{3}{4}\sqr{1}{3}\sqr{2}{4}\ ,
	\end{split}
\end{equation}
where
\begin{equation}
	\setlength{\jot}{7pt}
	\begin{split}
		\mathrm{c}^{[0]}_{3} &= \frac{s_{12}+s_{31}}{s_{23}}+\frac{s_{12}}{s_{34}}\left(1+\frac{s_{13}}{s_{14}}+\frac{s_{24}}{s_{23}}\right)-\frac{\sqr{1}{3}\sqr{2}{4}}{\sqr{1}{4}\sqr{2}{3}}\left[\frac{s_{123}(s_{12}+s_{31})-s_{13}^2}{s_{13}^2}-\frac{s_{12}}{s_{34}}\right]\\
		& - \frac{\sqr{1}{3}\sqr{2}{4}}{\sqr{1}{2}\sqr{3}{3}}\frac{s_{12}}{s_{13}^2 s_{23}}\left[s_{123}(s_{23}+s_{31})-2s_{31}^2\right] + \left(1,4\right)\longleftrightarrow \left(2,3\right)\ ,
	\end{split}
\end{equation}
while for the bubbles 
\begin{equation}
	\setlength{\jot}{7pt}
	\begin{split}
		B^{[0]}_1 &= 2 i \sqr{1}{3}^2 \sqr{2}{4}^2 \left(\frac{1}{s_{31}}-\frac{1}{s_{23}}+\frac{s_{12}}{s_{23}s_{31}}\right)+2i\sqr{1}{2}^2\sqr{3}{4}^2\left(\frac{2}{s_{23}}-\frac{2s_{12}}{s_{23}(s_{13}+s_{23})}\right)\\
		&+2 i \sqr{1}{2}\sqr{3}{4}\sqr{1}{3}\sqr{2}{4}\left(\frac{1}{s_{12}}+\frac{4}{s_{23}}-\frac{2}{s_{34}}-\frac{4s_{24}}{s_{23}s_{34}}\right) + \left(1,4\right)\longleftrightarrow \left(2,3\right)\ ,
	\end{split}
\end{equation}
and
\begin{equation}
	B^{[2]}_1 = \frac{4 i}{s_{12}^2}\sqr{1}{2}\sqr{3}{4}\left(\sqr{1}{3}\sqr{2}{4}+\sqr{2}{3}\sqr{1}{4}\right)\ .
\end{equation}
We have checked that  our result satisfies the  expected infrared consistency conditions. 
In particular, 
using the results for the coefficients $D_0$, $C_0$ and $C_1$, one immediately finds  that the coefficient of $\frac{(-s_{123})^{-\epsilon}}{\epsilon^2}$ vanishes, as required.
We have also confirmed that the coefficient of $\frac{(-s_{12})^{-\epsilon}}{\epsilon^2}$ is 
proportional to the corresponding tree-level non-minimal form factor derived in \cite{Brandhuber:2017bkg}, 
%Indeed by looking at the integrated expression of the integrals in \eqref{eq:TrF3nonminimalfinals12} one finds that the coefficient of the terms $\frac{(-s_{12})^{-\epsilon}}{\epsilon^2}$ 
\begin{equation}
	F^{(0)}_{\cO_{3}} (1^+,2^+,3^+,4^+;q) = -2\frac{\sqr{1}{2}\sqr{2}{3}\sqr{3}{4}\sqr{4}{1}}{s_{12}}\left(1 + \frac{\sqr{1}{3}\sqr{2}{4}}{\sqr{2}{3}\sqr{4}{1}}-\frac{s_{24}}{s_{41}}\right)+{\rm cyclic}\ .
\end{equation}

\subsection{The Minimal \texorpdfstring{$\Tr F^4$}{TrF4} Form Factors}

In this section we consider the form factors of $\Tr F^4$ in all possible helicity configurations. The case where all particles have the same  helicity is interesting since it admits an immediate generalisation to the minimal form factors of operators of the form $\Tr F^n$ defined in~\eqref{eq:TrFndef}. In this family, $\Tr F^4$ is the first operator whose minimal form factor contains rational terms. We are going to consider the quantities in the planar limit of the theory, \textit{i.e.} at one loop we will probe only the discontinuities in the Mandelstam invariants of adjacent momenta in the colour-ordered form factor. At this point it is important to stress that non-planar contributions behave differently: as one can see from \eqref{scalarF4} there is no non-planar scalar contribution, because in the operator the scalars can only appear next to each other, and then the complete four-dimensional contribution coincides with the diagrams with purely six-dimensional internal gluons.

\subsubsection*{All-Plus Helicity Configuration}

We begin by defining
\begin{equation}
	F^{(1)}_{\cO_{4}} (1^+,2^+,3^+,4^+;q) \coloneqq  2\sqr{1}{2}\sqr{2}{3}\sqr{3}{4}\sqr{4}{1}\cdot f^{(4)}\left(\{s_{ij}\}\right)\ .
\end{equation}
At one loop, we can make the following observations:
\begin{itemize}
	\item The cut-constructible part, coming from the form factor involving only gluons, has the same structure as $F^{(1)}_{\cO_{3}} (1^+,2^+,3^+;q)$, with both UV and IR divergences.
	\item Terms proportional to $\mu^2$ and $\mu^4$ now appear. As already mentioned, these could not arise for $n<4$ because of the limited kinematic, as we will show below. The new integrals are two triangles with $\mu^2$ and $\mu^4$ numerators\footnote{For analytic expressions of such integrals see for example Appendix~\ref{sec:integrals}.} and when expanded in powers of the dimensional regulator $\epsilon$ give a finite contribution in the $\epsilon \to 0$ limit. They are exactly the rational terms that cannot be seen by the completely four-dimensional cut construction, where clearly $\mu^2=0$.
\end{itemize}

\noindent
Following the procedure outlined in the previous sections, we find
\begin{equation}\label{eq:TrF4}
	\begin{array}{rl}
		f^4(\{s_{ij}\})\big|_{\rm s_{12}-disc} = & 
		-i \left(1+\frac{\sqr{1}{3}\sqr{2}{4}}{\sqr{1}{4}\sqr{2}{3}}\right)\cdot \begin{tikzpicture}[baseline={([yshift=-1mm]uno.base)},scale=15]
			\def\x{0}
			\def\y{0}

			\clip (-2.5pt,-2pt) rectangle (2.5pt,2pt);

			\def\x{0pt}
			\def\y{0}

			\node at (-0.75pt+\x,0pt+\y)(uno){};
			\node at (-1.5pt+\x,0.75pt+\y)(unoA){};
			\node at (-1.5pt+\x,-0.75pt+\y)(unoB){};
			\node at (-2pt+\x,0.75pt+\y){\footnotesize $q$};
			\node at (-2pt+\x,-0.75pt+\y){\footnotesize $p_3$};
			\node at (-1.5pt+\x,0+\y)(tre){};
			\node at (-2pt+\x,0+\y){\footnotesize $p_4$};
			\node at (0.75pt+\x,0pt+\y)(due){};
			\node at (1.5pt+\x,0.75pt+\y)(dueA){};
			\node at (1.5pt+\x,-0.75pt+\y)(dueB){};
			\node at (2pt+\x,0.75pt+\y){\footnotesize $p_1$};
			\node at (2pt+\x,-0.75pt+\y){\footnotesize $p_2$};
			\node at (0pt+\x,0+\y)[thick,scale=1.5]{};
			\node at (0+\x,-1.2pt+\y){\footnotesize $l$};

			\draw [thick] (0+\x,0+\y) circle (0.75pt);
			\draw [thick,double] (uno.center) -- (unoA.center);
			\draw [thick] (uno.center) -- (unoB.center);
			\draw [thick] (uno.center) -- (tre.center);
			\draw [thick] (due.center) -- (dueA.center);
			\draw [thick] (due.center) -- (dueB.center);
			%\draw [dotted,thick] (-1.3pt+\x,0.4pt+\y) arc (135:225:0.6pt);
		\end{tikzpicture}
		-i \, s_{12} \cdot \begin{tikzpicture}[baseline={([yshift=-1mm]uno.base)},scale=15]
			\def\x{0}
			\def\y{0}

			\clip (-1.8pt,-2pt) rectangle (3.5pt,2pt);

			\node at (0+\x,0+\y)(uno){};
			\node at (1.5pt+\x,1pt+\y)(due){};
			\node at (1.5pt+\x,-1pt+\y)(tre){};
			\node at (-0.75pt+\x,0.5pt+\y)(unoB){};
			\node at (-0.75pt+\x,-0.5pt+\y)(unoC){};
			\node at (2.25pt+\x,1.5pt+\y)(dueB){};
			\node [right=0.02pt of dueB, yshift=0.4pt]{\footnotesize$p_1$};
			\node at (2.25pt+\x,-1.5pt+\y)(treB){};
			\node [right=-0.02pt of treB, yshift=-0.4pt]{\footnotesize$p_2$};
			\node at (0.75pt+\x,-1pt+\y) (end){};
			\node at (-4pt+\x,-2pt+\y)(start){};
			\node at (-1.25pt+\x,0.75pt+\y){\footnotesize$q$};
			\node at (-1.2pt+\x,-0.75pt+\y){\footnotesize$p_3$};
			%		\node at (0.9pt+\x,0+\y)[thick,scale=1.5]{$\circlearrowright$};
			\node at (0.9pt+\x,0+\y)[thick,scale=1.5]{};
			\node at (-0.75pt+\x,0+\y)(cinque){};
			\node at (-1.25pt+\x,0+\y){\footnotesize $p_4$};

			\draw [thick] (uno.center) -- (due.center);
			\draw [thick] (due.center) -- (tre.center);
			\draw [thick] (uno.center) -- node [yshift=-0.2cm] {\footnotesize$l$} (tre.center);
			\draw [thick,double] (uno.center) -- (unoB.center);
			\draw [thick] (uno.center) -- (unoC.center);
			\draw [thick] (due.center) -- (dueB.center);
			\draw [thick] (tre.center) -- (treB.center);
			\draw [thick] (uno.center) -- (cinque.center);

			%\draw [dotted,thick](-0.6pt+\x,0.3pt+\y) arc (140:240:0.4pt);
		\end{tikzpicture}                                                       \\
		+ & i \frac{\sqr{1}{2}\sqr{3}{4}}{\sqr{2}{3}\sqr{4}{1}} \cdot
		\begin{tikzpicture}[baseline={([yshift=-1mm]uno.base)},scale=15]
			\def\x{0}
			\def\y{0}

			\clip (-1.8pt,-2pt) rectangle (3.5pt,2pt);

			\node at (0+\x,0+\y)(uno){};
			\node at (1.5pt+\x,1pt+\y)(due){};
			\node at (1.5pt+\x,-1pt+\y)(tre){};
			\node at (-0.75pt+\x,0.5pt+\y)(unoB){};
			\node at (-0.75pt+\x,-0.5pt+\y)(unoC){};
			\node at (2.25pt+\x,1.5pt+\y)(dueB){};
			\node [right=0.02pt of dueB, yshift=0.4pt]{\footnotesize$p_1$};
			\node at (2.25pt+\x,-1.5pt+\y)(treB){};
			\node [right=-0.02pt of treB, yshift=-0.4pt]{\footnotesize$p_2$};
			\node at (0.75pt+\x,-1pt+\y) (end){};
			\node at (-4pt+\x,-2pt+\y)(start){};
			\node at (-1.25pt+\x,0.75pt+\y){\footnotesize$q$};
			\node at (-1.2pt+\x,-0.75pt+\y){\footnotesize$p_3$};
			%		\node at (0.9pt+\x,0+\y)[thick,scale=1.5]{$\circlearrowright$};
			\node at (1pt+\x,0.1pt+\y){\footnotesize$\mu^2$};
			\node at (-0.75pt+\x,0+\y)(cinque){};
			\node at (-1.25pt+\x,0+\y){\footnotesize $p_4$};

			\draw [thick] (uno.center) -- (due.center);
			\draw [thick] (due.center) -- (tre.center);
			\draw [thick] (uno.center) -- node [yshift=-0.2cm] {\footnotesize$l$} (tre.center);
			\draw [thick,double] (uno.center) -- (unoB.center);
			\draw [thick] (uno.center) -- (unoC.center);
			\draw [thick] (due.center) -- (dueB.center);
			\draw [thick] (tre.center) -- (treB.center);
			\draw [thick] (uno.center) -- (cinque.center);

			%\draw [dotted,thick](-0.6pt+\x,0.3pt+\y) arc (140:240:0.4pt);
		\end{tikzpicture}
		-i\frac{\sqr{3}{4}}{\lbrack 3 | \slashed{p_2} \slashed{p_1} | 4 \rbrack}  \cdot
		\begin{tikzpicture}[baseline={([yshift=-1mm]uno.base)},scale=15]
			\def\x{0}
			\def\y{0}

			\clip (-1.8pt,-2pt) rectangle (3.5pt,2pt);

			\node at (0+\x,0+\y)(uno){};
			\node at (1.5pt+\x,1pt+\y)(due){};
			\node at (1.5pt+\x,-1pt+\y)(tre){};
			\node at (-0.75pt+\x,0.5pt+\y)(unoB){};
			\node at (-0.75pt+\x,-0.5pt+\y)(unoC){};
			\node at (2.25pt+\x,1.5pt+\y)(dueB){};
			\node [right=0.02pt of dueB, yshift=0.4pt]{\footnotesize$p_1$};
			\node at (2.25pt+\x,-1.5pt+\y)(treB){};
			\node [right=-0.02pt of treB, yshift=-0.4pt]{\footnotesize$p_2$};
			\node at (0.75pt+\x,-1pt+\y) (end){};
			\node at (-4pt+\x,-2pt+\y)(start){};
			\node at (-1.25pt+\x,0.75pt+\y){\footnotesize$q$};
			\node at (-1.2pt+\x,-0.75pt+\y){\footnotesize$p_3$};
			%		\node at (0.9pt+\x,0+\y)[thick,scale=1.5]{$\circlearrowright$};
			\node at (1pt+\x,0.1pt+\y){\footnotesize$\mu^4$};
			\node at (-0.75pt+\x,0+\y)(cinque){};
			\node at (-1.25pt+\x,0+\y){\footnotesize $p_4$};

			\draw [thick] (uno.center) -- (due.center);
			\draw [thick] (due.center) -- (tre.center);
			\draw [thick] (uno.center) -- node [yshift=-0.2cm] {\footnotesize$l$} (tre.center);
			\draw [thick,double] (uno.center) -- (unoB.center);
			\draw [thick] (uno.center) -- (unoC.center);
			\draw [thick] (due.center) -- (dueB.center);
			\draw [thick] (tre.center) -- (treB.center);
			\draw [thick] (uno.center) -- (cinque.center);

			%\draw [dotted,thick](-0.6pt+\x,0.3pt+\y) arc (140:240:0.4pt);
		\end{tikzpicture}
	\end{array}
\end{equation}
Notice that in the final result the integral $I^{4}_{3}[\mu^4]$ appears. In general, in a renormalisable gauge theory one would expect triangle integrals to appear with at most a third power of the loop momentum in the numerator, which allows for at most a $\mu^2$ triangle contribution. However we are considering an effective field theory with an operator of mass-dimension eight, hence the possibility of having also an $I^{4}_{3}[\mu^4]$ term. The last step of the calculation is the sum over all the possible channel discontinuities, as we did in \eqref{eq::sum_cuts} for $\Tr F^3$.

The above result can be immediately generalized to $\Tr F^n$ for arbitrary $n$ in the all-plus helicity configurations, where we define

\begin{equation}
	\Tr F^n (1^+,\ldots,n^+;q)\big|_{1\text{-loop}} \coloneqq  (-)^n \, 2 \prod_{k = 1}^{n} \sqr{k}{k+1} \cdot f^{(n)}\left(\{s_{ij}\}\right)\ ,
\end{equation}
and
\begin{equation}
	\label{eq::TrFncomp}
	\begin{array}{rl}
		f^{(n)}(\{s_{ij}\})\big|_{\rm s_{12}-disc}= &
		-i \left(1+\frac{\sqr{1}{3}\sqr{2}{n}}{\sqr{1}{n}\sqr{2}{3}}\right) \cdot \begin{tikzpicture}[baseline={([yshift=-1mm]uno.base)},scale=15]
			\def\x{0}
			\def\y{0}

			\clip (-2.5pt,-2pt) rectangle (2.5pt,2pt);

			\def\x{0pt}
			\def\y{0}

			\node at (-0.75pt+\x,0pt+\y)(uno){};
			\node at (-1.5pt+\x,0.75pt+\y)(unoA){};
			\node at (-1.5pt+\x,-0.75pt+\y)(unoB){};
			\node at (-2pt+\x,0.75pt+\y){\footnotesize $q$};
			\node at (-2pt+\x,-0.75pt+\y){\footnotesize $p_3$};
			\node at (-1.5pt+\x,0+\y)(tre){};
			%\node at (-2pt+\x,0+\y){\footnotesize $p_4$};
			\node at (0.75pt+\x,0pt+\y)(due){};
			\node at (1.5pt+\x,0.75pt+\y)(dueA){};
			\node at (1.5pt+\x,-0.75pt+\y)(dueB){};
			\node at (2pt+\x,0.75pt+\y){\footnotesize $p_1$};
			\node at (2pt+\x,-0.75pt+\y){\footnotesize $p_2$};
			\node at (0pt+\x,0+\y)[thick,scale=1.5]{};
			\node at (0+\x,-1.2pt+\y){\footnotesize $l$};

			\draw [thick] (0+\x,0+\y) circle (0.75pt);
			\draw [thick,double] (uno.center) -- (unoA.center);
			\draw [thick] (uno.center) -- (unoB.center);
			%\draw [thick] (uno.center) -- (tre.center);
			\draw [thick] (due.center) -- (dueA.center);
			\draw [thick] (due.center) -- (dueB.center);
			\draw [dotted,thick] (-1.3pt+\x,0.4pt+\y) arc (135:225:0.6pt);
		\end{tikzpicture}
		-i \, s_{12} \cdot \begin{tikzpicture}[baseline={([yshift=-1mm]uno.base)},scale=15]
			\def\x{0}
			\def\y{0}

			\clip (-1.8pt,-2pt) rectangle (3.5pt,2pt);

			\node at (0+\x,0+\y)(uno){};
			\node at (1.5pt+\x,1pt+\y)(due){};
			\node at (1.5pt+\x,-1pt+\y)(tre){};
			\node at (-0.75pt+\x,0.5pt+\y)(unoB){};
			\node at (-0.75pt+\x,-0.5pt+\y)(unoC){};
			\node at (2.25pt+\x,1.5pt+\y)(dueB){};
			\node [right=0.02pt of dueB, yshift=0.4pt]{\footnotesize$p_1$};
			\node at (2.25pt+\x,-1.5pt+\y)(treB){};
			\node [right=-0.02pt of treB, yshift=-0.4pt]{\footnotesize$p_2$};
			\node at (0.75pt+\x,-1pt+\y) (end){};
			\node at (-4pt+\x,-2pt+\y)(start){};
			\node at (-1.25pt+\x,0.75pt+\y){\footnotesize$q$};
			\node at (-1.2pt+\x,-0.75pt+\y){\footnotesize$p_3$};
			%		\node at (0.9pt+\x,0+\y)[thick,scale=1.5]{$\circlearrowright$};
			\node at (0.9pt+\x,0+\y)[thick,scale=1.5]{};
			\node at (-0.75pt+\x,0+\y)(cinque){};
			%\node at (-1.25pt+\x,0+\y){\footnotesize $p_4$};

			\draw [thick] (uno.center) -- (due.center);
			\draw [thick] (due.center) -- (tre.center);
			\draw [thick] (uno.center) -- node [yshift=-0.2cm] {\footnotesize$l$} (tre.center);
			\draw [thick,double] (uno.center) -- (unoB.center);
			\draw [thick] (uno.center) -- (unoC.center);
			\draw [thick] (due.center) -- (dueB.center);
			\draw [thick] (tre.center) -- (treB.center);
			%\draw [thick] (uno.center) -- (cinque.center);

			\draw [dotted,thick](-0.6pt+\x,0.3pt+\y) arc (140:240:0.4pt);
		\end{tikzpicture}                                                           \\
		                                            & +i \frac{\sqr{1}{2}\sqr{3}{n}}{\sqr{2}{3}\sqr{n}{1}} \cdot
		\begin{tikzpicture}[baseline={([yshift=-1mm]uno.base)},scale=15]
			\def\x{0}
			\def\y{0}

			\clip (-1.8pt,-2pt) rectangle (3.5pt,2pt);

			\node at (0+\x,0+\y)(uno){};
			\node at (1.5pt+\x,1pt+\y)(due){};
			\node at (1.5pt+\x,-1pt+\y)(tre){};
			\node at (-0.75pt+\x,0.5pt+\y)(unoB){};
			\node at (-0.75pt+\x,-0.5pt+\y)(unoC){};
			\node at (2.25pt+\x,1.5pt+\y)(dueB){};
			\node [right=0.02pt of dueB, yshift=0.4pt]{\footnotesize$p_1$};
			\node at (2.25pt+\x,-1.5pt+\y)(treB){};
			\node [right=-0.02pt of treB, yshift=-0.4pt]{\footnotesize$p_2$};
			\node at (0.75pt+\x,-1pt+\y) (end){};
			\node at (-4pt+\x,-2pt+\y)(start){};
			\node at (-1.25pt+\x,0.75pt+\y){\footnotesize$q$};
			\node at (-1.2pt+\x,-0.75pt+\y){\footnotesize$p_3$};
			%		\node at (0.9pt+\x,0+\y)[thick,scale=1.5]{$\circlearrowright$};
			\node at (1pt+\x,0.1pt+\y){\footnotesize$\mu^2$};
			\node at (-0.75pt+\x,0+\y)(cinque){};
			%\node at (-1.25pt+\x,0+\y){\footnotesize $p_4$};

			\draw [thick] (uno.center) -- (due.center);
			\draw [thick] (due.center) -- (tre.center);
			\draw [thick] (uno.center) -- node [yshift=-0.2cm] {\footnotesize$l$} (tre.center);
			\draw [thick,double] (uno.center) -- (unoB.center);
			\draw [thick] (uno.center) -- (unoC.center);
			\draw [thick] (due.center) -- (dueB.center);
			\draw [thick] (tre.center) -- (treB.center);
			%\draw [thick] (uno.center) -- (cinque.center);

			\draw [dotted,thick](-0.6pt+\x,0.3pt+\y) arc (140:240:0.4pt);
		\end{tikzpicture}
		-i\frac{\sqr{3}{n}}{\lbrack 3 | \slashed{p_2} \slashed{p_1} | n \rbrack} \cdot
		\begin{tikzpicture}[baseline={([yshift=-1mm]uno.base)},scale=15]
			\def\x{0}
			\def\y{0}

			\clip (-1.8pt,-2pt) rectangle (3.5pt,2pt);

			\node at (0+\x,0+\y)(uno){};
			\node at (1.5pt+\x,1pt+\y)(due){};
			\node at (1.5pt+\x,-1pt+\y)(tre){};
			\node at (-0.75pt+\x,0.5pt+\y)(unoB){};
			\node at (-0.75pt+\x,-0.5pt+\y)(unoC){};
			\node at (2.25pt+\x,1.5pt+\y)(dueB){};
			\node [right=0.02pt of dueB, yshift=0.4pt]{\footnotesize$p_1$};
			\node at (2.25pt+\x,-1.5pt+\y)(treB){};
			\node [right=-0.02pt of treB, yshift=-0.4pt]{\footnotesize$p_2$};
			\node at (0.75pt+\x,-1pt+\y) (end){};
			\node at (-4pt+\x,-2pt+\y)(start){};
			\node at (-1.25pt+\x,0.75pt+\y){\footnotesize$q$};
			\node at (-1.2pt+\x,-0.75pt+\y){\footnotesize$p_3$};
			%		\node at (0.9pt+\x,0+\y)[thick,scale=1.5]{$\circlearrowright$};
			\node at (1pt+\x,0.1pt+\y){\footnotesize$\mu^4$};
			\node at (-0.75pt+\x,0+\y)(cinque){};
			%\node at (-1.25pt+\x,0+\y){\footnotesize $p_4$};

			\draw [thick] (uno.center) -- (due.center);
			\draw [thick] (due.center) -- (tre.center);
			\draw [thick] (uno.center) -- node [yshift=-0.2cm] {\footnotesize$l$} (tre.center);
			\draw [thick,double] (uno.center) -- (unoB.center);
			\draw [thick] (uno.center) -- (unoC.center);
			\draw [thick] (due.center) -- (dueB.center);
			\draw [thick] (tre.center) -- (treB.center);
			%\draw [thick] (uno.center) -- (cinque.center);

			\draw [dotted,thick](-0.6pt+\x,0.3pt+\y) arc (140:240:0.4pt);
		\end{tikzpicture}
	\end{array}
\end{equation}

This simple generalisation is due to the fact that, upon properly normalising with the corresponding four-dimensional quantities, the six-dimensional minimal tree-level form factor of $\Tr F^n$ is identical to that of $\Tr F^4$ up to the replacement $4\mapsto n$, as can be seen from~\eqref{trFn}. As a final remark, notice that we can a posteriori explain the absence of rational terms for  $\Tr F^3$: indeed we can recover  \eqref{eq::TrF3comp} by simply replacing  $n\mapsto 3$ in \eqref{eq::TrFncomp}. Then, rational terms vanish since they are proportional to $\sqr{3}{n}$.

\subsubsection*{MHV Configuration: the Alternate and Split-Helicity Colour Ordered Form Factors}

We define the MHV colour-ordered form factor with alternate-helicity gluons as follows:
\begin{equation}
	F^{(1)}_{\cO_{4}} (1^+,2^-,3^+,4^-;q) \coloneqq  \agl{2}{4}^2 \sqr{1}{3}^2 \cdot f_a^{(4)}\left(\{ s_{ij}\}\right)\ .
\end{equation}

Since this case presents some peculiarities in the calculations, we will give more details about it. In particular, the cut of the form factor with six-dimensional internal gluons  in  the $s_{12}$-channel  is given by
\begin{equation}
	\label{6D_MHValternate}
	\begin{split}
		f_{a,6D}^{(4)} \left( \{ s_{ij} \} \right) \big|_{s_{12}\rm{-cut}} =& - \int \dd{\rm LIPS} \frac{i}{s_{1 2} s_{2 l_2}} \frac{\mathcal{I}_{6D}^{\, 2}}{\agl{2}{4}^2 \sqr{1}{3}^2}\\
		=& - \int \dd{\rm LIPS} \frac{i}{s_{1 2} s_{2 l_2}} (2 k \cdot l_2)^2\ ,
	\end{split}
\end{equation}
where
\begin{equation}
	\mathcal{I}_{6D} =  2 \mu^2 \agl{2}{4} \sqr{1}{3} + \langle 2 | \slashed{l}^{(4)}_1 | 3 \rbrack \langle 4 | \slashed{l}^{(4)}_2 | 1 \rbrack + \langle 2 | \slashed{l}^{(4)}_2 | 3 \rbrack \langle 4 | \slashed{l}^{(4)}_1 | 1 \rbrack
\end{equation}
and in the last step we removed terms proportional to $\langle 2 | \slashed{l}^{(4)}_2 | 1 \rbrack$ that vanish upon integration. Also $k_{\mu}$ is a massive momentum defined by
\begin{equation}
	k_{\alpha \dot{\alpha}} = \frac{\sqr{1}{2}}{\sqr{1}{3}} \lambda_{2 \alpha} \tilde{\lambda}_{3 \dot{\alpha}} - \frac{\agl{1}{2}}{\agl{2}{4}} \lambda_{4 \alpha} \tilde{\lambda}_{1 \dot{\alpha}}\ ,
\end{equation}
and it is easy to prove that it satisfies the following relations:
\begin{equation}
	k^2 = 2 p_{1} \cdot k = 2 p_{2} \cdot k = s_{1 2}\ .
\end{equation}

Surprisingly, the scalar contribution is identically zero after integration:
\begin{equation}
	f_{a,\phi}^{(4)} \left( \{ s_{ij} \} \right) \big|_{s_{12}\rm{-cut}} = \int \dd{\rm LIPS} \frac{i}{s_{1 2} s_{2 l_2}} \frac{\langle 4 | \slashed{l}^{(4)}_1 | 3 \rbrack \langle 4 | \slashed{l}^{(4)}_2 | 3 \rbrack \langle 2 | \slashed{l}^{(4)}_2 | 1 \rbrack ^ 2}{\agl{2}{4}^2 \sqr{1}{3}^2} = 0\ ,
\end{equation}
because of the presence of the term $\langle 2 | \slashed{l}^{(4)}_2 | 1 \rbrack^2$. Thus the discontinuity in the $s_{12}$-channel is completely given by the pure six-dimensional contribution \eqref{6D_MHValternate}, which after the integral reduction can be written as

\begin{equation}
	f_{a}^{(4)} \left( \{ s_{ij} \} \right) \big|_{s_{12}\rm{-disc}} = - i s_{1 2}\cdot
	\begin{tikzpicture}[baseline={([yshift=-1mm]uno.base)},scale=15]
		\def\x{0}
		\def\y{0}

		\clip (-1.8pt,-2pt) rectangle (3.5pt,2pt);

		\node at (0+\x,0+\y)(uno){};
		\node at (1.5pt+\x,1pt+\y)(due){};
		\node at (1.5pt+\x,-1pt+\y)(tre){};
		\node at (-0.75pt+\x,0.5pt+\y)(unoB){};
		\node at (-0.75pt+\x,-0.5pt+\y)(unoC){};
		\node at (2.25pt+\x,1.5pt+\y)(dueB){};
		\node [right=0.02pt of dueB, yshift=0.4pt]{\footnotesize$p_1$};
		\node at (2.25pt+\x,-1.5pt+\y)(treB){};
		\node [right=-0.02pt of treB, yshift=-0.4pt]{\footnotesize$p_2$};
		\node at (0.75pt+\x,-1pt+\y) (end){};
		\node at (-4pt+\x,-2pt+\y)(start){};
		\node at (-1.25pt+\x,0.75pt+\y){\footnotesize$q$};
		\node at (-1.2pt+\x,-0.75pt+\y){\footnotesize$p_3$};
		%		\node at (0.9pt+\x,0+\y)[thick,scale=1.5]{$\circlearrowright$};
		\node at (-0.75pt+\x,0+\y)(cinque){};
		%\node at (-1.25pt+\x,0+\y){\footnotesize $p_4$};

		\draw [thick] (uno.center) -- (due.center);
		\draw [thick] (due.center) -- (tre.center);
		\draw [thick] (uno.center) -- node [yshift=-0.2cm] {\footnotesize$l$} (tre.center);
		\draw [thick,double] (uno.center) -- (unoB.center);
		\draw [thick] (uno.center) -- (unoC.center);
		\draw [thick] (due.center) -- (dueB.center);
		\draw [thick] (tre.center) -- (treB.center);
		%\draw [thick] (uno.center) -- (cinque.center);

		\draw [dotted,thick](-0.6pt+\x,0.3pt+\y) arc (140:240:0.4pt);
	\end{tikzpicture}\ .
\end{equation}

It is worth stressing that all the other planar contributions can be obtained from the previous one easily by symmetry arguments.

As usual, for the split-helicity configuration we factorise the tree-level form factor:
\begin{equation}
	F^{(1)}_{\cO_{4}} (1^+,3^+,2^-,4^-;q) \coloneqq \sqr{1}{3}^2 \agl{2}{4}^2 \cdot f_s^{(4)}\left(\{ s_{ij}\}\right)\ .
\end{equation}

Unlike the previous case, in the planar limit we have two different cuts which cannot be related by symmetry: in particular, we can perform the cut in channels with same or opposite helicity gluons. The discontinuity in the $s_{12}$-channel, after the scalar subtraction, is given by
\begin{equation}
	f_{s}^{(4)} \left( \{ s_{ij} \} \right) \big|_{s_{13}\rm{-disc}} =
	- i s_{1 3}\cdot
	\begin{tikzpicture}[baseline={([yshift=-1mm]uno.base)},scale=15]
		\def\x{0}
		\def\y{0}

		\clip (-1.8pt,-2pt) rectangle (3.5pt,2pt);

		\node at (0+\x,0+\y)(uno){};
		\node at (1.5pt+\x,1pt+\y)(due){};
		\node at (1.5pt+\x,-1pt+\y)(tre){};
		\node at (-0.75pt+\x,0.5pt+\y)(unoB){};
		\node at (-0.75pt+\x,-0.5pt+\y)(unoC){};
		\node at (2.25pt+\x,1.5pt+\y)(dueB){};
		\node [right=0.02pt of dueB, yshift=0.4pt]{\footnotesize$p_1$};
		\node at (2.25pt+\x,-1.5pt+\y)(treB){};
		\node [right=-0.02pt of treB, yshift=-0.4pt]{\footnotesize$p_3$};
		\node at (0.75pt+\x,-1pt+\y) (end){};
		\node at (-4pt+\x,-2pt+\y)(start){};
		\node at (-1.25pt+\x,0.75pt+\y){\footnotesize$q$};
		\node at (-1.2pt+\x,-0.75pt+\y){\footnotesize$p_2$};
		%		\node at (0.9pt+\x,0+\y)[thick,scale=1.5]{$\circlearrowright$};
		\node at (-0.75pt+\x,0+\y)(cinque){};
		%\node at (-1.25pt+\x,0+\y){\footnotesize $p_4$};

		\draw [thick] (uno.center) -- (due.center);
		\draw [thick] (due.center) -- (tre.center);
		\draw [thick] (uno.center) -- node [yshift=-0.2cm] {\footnotesize$l$} (tre.center);
		\draw [thick,double] (uno.center) -- (unoB.center);
		\draw [thick] (uno.center) -- (unoC.center);
		\draw [thick] (due.center) -- (dueB.center);
		\draw [thick] (tre.center) -- (treB.center);
		%\draw [thick] (uno.center) -- (cinque.center);

		\draw [dotted,thick](-0.6pt+\x,0.3pt+\y) arc (140:240:0.4pt);
	\end{tikzpicture}\ .
\end{equation}

The cut in the $s_{23}$-channel is reminiscent of the alternate-helicity case, with vanishing scalar contribution up to integration:
\begin{equation}
	f_{s}^{(4)} \left( \{ s_{ij} \} \right) \big|_{s_{23}\rm{-cut}} \simeq - \int \dd{\rm LIPS} \frac{i}{s_{1 3} s_{3 l_2}} (2 k \cdot l_2)^2\ ,
\end{equation}
where the momentum $k_{\mu}$ is defined by
\begin{equation}
	k_{\alpha \dot{\alpha}} = \frac{\sqr{2}{3}}{\sqr{1}{3}} \lambda_{2 \alpha} \tilde{\lambda}_{1 \dot{\alpha}} + \frac{\agl{2}{3}}{\agl{2}{4}} \lambda_{4 \alpha} \tilde{\lambda}_{3 \dot{\alpha}}\ ,
\end{equation}
and it satisfies the following relations:
\begin{equation}
	k^2 = 2 p_{2} \cdot k = 2 p_{3} \cdot k = s_{2 3}\ .
\end{equation}
\noindent
The cut in the $s_{23}$ channel is
\begin{equation}
	f_{s}^{(4)} \left( \{ s_{ij} \} \right) \big|_{s_{23}\rm{-disc}} = - i s_{2 3}\cdot
	\begin{tikzpicture}[baseline={([yshift=-1mm]uno.base)},scale=15]
		\def\x{0}
		\def\y{0}

		\clip (-1.8pt,-2pt) rectangle (3.5pt,2pt);

		\node at (0+\x,0+\y)(uno){};
		\node at (1.5pt+\x,1pt+\y)(due){};
		\node at (1.5pt+\x,-1pt+\y)(tre){};
		\node at (-0.75pt+\x,0.5pt+\y)(unoB){};
		\node at (-0.75pt+\x,-0.5pt+\y)(unoC){};
		\node at (2.25pt+\x,1.5pt+\y)(dueB){};
		\node [right=0.02pt of dueB, yshift=0.4pt]{\footnotesize$p_2$};
		\node at (2.25pt+\x,-1.5pt+\y)(treB){};
		\node [right=-0.02pt of treB, yshift=-0.4pt]{\footnotesize$p_3$};
		\node at (0.75pt+\x,-1pt+\y) (end){};
		\node at (-4pt+\x,-2pt+\y)(start){};
		\node at (-1.25pt+\x,0.75pt+\y){\footnotesize$q$};
		\node at (-1.2pt+\x,-0.75pt+\y){\footnotesize$p_4$};
		%		\node at (0.9pt+\x,0+\y)[thick,scale=1.5]{$\circlearrowright$};
		\node at (-0.75pt+\x,0+\y)(cinque){};
		%\node at (-1.25pt+\x,0+\y){\footnotesize $p_4$};

		\draw [thick] (uno.center) -- (due.center);
		\draw [thick] (due.center) -- (tre.center);
		\draw [thick] (uno.center) -- node [yshift=-0.2cm] {\footnotesize$l$} (tre.center);
		\draw [thick,double] (uno.center) -- (unoB.center);
		\draw [thick] (uno.center) -- (unoC.center);
		\draw [thick] (due.center) -- (dueB.center);
		\draw [thick] (tre.center) -- (treB.center);
		%\draw [thick] (uno.center) -- (cinque.center);

		\draw [dotted,thick](-0.6pt+\x,0.3pt+\y) arc (140:240:0.4pt);
	\end{tikzpicture}
\end{equation}

Let us emphasise some relevant features of the result:
\begin{itemize}
	\item The final result  is free of rational terms. Thus we would have found the same, complete, quantity even with four-dimensional unitarity-cuts.
	\item The only operator that contributes in four dimensions is $\Tr \left(F^{2}_{\rm SD} F^{2}_{\rm ASD}\right)$, which is a descendant of $\Tr \phi^4$ in $\mathcal{N} = 4$ SYM%
	      \footnote{See for example Table 7 in \cite{DHoker:2002nbb}.}.
	\item  We note the absence of bubbles in the final result for this (unrenormalised) form factor. This may be related to the independence of the bare quantity on the matter content of the theory. One could then regard the computation as if it was performed in $\mathcal{N} = 4$ SYM, where the operator under consideration belongs to a protected multiplet.
	\item An unrelated observation is that the colour-ordered form factors with alternate and split-helicity configurations are the same:
	      \begin{equation}\label{eq:accident}
		      F_{\cO_4}^{(1)}(1^+,2^-,3^+,4^-;q) = F_{\cO_4}^{(1)}(1^+,3^+,2^-,4^-;q)\ .
	      \end{equation}
	      This is an accident due to the simple topology of the integral basis combined with the fact that bubbles do not appear. At first, the equality~\eqref{eq:accident} could appear as a consequence of the {photon decoupling} identities which hold in Yang-Mills theory. However these identities are no longer valid when one considers  interactions with higher powers of the  field strength.
\end{itemize}

%\section{The two-loop sub-minimal \texorpdfstring{$\Tr F^3$}{TrF3} form factor}

\section*{Acknowledgements}
We would like to thank Lorenzo Coccia, Harald Ita, Rodolfo Panerai, Sebastian P\"{o}gel, and in particular Lorenzo Bianchi for interesting discussions. This work  was supported by the Science and Technology Facilities Council (STFC) Consolidated Grant ST/P000754/1 \textit{``String theory, gauge theory \& duality''}, and by the European Union's Horizon 2020 research and innovation programme under the Marie Sk\l{}odowska-Curie grant agreement No.~764850 {\it ``\href{https://sagex.ph.qmul.ac.uk}{SAGEX}''}.

\newpage

\appendix

\section{Four-dimensional Spinor Helicity Formalism}\label{sec:spinorhel4D}

In this section we briefly review the four-dimensional Spinor Helicity Formalism (SHF) \cite{DeCausmaecker:1981jtq,Berends:1981uq,Kleiss:1985yh,Xu:1986xb}, having as a main goal to present our notation and conventions.

%Spinor Helicity Formalism (SHF) \cite{DeCausmaecker:1981jtq,Berends:1981uq,Kleiss:1985yh,Xu:1986xb} is a largely used and compact formalism to express amplitudes, because it allows to describe uniformly on-shell all the degrees of freedom, momenta and helicities, of particles with any spin.

SL(2,$\CC$) is the {\it universal covering} of the Lorentz group SO$^{+}$(1,3). This means that the projective representations of SO(1,3) on the Hilbert space are in one-to-one correspondence to the {\it unitary representations} of SL(2,$\CC$). Furthermore, these infinite-dimensional unitary representations on the states naturally induces finite-dimensional representations on the operators ({\it e.g.} the fields) of the theory. Moreover, group theory ensures that all finite dimensional irreducible representations of SL(2,$\CC$) can be found by taking the totally symmetric tensor product of a finite number of its fundamental and anti-fundamental representations. The finite dimensional irreducible representations, labelled by two semi-integer numbers $(m,n)$\footnote{In our conventions, $(m,n)$ has dimension $(2m+1)(2n+1)$.}, are obtained by the symmetric tensor product of the representations $\left(\frac{1}{2},0\right)$ and $\left(0,\frac{1}{2}\right)$, respectively $2m$ and $2n$ times.

%Let us introduce some notation and how the SHF works. The so-called undotted indices $\alpha,\beta,\dots$ are the ones of $\left(\frac{1}{2},0\right)$ and the dotted $\dot{\alpha}, \dot{\beta},\dots$ are the $\left(0,\frac{1}{2}\right)$ indices.
The fundamental objects transforming in the $\left(\frac{1}{2},0\right)$ representation will be labelled by $\lambda$ and the associated indices will be undotted Greek letters as $\alpha,\beta,\dots$. The fundamental objects transforming in the $\left(0,\frac{1}{2}\right)$ will be labelled
by $\tilde{\lambda}$ with associated dotted Greek indices $\dot{\alpha}, \dot{\beta},\dots$. We will refer to $\lambda$ and $\tilde{\lambda}$ collectively as {\it helicity spinors}.
In real momentum space, these two representations are related by complex conjugation.

What group theory states is that we can construct any quantity transforming in some representation of the Lorentz group in a uniform way in terms of objects transforming in these two representations. In general any scattering amplitude can be written as a function of the set $\{ \lambda_{i\, \alpha}, \widetilde{\lambda}_{i \, \dot{\alpha}}\}$, with the index $i$ running over all in- and outgoing particles of the considered process. In order to level out the description we will take all particles as outgoing from now on.

Dotted and undotted indices are raised and lowered through the Levi-Civita tensor $\epsilon$ and we adopt the following conventions:
\begin{equation}
	\lambda^{\alpha} = \epsilon^{\alpha \beta} \lambda_{\beta} = \epsilon^{\alpha \beta} \epsilon_{\beta \gamma} \lambda^{\gamma} \hspace{0.5cm} \rightarrow \hspace{0.5cm} \epsilon_{\alpha \gamma} \epsilon^{\gamma \beta} = \delta_{\alpha}^{\beta} \ .
\end{equation}
The same is true for dotted indices as well. Lorentz singlets can be build out of contractions of helicity spinors. Namely, our conventions for the angle and the square Lorentz invariant brackets are as follows:
\begin{equation}
	\setlength{\jot}{7pt}
	\begin{split}
		\agl{i}{j} &\coloneqq \agl{\lambda_{i}}{\lambda_{j}}\coloneqq \lambda_{i}^{\alpha} \lambda_{j \alpha} = - \agl{\lambda_{j}}{\lambda_{i}}\ ,\\
		\sqr{i}{j} &\coloneqq \sqr{\lambda_{i}}{\lambda_{j}}\coloneqq \widetilde{\lambda}_{i \dot{\alpha}} \widetilde{\lambda}_{j}^{\dot{\alpha}} = - \sqr{\lambda_{j}}{\lambda_{i}}\ .\\
	\end{split}
\end{equation}
It is simple to convince oneself that the so called {\it Schouten identity} holds:
\begin{equation}
	\agl{i}{j}\lambda_{k \alpha} + \agl{j}{k}\lambda_{i \alpha} + \agl{k}{i}\lambda_{j \alpha} = 0\ .
\end{equation}
An similar identity can be written for the $\widetilde{\lambda}$'s as well.

\subsection*{Massless momenta}

Considering Lorentz vectors, one has that the vector representation of SO(1,3) corresponds to the $\left(\frac{1}{2},\frac{1}{2}\right)$ representation of SL(2,$\CC$). This correspondence is explicitly given~by
\begin{equation}
	p^{\mu} \longrightarrow p_{\alpha \dot{\alpha}} \coloneqq p_{\mu} \sigma^{\mu}_{\alpha \dot{\alpha}}\ .
\end{equation}
The sigma matrices satisfy the Clifford algebra
\begin{equation}
	\{ \sigma_{\mu}, \bar{\sigma}_{\nu} \} =2 \eta_{\mu \nu}\ ,
\end{equation}
where $\bar{\sigma}^{\mu \dot{\alpha} \alpha}\coloneqq\epsilon^{\alpha \beta} \epsilon^{\dot{\alpha} \dot{\beta}} \sigma^{\mu}_{\beta \dot{\beta}}$.
Then, if we further define $p^{\dot{\alpha} \alpha} \coloneqq p^{\mu} \bar{\sigma}_{\mu}^{\dot{\alpha} \alpha}$, we have that
\begin{equation}
	\label{scalarproduct}
	p_{i}^{\dot{\alpha} \alpha} p_{j \alpha \dot{\alpha}} = 2 p_{i} \cdot p_{j}\ .
\end{equation}
For a massive particle the mass-shell condition can then be written as
\begin{equation}
	p^2 = m^2 \longrightarrow p^{\dot{\alpha} \alpha} p_{\alpha \dot{\alpha}}= \det(p_{\alpha \dot{\alpha}}) = m^2\ .
\end{equation}
The power of the SHF becomes manifest when we consider massless momenta. Indeed, the massless condition can be trivialised by picking
\begin{equation}
	\label{masslessmomentum}
	p_{i \alpha \dot{\alpha}} = \lambda_{i \alpha} \widetilde{\lambda}_{i \dot{\alpha}}\ .
\end{equation}

There is an ambiguity in the definition of $p_i$ in terms of $\{ \lambda_{i},\widetilde{\lambda}_{i}\}$, which is represented by a phase
\begin{equation}
	\lambda_{i \alpha} \longrightarrow e^{- i \varphi_i} \lambda_{i \alpha}\ , \hspace{1.5cm} \widetilde{\lambda}_{i \dot{\alpha}}\longrightarrow e^{i \varphi_i} \widetilde{\lambda}_{i \dot{\alpha}}\ .
\end{equation}
This rescaling leaves momentum invariant and is thus a {\it little group} transformation. The little group in four dimensions is the double covering of $\mathrm{SO}(2)\simeq\mathrm{U}(1)$ and we choose to assign helicity $-\frac{1}{2}$ to $\lambda$ and $+\frac{1}{2}$ to $\widetilde{\lambda}$. Thus it is now manifest how the new variables can carry information about both the momentum and the helicity of an associated particle.

\subsection*{Massive momenta}

At this point we turn our attention to the spinor helicity description of massive momenta, of which we make extensive use. One can always write a massive momentum $L$ as~\cite{Kosower:2004yz}
\begin{equation}
	\label{defmassivemomentum}
	L^{\mu}=l^{\mu} + \frac{L^2}{2 l\cdot \eta} \eta^{\mu}\ ,
\end{equation}
where both $l$ and $\eta$ are massless momenta and $L^2=m^2$ is the mass associated to this momentum. The previous expression fixes $l^{\mu}$ in terms of the massive momentum $L^{\mu}$ completely once we have chosen the \textit{arbitrary} $\eta^{\mu}$.
%\begin{equation}
%	l^{\mu}=L^{\mu}-\frac{L^2}{2 L\cdot \eta} \eta^{\mu}\ .
%\end{equation}
We can write \eqref{defmassivemomentum} in terms of helicity spinors as
\begin{equation}
	\label{massivemomentum}
	p_{i \alpha \dot{\alpha}} = \lambda_{i \alpha} \widetilde{\lambda}_{i \dot{\alpha}} + \frac{m^2}{\agl{\lambda_{i} }{\mu_{i}}\sqr{\widetilde{\mu}_{i}}{\widetilde{\lambda}_{i}}} \mu_{i \alpha} \widetilde{\mu}_{i \dot{\alpha}}\ .
\end{equation}
Focusing on the number of degrees of freedom (d.o.f.) we expect to have 3 d.o.f. from the spinor variables, plus an additional d.o.f. from the mass squared $m^2$. The quantity $\lambda_\alpha \widetilde{\lambda}_{\dot{\alpha}}$ already carries by itself 3 d.o.f., but $\mu_{\alpha}$ and $\widetilde{\mu}_{\dot{\alpha}}$ apparently carry two additional complex degrees, which coincide with their direction, while the momentum is invariant under the rescaling
\begin{equation}
	\label{nomodules}
	\mu_\alpha \longrightarrow a\ \mu_\alpha\ , \hspace{1.5cm}
	\tilde{\mu}_{\dot{\alpha}}\longrightarrow b\ \tilde{\mu}_{\dot{\alpha}}\ ,
\end{equation}
where $a,b\in \mathbb{C}$. The redundancy is taken into account by the four-dimensional {\it massive} little group $\widetilde{\mathrm{SO}}(3) \simeq \mathrm{SU}(2)$, which has two additional generators, with respect to the massless one.
Indeed we can write the massive momentum in terms of the irreducible SU(2) helicity spinors \cite{Arkani-Hamed:2017jhn}
\begin{equation}
	\lambda_{\alpha}^{I} = \mqty( \lambda_{\alpha} & \frac{m}{\agl{\lambda}{\mu}} \mu_{\alpha})\ ,
\end{equation}
where $I$ is an index in the fundamental of SU$(2)$ and
\begin{equation}
	p_{\alpha \dot{\alpha}} = \lambda_{\alpha}^{I} \widetilde{\lambda}_{\dot{\alpha} I} = \epsilon_{I J} \lambda_{\alpha}^{I} \widetilde{\lambda}_{\dot{\alpha}}^{J} \ ,
\end{equation}
where $\widetilde{\lambda}_{\dot{\alpha} I} =\pm (\lambda_{\alpha}^{I})^{\dagger}$, according to the sign of the $p_0$ component of momentum. In this form it is obvious that any SU(2) transformation
\begin{equation}
	\lambda_{\alpha}^{I} \longrightarrow \lambda_{\alpha}^{J} U_{J}\,^{I}
\end{equation}
leaves the momentum invariant.

\section{Six-Dimensional Spinor Helicity Formalism}
\label{sec:spinorhel6D}

In this section we give a concise overview of the six-dimensional spinor helicity formalism. In particular we are interested in how it can be broken down in terms of a four-dimensional subspace and the associated four-dimensional spinors. For a more detailed discussion see \cite{Bern:2010qa,Cheung:2009dc,Brandhuber:2010mm,Dennen:2009vk}.

\subsection{Helicity Spinors in Six Dimensions}

In six-dimensional Minkowski spacetime, the Lorentz group is SO(1,5), whose universal covering group is  SL(2,$\mathbb{H}$), and we will denote it as SU$^*$(4). Indeed, its representations are in one-to-one correspondence to those of SU(4), which is the universal covering of SO(6). The six-dimensional little group is $\widetilde{\rm SO}(4)\simeq {\rm SU}(2) \times {\rm SU}(2)$.

Let us denote with $\square^A$ and $\square_A$ the objects transforming respectively in the fundamental and anti-fundamental representations of the  Lorentz group SU$^*(4)$ and $(a,\dot{a})$ the indices of the bi-fundamental representations of the two components of the little group. The Clifford algebra is defined by
\begin{equation}
	\{\gamma^{\mu}, \widetilde{\gamma}^{\nu}\}_{A}\,^{B}\coloneqq \gamma^{\mu}_{A C} \widetilde{\gamma}^{\nu C B} + \gamma^{\nu}_{A C} \widetilde{\gamma}^{\mu C B} = 2\eta^{\mu \nu} \delta_{A}^{B}\ ,
\end{equation}
where $\mu=0,\dots ,6$, $\gamma^{\mu}_{A B} \equiv \gamma^{\mu}_{[A B]}$ and $\widetilde{\gamma}^{\mu A B} \equiv \widetilde{\gamma}^{\mu [A B]}$. These gamma matrices transform in the pseudo-real representation $\mathbf{6}=\mathbf{4}\,\wedge\, \mathbf{4}$ of SU$^*$(4) and are related by
\begin{equation}
	\widetilde{\gamma}^{\mu A B}=\left(\gamma^{\mu}_{A B}\right)^*=\frac{1}{2} \epsilon^{A B C D} \gamma^{\mu}_{A B} \> .
\end{equation}
Six-dimensional momenta can be written as
\begin{equation}
	p_{A B} \coloneqq p_{\mu} \gamma^{\mu}_{A B}\ ,
\end{equation}
and also transform in the $\mathbf{6}$ representation. The massless condition on the momenta reads
\begin{equation}
	p^2 \sim \epsilon^{A B C D} p_{A B} p_{C D} = 0\ ,
\end{equation}
which can be solved by expressing the momentum as the bi-spinor
\begin{equation}
	p_{A B} = \epsilon^{\dot{a} \dot{b}} \widetilde{\lambda}_{\dot{a} A} \widetilde{\lambda}_{\dot{b} B} = \widetilde{\lambda}_{\dot{a} A} \widetilde{\lambda}^{\dot{a}}_{B}\ ,
\end{equation}
where $\widetilde{\lambda}_{\dot{a} A}$ is a pseudo-real spinor. Analogously, we can write
\begin{equation}
	p^{A B} = \lambda^{a A} \lambda_{a}^{B} = -\epsilon^{a b} \lambda_{a}^{A} \lambda_b^B\ ,
\end{equation}
which satisfies
\begin{equation}
	p^{A B} = \left(p_{A B}\right)^* = -\frac{1}{2} \epsilon^{A B C D} p_{C D}\ .
\end{equation}
Notice that, given the above definitions, the spinors $\lambda_{a A}$ and $\widetilde{\lambda}_{\dot{a} A}$ {\it automatically} satisfy the Dirac equation:
\begin{equation}
	p_{A B} \lambda_{a}^{B} = -\frac{1}{2} \epsilon_{A B C D} \lambda_{a}^{B} \lambda^{b C} \lambda_{b}^{D} =-\epsilon_{A B C D} \lambda_{a}^{B} \lambda_{1}^{C} \lambda_{2}^{D} = 0\ ,
\end{equation}
and similarly for $\widetilde{\lambda}_{\dot{a} A}$. The Dirac equation can be also written equivalently as a relation between $\lambda$ and $\widetilde{\lambda}$:
\begin{equation}
	\label{Diracequation}
	0=\lambda^{a A} \lambda^{B}_{a} \widetilde{\lambda}_{B \dot{a}} = - \lambda_{1}^{A} \lambda^{B}_{2} \widetilde{\lambda}_{B \dot{a}} + \lambda_{2}^{A} \lambda^{B}_{1} \widetilde{\lambda}_{B \dot{a}}\ ,
\end{equation}
which implies
\begin{equation}
	\label{6DDirac}
	\lambda^{A}_{a} \widetilde{\lambda}_{A \dot{a}} = 0\ .
\end{equation}

Finally we need to define polarisation vectors in terms of the spinors. Just as in four dimensions one has to introduce a reference spinor to do so, which we call $q$. Then a good definition is given by
\begin{equation}\label{eq:pol6D}
	\def\arraystretch{2.5}
	\begin{array}{rl}
		\varepsilon^{A B}_{a \dot{a}}(p,q) & = \dfrac{\sqrt{2}}{s_{p q}}| p_{a} \rangle^{[A} \lbrack p_{\dot{a}} | \slashed{q}^{B]}\ ,   \\
		\varepsilon_{a \dot{a} A B}(p,q)   & = - \dfrac{\sqrt{2}}{s_{p q}}| p_{\dot{a}} \rbrack_{[A} \langle p_{a} | \slashed{q}_{B]}\ .
	\end{array}
\end{equation}

\subsection{\texorpdfstring{SU$^*$\hspace{-.2mm}(4) Spinor Identities}{SU(4) Spinor Identities}}

In this subsection we present some useful identities for six-dimensional spinors. We focus on the ${\rm SU}^*(4)$ structure of the spinors and keep the little group indices implicit for the sake of clarity. Of course little-group indices can be restored at any time because they are unambiguously related to each spinor.

Consider a certain number of spinors $\lambda_{i}^{A}$ (and $\widetilde{\lambda}_{i A}$), with labels $i=1,\ldots ,n$. The Lorentz invariant objects which can be built out of these spinors are of three types:
\begin{itemize}
	\item Bi-spinor invariant objects:
	      \begin{equation}
		      \lambda_{i}^{A} \widetilde{\lambda}_{j A} \coloneqq \langle i j \rbrack
	      \end{equation}
	\item Two distinct four-spinors invariant objects:
	      \begin{equation}
		      \epsilon_{A B C D} \lambda_{i}^{A}\lambda_{j}^{B}\lambda_{k}^{C}\lambda_{l}^{D} \coloneqq \langle i j k l \rangle\ , \hspace{.5cm}
		      \epsilon^{A B C D} \widetilde{\lambda}_{i A}\widetilde{\lambda}_{j B}\widetilde{\lambda}_{k C}\widetilde{\lambda}_{l D} \coloneqq \lbrack i j k l \rbrack\ .
	      \end{equation}
\end{itemize}
The spinors transform in the fundamental representation of ${\rm SU}^*(4)$, thus $A =1,\ldots , 4$. Two identities (and their two complex conjugate) follow immediately from this:

\begin{equation}
	\label{4id1}
	\lambda_{1}^{[A} \lambda_{2}^{B} \lambda_{3}^{C} \lambda_{4}^{D} \lambda_{5}^{E ]} = 0\ ,
\end{equation}
and
\begin{equation}
	\label{Sid1}
	\lambda_{1}^{[A} \lambda_{2}^{B} \lambda_{3}^{C} \lambda_{4}^{D]} = \frac{1}{4!} \epsilon^{A B C D} \langle 1 2 3 4\rangle \ , \\
\end{equation}
and analogous relations hold for $\widetilde{\lambda}_{i A}$. Equations \eqref{4id1} and \eqref{Sid1} can be combined to give the six-dimensional Schouten identity:

\begin{equation}
	\label{identity1}
	\sum_{\mathrm{cyclic}}\langle 1 2 3 4 \rangle \lambda_{5}^{A} = 0\ .
\end{equation}

\subsection{From Six-Dimensional to Four-Dimensional Quantities}
\label{sec::from_six_to_four}

%The main goal of the dimensional reconstruction procedure discussed in the main body of this paper, is to compute four-dimensional amplitudes
%from higher-dimensional ones.
For our purposes, we find it convenient to write six-dimensional spinors in terms of four-dimensional ones, allowing amplitudes to be expressed in terms of the more familiar four-dimensional spinors. We can view six-dimensional null vectors as four-dimensional massive ones, by defining the two complex mass parameters
\begin{equation}
	\label{eq:def_extramass}
	m\coloneqq p_4 + i p_5\ , \hspace{1.5cm}
	\widetilde{m} \coloneqq p_4 - i p_5\ ,
\end{equation}
where $p_4$ and $p_5$ are the fifth and the sixth components of the 6D momentum $p_{\mu}$. The six-dimensional massless condition becomes then
\begin{equation}
	p^2 = (p^{(4)})^2 - m \widetilde{m} =0\ .
\end{equation}
where $(p^{(4)})^2= p_0^2-p_1^2-p_2^2-p_3^2$ is the four-dimensional massive momentum associated to $p_{\mu}$.
We found it more efficient for our calculation to describe these momenta as a combination of two massless momenta, as in \eqref{massivemomentum}. We can decompose 6D helicity spinors in terms of 4D spinors as
\begin{equation}
	\label{6Dfermions}
	\lambda^{A}_{a}=\mqty(-\frac{m}{\langle\lambda \mu \rangle}\mu_\alpha & \lambda_\alpha  \vspace{1.5mm}\\ \widetilde{\lambda}^{\dot{\alpha}} & \frac{\widetilde{m}}{[\mu \lambda]}\widetilde{\mu}^{\dot{\alpha}})\ , \hspace{1cm}
	\widetilde{\lambda}_{A \dot{a}}=\mqty(\frac{\widetilde{m}}{\langle\lambda \mu \rangle}\mu^\alpha & \lambda^\alpha \vspace{1.5mm}\\ -\widetilde{\lambda}_{\dot{\alpha}} & \frac{m}{[\mu \lambda]}\widetilde{\mu}_{\dot{\alpha}})\ ,
\end{equation}
where the little group indices label the columns and the SU$^*$(4) indices label the rows. The SU$^*$(4) index structure can be broken down into two SL(2,$\CC$) complex conjugated indices:
\begin{equation}
	\label{splitting6D4D}
	\square^A = \mqty(\square_{\alpha} \\ \square^{\dot{\alpha}})\ , \hspace{1.5cm}
	\square_A = \mqty(\square^{\alpha} \\ \square_{\dot{\alpha}})\ .
\end{equation}
This embedding is specific of our choice of gamma matrices: indeed, we choose them such that the $\gamma$-matrices restricted to $\mu=0,\dots ,3$ reduce to the familiar chiral representation in four dimensions\footnote{For the explicit basis of gamma matrices see Appendix A of~\cite{Cheung:2009dc}.}.

$p^{A B}$ and $p_{A B}$ are invariant under the little group ${\rm SU}(2)\times {\rm SU}(2)$ transformations
\begin{equation}
	{\lambda'}^{A}_{a} = U_{a}\,^{b} \lambda^{A}_{b},\hspace{5mm} {\widetilde{\lambda}'}_{A \dot{a}} = U_{\dot{a}}\,^{\dot{b}} \widetilde{\lambda}_{A \dot{b}}, \hspace{5mm} (U_{a}\,^{b}, U_{\dot{a}}\,^{\dot{b}}) \in {\rm SU}(2)\times {\rm SU}(2)\ .
\end{equation}
The 6D momentum in 4D components reads:
\begin{equation}
	\label{eq:momentum6D}
	p^{A B} = \mqty(-m\epsilon_{\alpha \beta} & \lambda_\alpha \widetilde{\lambda}^{\dot{\beta}} + \rho \mu_{\alpha} \mu^{\dot{\beta}}\\ -\widetilde{\lambda}^{\dot{\alpha}}\lambda_\beta -\rho \widetilde{\mu}^{\dot{\alpha}} \mu_{\beta} & \widetilde{m}\epsilon^{\dot{\alpha} \dot{\beta}})\ ,
\end{equation}
where $\rho = \frac{m \widetilde{m}}{\langle\lambda \mu \rangle [\mu \lambda ]}$. We notice that $m$ and $\widetilde{m}$ completely fix the diagonal components, thus they are little group invariant objects (this was obvious from their definitions). In our choice of gamma matrices, the off-diagonal components precisely coincide with the 4D massive momentum:
\begin{equation}\label{eq:massivemom}
	p_{\alpha \dot{\alpha}}^{(4)} = \lambda_\alpha \widetilde{\lambda}_{\dot{\alpha}} + \rho \mu_{\alpha} \widetilde{\mu}_{\dot{\alpha}}\ ,\hspace{1cm}  (p^{(4)})^2 = m \widetilde{m}\ .
\end{equation}
It is easy to see that the two copies of ${\rm SU}(2)$ of the 6D little group act in an identical way on the  4D momenta and we recover the usual massive little group: indeed, they depend only on the combination $m \widetilde{m}$ and we can obtain dotted transformations from the undotted by simply replacing
\begin{equation}
	m \longrightarrow -\widetilde{m}\ ,\hspace{1cm} \widetilde{m} \longrightarrow - m\ .
\end{equation}

The Lorentz invariant quantities $\langle i_{a} j_{\dot{a}}\rbrack$, $\langle i_{a} j_{b} k_{c} l_{d} \rangle$, $\lbrack i_{\dot{a}} j_{\dot{b}} k_{\dot{c}} l_{\dot{d}} \rbrack$ can be written in terms of four-dimensional angle and square brackets, once the helicity indices are fixed ($a,b,c,d=1,2$ and $\dot{a},\dot{b},\dot{c},\dot{d}= \dot{1},\dot{2}$), by using the decomposition given in \eqref{splitting6D4D} and decomposing $\epsilon_{A B C D} \sim \sum \epsilon^{\alpha \beta} \epsilon_{\dot{\alpha} \dot{\beta}}$ and $\delta^{A}_{B}=\mathrm{diag}(\delta^{\beta}_{\alpha},\delta^{\dot{\alpha}}_{\dot{\beta}})$.

\section{Six-Dimensional Scattering Amplitudes}
\label{sec:SixDAmp}

Six-dimensional tree-level amplitudes are the basic ingredients of our unitarity-based recipe. In this section we give the analytic expressions needed for our calculations and comment on how to recover four-dimensional expressions in a specific limit.

As we already mentioned, in six dimensions the notion of helicity is encoded in a tensorial structure, which must be reflected by the amplitudes. The advantage of this tensorial nature of helicity is that a single (tensorial) expression of the amplitude contains all the possible four-dimensional helicity configurations, when dimensional reduced. The drawback however is that one looses some of the simplicity which was peculiar to specific helicity configurations. In particular there is no concept of MHV amplitudes.

In the previous section we have chosen the embedding of the four dimensions into the six-dimen\-sional space. Thus the four-dimensional helicity structure is embedded in the six-dimensional amplitudes. In general this represents a good consistency check for six-dimensional results. In fact for an appropriate limit these results must return their four-dimensional counterparts. More specifically, accordingly to our embedding, it turns out that states characterised by little-group indices $(1,1)$ and $(2,2)$ correspond to the positive and the negative helicity states in the four-dimensional limit ($m,\tilde{m} \to 0$), because of representation we chose for the gamma matrices. On the other hand, in four dimensions the additional $(1,2)$ and $(2,1)$ components coincide with two 4D scalars.

The four-gluon amplitude, computed in~\cite{Cheung:2009dc}, is
\begin{equation}
	\mathcal{A}_g(1_{a \dot{a}} , 2_{b \dot{b}} , 3_{c \dot{c}} , 4_{d \dot{d}})=
	\begin{tikzpicture}[scale=14,auto,cross/.style={path picture={
							\draw[black]
							(path picture bounding box.south east) -- (path picture bounding box.north west) (path picture bounding box.south west) -- (path picture bounding box.north east);
						}},baseline={([yshift=-1mm]tree.base)}	]

		\clip (-3pt,-2.5pt) rectangle (4.5pt,2.5pt);

		%The four point amplitude with gluons

		\def\x{0pt}
		\def\y{0pt}

		\node at (0+\x,0+\y) (one){};
		\node at (-2pt+\x,2pt+\y)(oneA){$1_{a\dot{a}}$};
		\node at (-2pt+\x,-2pt+\y)(oneB){$4_{d\dot{d}}$};
		\node at (1.5pt+\x,0+\y) (two){};
		\node at (0.75pt+\x,0+\y)[draw,circle,fill=gray!75,inner sep=3pt](center){};
		\node at (3.5pt+\x,2pt+\y)(twoA){$2_{b\dot{b}}$};
		\node at (3.5pt+\x,-2pt+\y)(twoB){$3_{c\dot{c}}$};

		\draw [thick,decorate, decoration={coil, amplitude=2.3pt, segment length=3pt}](center.center) -- (oneA);
		\draw [thick,decorate, decoration={coil, amplitude=2.3pt, segment length=3pt}](center.center) -- (oneB);
		\draw [thick,decorate, decoration={coil, amplitude=2.3pt, segment length=3pt}](center.center) -- (twoA);
		\draw [thick,decorate, decoration={coil, amplitude=2.3pt, segment length=3pt}](center.center) -- (twoB);

		\node at (0.75pt+\x,0+\y)[draw,circle,fill=gray!75,inner sep=9pt]{};
		\node at (0.75pt+\x,0+\y)(tree){tree};

	\end{tikzpicture}
	= -\dfrac{i}{s_{12}s_{23}}\langle 1_a \; 2_b \; 3_c \; 4_d \rangle [1_{\dot{a}},2_{\dot{b}},3_{\dot{c}},4_{\dot{d}}]\>.
\end{equation}

According to our embedding, we expect $\mathcal{A}_g(1_{22},2_{22},3_{11},4_{11})$ to reproduce the MHV amplitude $\mathcal{A}(1^-,2^-,3^+,4^+)$ in the limit $m_i,\tilde{m_i} \to 0$ for $i=1,\ldots,4$, which is indeed the case:

\begin{equation}
	\def\arraystretch{2}
	\begin{array}{rl}
		\mathcal{A}_g(1_{22},2_{22},3_{11},4_{11}) \bigg|_{\text{4D}}%&= -\dfrac{i\; [3\medspace 4]^2 \langle 1\medspace 2\rangle }{[1\medspace 2] [2\medspace 3]\langle 2\medspace 3\rangle } \\
		%&=i\; \dfrac{[3 \; 4]^4}{[1 \; 2][2 \; 3][3 \; 4][4 \; 1]} \\
		 & =i \, \dfrac{\agl{1}{2}^4}{\agl{1}{2} \agl{2}{3} \agl{3}{4} \agl{4}{1}} \ .
	\end{array}
\end{equation}

While $\mathcal{A}_g(1_{12},2_{21},3_{11},4_{22})$ reproduces the four-point amplitude with two scalars and two oppo\-site-helicity gluons $\mathcal{A}(1_{\phi},2_{\bar{\phi}},3^+,4^-)$:

\begin{equation}
	\mathcal{A}_g(1_{12},2_{21},3_{11},4_{22}) = i \, \dfrac{\agl{1}{4}^2 \agl{2}{4}^2}{\agl{1}{2} \agl{2}{3} \agl{3}{4} \agl{4}{1}} \ .
\end{equation}

Another amplitude of which we make frequent use is the six-dimensional four-point amplitude with two gluons and two scalars~\cite{Davies:2011vt}

\begin{equation}
	\mathcal{A}_s(1_{a \dot{a}} , 2_{b \dot{b}} , 3 , 4)=
	\begin{tikzpicture}[scale=14,auto,cross/.style={path picture={
							\draw[black]
							(path picture bounding box.south east) -- (path picture bounding box.north west) (path picture bounding box.south west) -- (path picture bounding box.north east);
						}},baseline={([yshift=-1mm]tree.base)}	]

		\clip (-3pt,-2.5pt) rectangle (4.5pt,2.5pt);

		%The four point amplitude with gluons

		\def\x{0pt}
		\def\y{0pt}

		\node at (0+\x,0+\y) (one){};
		\node at (-2pt+\x,2pt+\y)(oneA){$1_{a\dot{a}}$};
		\node at (-2pt+\x,-2pt+\y)(oneB){$4$};
		\node at (1.5pt+\x,0+\y) (two){};
		\node at (0.75pt+\x,0+\y)[draw,circle,fill=gray!75,inner sep=3pt](center){};
		\node at (3.5pt+\x,2pt+\y)(twoA){$2_{b\dot{b}}$};
		\node at (3.5pt+\x,-2pt+\y)(twoB){$3$};

		\draw [thick,decorate, decoration={coil, amplitude=2.3pt, segment length=3pt}](center.center) -- (oneA);
		\draw [thick,dashed](center.center) -- (oneB);
		\draw [thick,decorate, decoration={coil, amplitude=2.3pt, segment length=3pt}](center.center) -- (twoA);
		\draw [thick,dashed](center.center) -- (twoB);

		\node at (0.75pt+\x,0+\y)[draw,circle,fill=gray!75,inner sep=9pt]{};
		\node at (0.75pt+\x,0+\y)(tree){tree};

	\end{tikzpicture}
	= -\dfrac{i}{4s_{12}s_{23}}\langle 1_a \; 2_b \; 3_c \; 3^c \rangle [1_{\dot{a}},2_{\dot{b}},4^{\dot{d}},4_{\dot{d}}]\>.
\end{equation}

The massless scalars in six dimensions behave as massive scalars when reduced to four dimensions. Taking the limits $m_1,m_2,\widetilde{m}_1,\widetilde{m}_2 \to 0$ and choosing the helicity components we found the four-point amplitudes for gluons and massive scalars in four dimensions:
\begin{equation}
	\def\arraystretch{2}
	\begin{array}{rl}
		\mathcal{A}_s(1_{22},2_{11},3,4)\bigg|_{4D} & = - i \dfrac{\langle 1 | \slashed{p}_{3}^{(4)} | 2 \rbrack}{s_{12} s_{23}} \ , \\
		\mathcal{A}_s(1_{11},2_{11},3,4)\bigg|_{4D} & = i \mu^2 \dfrac{\sqr{1}{2}^2}{s_{12} s_{23}}\ ,
	\end{array}
\end{equation}
where $\mu^2$ coincides in this case with the mass of the scalar squared.

%It is worth stressing once more that, despite being different and independent objects in the full six-dimensional space, when projected onto the physical four-dimensional subspace both $\mathcal{A}_s(1_{a \dot{a}} , 2_{b \dot{b}} , 3 , 4)$ and (for example) $\mathcal{A}_g(1_{a \dot{a}} , 2_{b \dot{b}} , 3_{12} , 4_{12})$ describe an amplitude involving scalars. Since this amplitude is unique the two expressions must match, and it is indeed so. This argument can be used to check that the relative normalisations of the two amplitudes are correct, or even to fix the normalisation of one once the other is given. We made extensive use of this technique to fix the normalisations of the scalar form factors once the normalisation of the corresponding form factor for gluons is known.

Finally, the last amplitude one needs %or two-loop calculations
is the five-point tree-level amplitude. The amplitude with five-gluons has first been computed in~\cite{Cheung:2009dc}. In \cite{Dennen:2009vk,Bern:2010qa} this result has been extended to the five-point superamplitude in the $\cN=(1,1)$ theory. This superamplitude also contains information about the amplitude with scalar fields which is needed for the scalar subtraction when doing dimensional reconstruction. The amplitude with five gluons is
\begin{equation}
	\mathcal{A}_g(1_{a\dot{a}},2_{b\dot{b}},3_{c\dot{c}},4_{d\dot{d}},5_{e\dot{e}})=\frac{i}{s_{12}s_{23}s_{34}s_{45}s_{51}} \; (-\mathcal{M}_{a\dot{a}b\dot{b}c\dot{c}d\dot{d}e\dot{e}} + \mathcal{D}_{a\dot{a}b\dot{b}c\dot{c}d\dot{d}e\dot{e}})
\end{equation}
with
\begin{equation}
	\mathcal{M}_{a\dot{a}b\dot{b}c\dot{c}d\dot{d}e\dot{e}}=\langle 1_a | \slashed{p}_2 \slashed{p}_3\slashed{p}_4\slashed{p}_5|1_{\dot{a}}]\langle 2_b \; 3_c \; 4_d \; 5_e \rangle [2_{\dot{b}} \; 3_{\dot{c}} \; 4_{\dot{d}} \; 5_{\dot{e}}] +\; \text{cyclic} \>,
\end{equation}
and
\begin{equation}
	\def\arraystretch{1.5}
	\begin{array}{rl}
		2 \mathcal{D}_{a\dot{a}b\dot{b}c\dot{c}d\dot{d}e\dot{e}}  = & \langle 1_a \, \tilde{\Sigma}_{2 \dot{b}}]\langle 2_b \, 3_c \, 4_d \, 5_e \rangle [1_{\dot{a}}\, 3_{\dot{c}} \, 4_{\dot{d}} \, 5_{\dot{e}}]+ \langle 3_{c} \, \tilde{\Sigma}_{4 \dot{d}}]\langle 1_a \, 2_b \, 4_d \, 5_e \rangle [1_{\dot{a}} \, 2_{\dot{b}} \, 3_{\dot{c}} \, 5_{\dot{e}}]              \\
		+                                                           & \langle 4_d \, \tilde{\Sigma}_{5 \dot{e}}] \langle 1_a \, 2_b \, 3_c \, 5 _e \rangle [1_{\dot{a}} \, 2_{\dot{b}} \, 3_{\dot{c}} \, 4_{\dot{d}}] - \langle 3_c \, \tilde{\Sigma}_{5 \dot{e}}]\langle 1_a \, 2_b \, 4_d \, 5_e \rangle [1_{\dot{a}} \, 2_{\dot{b}} \, 3_{\dot{c}} \, 4_{\dot{d}}]            \\
		-                                                           & \lbrack 1_{\dot{a}} \, \Sigma_{2 b}\rangle \langle 1_a \, 3_c \, 4_d \, 5_e \rangle [2_{\dot{b}}\, 3_{\dot{c}} \, 4_{\dot{d}} \, 5_{\dot{e}}] - \lbrack 3_{\dot{c}} \, \Sigma_{4 d}\rangle\langle 1_a \, 2_b \, 3_c \, 5_e \rangle [1_{\dot{a}} \, 2_{\dot{b}} \, 4_{\dot{d}} \, 5_{\dot{e}}]              \\
		-                                                           & \lbrack 4_{\dot{d}} \, \Sigma_{5 e}\rangle \langle 1_a \, 2_b \, 3_c \, 4_d \rangle [1_{\dot{a}} \, 2_{\dot{b}} \, 3_{\dot{c}} \, 5_{\dot{e}}] + \lbrack 3_{\dot{c}} \, \tilde{\Sigma}_{5 e}\rangle \langle 1_a \, 2_b \, 3_c \, 4_d \rangle [1_{\dot{a}} \, 2_{\dot{b}} \, 4_{\dot{d}} \, 5_{\dot{e}}]\>.
	\end{array}
\end{equation}

The amplitude with two scalars and three gluons is
\begin{equation}
	\mathcal{A}_g(1_{\phi},2_{\bar{\phi}},3_{c\dot{c}},4_{d\dot{d}},5_{e\dot{e}})= - \frac{i}{s_{12}s_{23}s_{34}s_{45}s_{51}} \; (\mathcal{M}^{s}_{c\dot{c}d\dot{d}e\dot{e}} + \mathcal{D}^{s}_{c\dot{c}d\dot{d}e\dot{e}})\ ,
\end{equation}
with
\begin{equation}
	\def\arraystretch{1.5}
	\begin{array}{rl}
		\mathcal{M}^{s}_{c\dot{c}d\dot{d}e\dot{e}} = & \langle 3_{c} | \slashed{p}_{1} | 4_{d} \rangle \lbrack 3_{\dot{c}} | \slashed{p}_{2} | 4_{\dot{d}} \rbrack \langle 5_{e} | \slashed{p}_1 \slashed{p}_{2} \slashed{p}_{3} \slashed{p}_{4} | 5_{\dot{e}} \rbrack + \langle 4_{d} | \slashed{p}_{1} | 5_{e} \rangle \lbrack 4_{\dot{d}} | \slashed{p}_{2} | 5_{\dot{e}} \rbrack \langle 3_{c} | \slashed{p}_4 \slashed{p}_{5} \slashed{p}_{1} \slashed{p}_{2} | 3_{\dot{c}} \rbrack \\
		+                                            & \langle 3_{c} | \slashed{p}_{1} | 5_{e} \rangle \lbrack 3_{\dot{c}} | \slashed{p}_{2} | 5_{\dot{e}} \rbrack \langle 4_{d} | \slashed{p}_5 \slashed{p}_{1} \slashed{p}_{2} \slashed{p}_{3} | 4_{\dot{d}} \rbrack + \frac{1}{2} \langle 3_{c} \, 4_{d} \, 5_{e} \, 1^{a} \rangle \lbrack 3_{\dot{c}} \, 4_{\dot{d}} \, 5_{\dot{e}} \, 2_{\dot{b}} \rbrack \langle 1_{a} \, \tilde{\Sigma}_{2}^{\dot{b}} \rbrack\ ,
	\end{array}
\end{equation}
and
\begin{equation}
	\def\arraystretch{1.5}
	\begin{array}{rl}
		2 \mathcal{D}^{s}_{c\dot{c}d\dot{d}e\dot{e}} = & -\langle 4_{d} | \slashed{p}_{1} | 5_{e} \rangle \lbrack 3_{\dot{c}} | \slashed{p}_{2} | 5_{\dot{e}} \rbrack \langle 3_{c} \, \tilde{\Sigma}_{4 \dot{d}} \rbrack + \langle 4_{d} | \slashed{p}_{1} | 5_{e} \rangle \lbrack 3_{\dot{c}} | \slashed{p}_{2} | 4_{\dot{d}} \rbrack \langle 3_{c} \, \tilde{\Sigma}_{5 \dot{e}} \rbrack \\
		                                               & - \langle 3_{c} | \slashed{p}_{1} | 5_{e} \rangle \lbrack 3_{\dot{c}} | \slashed{p}_{2} | 4_{\dot{d}} \rbrack \langle 4_{d} \, \tilde{\Sigma}_{5 \dot{e}} \rbrack + \langle 3_{c} | \slashed{p}_{1} | 5_{e} \rangle \lbrack 4_{\dot{d}} | \slashed{p}_{2} | 5_{\dot{e}} \rbrack \lbrack 3_{\dot{c}} \, \Sigma_{4 d} \rangle        \\
		                                               & - \langle 3_{c} | \slashed{p}_{1} | 4_{d} \rangle \lbrack 4_{\dot{d}} | \slashed{p}_{2} | 5_{\dot{e}} \rbrack \lbrack 3_{\dot{c}} \, \Sigma_{5 e} \rangle + \langle 3_{c} | \slashed{p}_{1} | 4_{d} \rangle \lbrack 3_{\dot{c}} | \slashed{p}_{2} | 5_{\dot{e}} \rbrack \lbrack 4_{\dot{d}} \, \Sigma_{5 e} \rangle\ .
	\end{array}
\end{equation}

The $\Sigma$ and $\tilde{\Sigma}$ that appear in the previous formulae are defined as
\begin{equation}
	\def\arraystretch{1.5}
	\begin{array}{rl}
		| \Sigma_{i a}\rangle         & = \left(\slashed{p}_{i} \slashed{p}_{i+1} \slashed{p}_{i+2} \slashed{p}_{i+3} - \slashed{p}_{i} \slashed{p}_{i+3} \slashed{p}_{i+2} \slashed{p}_{i+1} \right) | i_{a} \rangle \\
		| \tilde{\Sigma}_{i a}\rbrack & = \left(\slashed{p}_{i} \slashed{p}_{i+1} \slashed{p}_{i+2} \slashed{p}_{i+3} - \slashed{p}_{i} \slashed{p}_{i+3} \slashed{p}_{i+2} \slashed{p}_{i+1} \right) | i_{a} \rbrack
	\end{array}
\end{equation}
where we define $\slashed{p}_{6} \equiv \slashed{p}_{1}$.

\section{Non-Minimal Form Factors}\label{sec:nonminimal}

In this section we will address the computation of six-dimensional tree-level building blocks using BCFW recursion relations%
\footnote{For a more detailed account of six-dimensional BCFW see~\cite{Cheung:2009dc}.}. In particular we briefly comment on the main steps of the calculation of $\Tr F^2$ in the non-minimal configuration.

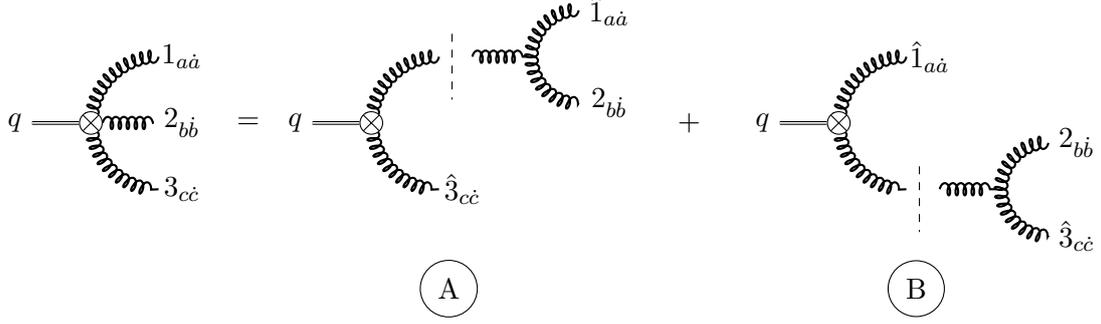
\begin{figure}[H]
	\centering
	\begin{tikzpicture}[scale=25,auto,cross/.style={path picture={
							\draw[black]
							(path picture bounding box.south east) -- (path picture bounding box.north west) (path picture bounding box.south west) -- (path picture bounding box.north east);
						}}]

		%One loop the form factors

		\def\y{0}
		\def\x1{0}

		\node at (0.5pt+\x1,2pt+\y) (one){};
		\draw [thick,decorate, decoration={coil, amplitude=2.3pt, segment length=3pt}](one.east) arc (90:270:1pt);
		\node [draw,circle,cross, inner sep=3pt,fill=white] at (-0.36pt+\x1,1pt+\y)(cross){};
		\node at (-1.5pt+\x1,1pt+\y)(Q){$q$};

		\draw [double](Q) -- (cross);

		\node at (1pt+\x1,2pt+\y) (oneB){$1_{a \dot{a}}$};
		\node at (1pt+\x1,1pt+\y) (two){$2_{b \dot{b}}$};
		\node at (1pt+\x1,0pt+\y) (three){$3_{c \dot{c}}$};

		\draw [thick,decorate, decoration={coil, amplitude=2.3pt, segment length=3pt}](cross) -- (two);

		\node at (2pt+\x1,1pt+\y){$=$};

		\def\y{0}
		\def\x1{4.2pt}

		\node at (0.5pt+\x1,2pt+\y) (one){};
		\draw [thick,decorate, decoration={coil, amplitude=2.3pt, segment length=3pt}](one.east) arc (90:270:1pt);
		\node [draw,circle,cross, inner sep=3pt,fill=white] at (-0.36pt+\x1,1pt+\y)(cross){};
		\node at (-1.5pt+\x1,1pt+\y)(Q){$q$};

		\draw [double](Q) -- (cross);

		\node at (1pt+\x1,2pt+\y) (oneB){};
		\node at (1pt+\x1,0pt+\y) (three){$\hat{3}_{c \dot{c}}$};

		%the cut

		\node at (0.85pt+\x1,2.5pt+\y) (cut1){};
		\node at (0.85pt+\x1,1.2pt+\y) (cut2){};
		\draw [dashed] (cut1) -- (cut2);
		%\node at (0.95pt+\x1,2.4pt+\y)[rotate around={-90:(1pt,2.5pt)}]{\Cutright};

		%The tree-level three-point amplitude

		\node at (1pt+\x1,2pt+\y)(quattro){};
		\node at (2pt+\x1,2pt+\y)(center){};
		\node at (3.2pt+\x1,2.7pt+\y)(out1){$\hat{1}_{a \dot{a}}$};
		\node at (3.2pt+\x1,1.3pt+\y)(out2){$2_{b\dot{b}}$};

		\draw [thick,decorate, decoration={coil, amplitude=2.3pt, segment length=3pt}] (quattro) -- (center.center);
		%\draw [thick,decorate, decoration={coil, amplitude=2.3pt, segment length=3pt}] (out1) -- (center.center);
		%\draw [thick,decorate, decoration={coil, amplitude=2.3pt, segment length=3pt}] (out2) -- (center.center);

		\draw [thick,decorate, decoration={coil, amplitude=2.3pt, segment length=3pt}](out1.west) arc (90:270:0.7pt);

		\node at (8.6pt,1pt){$+$};

		\def\y{0}
		\def\x1{11.2pt}

		\node at (0.5pt+\x1,2pt+\y) (one){};
		\draw [thick,decorate, decoration={coil, amplitude=2.3pt, segment length=3pt}](one.east) arc (90:270:1pt);
		\node [draw,circle,cross, inner sep=3pt,fill=white] at (-0.36pt+\x1,1pt+\y)(cross){};
		\node at (-1.5pt+\x1,1pt+\y)(Q){$q$};

		\draw [double](Q) -- (cross);

		\node at (1pt+\x1,2pt+\y) (oneB){$\hat{1}_{a\dot{a}}$};
		\node at (1pt+\x1,0pt+\y) (three){};

		%the cut

		\node at (0.85pt+\x1,0.5pt+\y) (cut1){};
		\node at (0.85pt+\x1,-0.8pt+\y) (cut2){};
		\draw [dashed] (cut1) -- (cut2);
		%\node at (0.95pt+\x1,0.4pt+\y)[rotate around={-90:(1pt,2.5pt)}]{\Cutright};

		%The tree-level three-point amplitude

		\node at (1pt+\x1,0pt+\y)(quattro){};
		\node at (2pt+\x1,0pt+\y)(center){};
		\node at (3.2pt+\x1,0.7pt+\y)(out1){$2_{b \dot{b}}$};
		\node at (3.2pt+\x1,-0.7pt+\y)(out2){$\hat{3}_{c\dot{c}}$};

		\draw [thick,decorate, decoration={coil, amplitude=2.3pt, segment length=3pt}] (quattro) -- (center.center);
		%\draw [thick,decorate, decoration={coil, amplitude=2.3pt, segment length=3pt}] (out1) -- (center.center);
		%\draw [thick,decorate, decoration={coil, amplitude=2.3pt, segment length=3pt}] (out2) -- (center.center);

		\draw [thick,decorate, decoration={coil, amplitude=2.3pt, segment length=3pt}](out1.west) arc (90:270:0.7pt);

		%the labels of the two terms

		\node at (5pt, -1.5pt) [draw, circle] {A};
		\node at (12pt, -1.5pt) [draw, circle] {B};

	\end{tikzpicture}
	\caption{BCFW construction of the tree-level non-minimal $\Tr F^2$ form factor in six dimensions.}\label{fig:TrF2nonminimal}
\end{figure}

Diagrammatically the terms we need to compute are represented in Figure~\ref{fig:TrF2nonminimal}. In this computation one needs to make use of the
three-point on-shell amplitudes in six-dimensions. These are most conveniently defined in terms of a set of auxiliary SU(2) spinors which we
denote by $u_a,\tilde{u}_{\dot{a}},w_a$ and $\tilde{w}_{\dot{a}}$, following the conventions of~\cite{Cheung:2009dc}. These objects are not Lorentz invariants in six dimensions and thus are not allowed to appear in the final expression, however they enjoy useful properties which simplify the calculation. The on-shell three-point amplitude cleanly expressed in terms of the above mentioned spinors:
\begin{equation}\label{eq:threepoint6D}
	\mathcal{A}_3 (1_{a \dot{a}},2_{b \dot{b}},3_{c \dot{c}}) = i \, \Gamma_{abc}(1,2,3)\,\widetilde{\Gamma}_{\dot{a}\dot{b}\dot{c}}(1,2,3) \> ,
\end{equation}
with
\begin{equation}
	\def\arraystretch{1.5}
	\begin{array}{c}
		\Gamma_{abc}(1,2,3)=u_{1 \, a}u_{2 \, b}w_{3 \, c} + u_{1 \, a}w_{2 \, b} u_{3 \, c}+ w_{1 \, a} u_{2 \, b} u_{3 \, c} \ , \\
		\widetilde{\Gamma}_{\dot{a}\dot{b}\dot{c}}(1,2,3)=\tilde{u}_{1 \, \dot{a}}\tilde{u}_{2 \, \dot{b}}\tilde{w}_{3 \, \dot{c}} + \tilde{u}_{1 \, \dot{a}}\tilde{w}_{2 \, \dot{b}} \tilde{u}_{3 \, \dot{c}}+ \tilde{w}_{1 \, \dot{a}} \tilde{u}_{2 \, \dot{b}} \tilde{u}_{3 \, \dot{c}} \ .
	\end{array}
\end{equation}

Consider now applying six-dimensional BCFW as in Figure~\ref{fig:TrF2nonminimal}. The hatted momenta are shifted by a quantity proportional to the complex parameter $z$ as
\begin{equation}\label{eq:momshift6D}
	\begin{array}{c}
		\hat{p}_1=p_1+z \, X^{a \dot{a}}\,\varepsilon_{1 \, a\dot{a}} \> , \\
		\hat{p}_3=p_3-z \, X^{a \dot{a}}\,\varepsilon_{1 \, a\dot{a}} \> ,
	\end{array}
\end{equation}
where $X^{a \dot{a}}$ is an arbitrary tensor needed to saturate the little group indices. This tensor, which also multiplies~\ref{eq:termA}, will be removed at the end of the calculation. The on-shell condition $\hat{p}_{1,2}^2=0$ implies $\text{det}\,X=0$, which allows to express $X$ as
\begin{equation}
	X^{a \dot{a}}=x^a \tilde{x}^{\dot{a}} \>.
\end{equation}
Furthermore we can define the quantities
\begin{equation}
	y^b=\tilde{x}^{\dot{a}}\langle 3_b \, 1^{\dot{a}}]^{-1} \> , \hspace{0.5cm} \tilde{y}_{\dot{b}}=x^a \langle 1^a \, 3^{\dot{b}} ]^{-1} \>,
\end{equation}

which allow us to rewrite the momentum shift~\ref{eq:momshift6D} in terms of the spinor shifts
\begin{equation}\label{eq:spinorshift6D}
	\def\arraystretch{1.5}
	\begin{array}{l}
		|\hat{1}^a \rangle = |1^a \rangle + z \, x^a y_b \, |3^b \rangle \>,                                  \\
		|\hat{3}^b \rangle = |3^b \rangle + z \, y^b x_a \, |1^a \rangle \>,                                  \\
		|\hat{1}^{\dot{a}}]= |1^{\dot{a}}] - z \, \tilde{x}^{\dot{a}}\tilde{y}_{\dot{b}} \, |3^{\dot{b}}] \>, \\
		|\hat{3}^{\dot{b}}]= |3^{\dot{b}}] - z \, \tilde{y}^{\dot{b}}\tilde{x}_{\dot{a}} \, |1^{\dot{a}}] \>.
	\end{array}
\end{equation}

% The reader can easily convince himself that these spinor-shifts give the appropriate momentum-shifts by writing for example $\hat{p}_1$ as
% \begin{equation}
% 	\hat{p}_1^{\mu}=-\frac{1}{4}\langle \hat{1}^a \, \sigma^{\mu} \, \hat{1}^b \rangle \, \epsilon_{ab} \>.
% \end{equation}

Considering now for example term A in Figure~\ref{fig:TrF2nonminimal} one has
\begin{equation}\label{eq:termA}
	\def\arraystretch{1.5}
	\begin{array}{rl}
		(\text{A}) & =X^{a \dot{a}}\mathcal{A}_{3}(\hat{1}_{a \dot{a}},2_{b \dot{b}},\hat{k}_{d \dot{d}})\, \dfrac{-i}{s_{12}} \, F^{(0)}_{\mathcal{O}_2}(-\hat{k}^{d \dot{d}},\hat{3}_{c \dot{c}};q)                                                    \\
		           & = \dfrac{i}{s_{12}} \, X^{a \dot{a}} \, \Gamma_{abd}(\hat{1},2,\hat{k})\, \widetilde{\Gamma}_{\dot{a}\dot{b}\dot{d}}(\hat{1},2,\hat{k}) \, \langle \hat{k}^{d} \, \hat{3}_{\dot{c}}] \langle \hat{3}_{c} \, \hat{k}^{\dot{d}}] \> .
	\end{array}
\end{equation}

Before substituting the definitions \eqref{eq:spinorshift6D} in \eqref{eq:termA}, we make use of the properties of $u,\tilde{u},w,\tilde{w}$ to simplify this expression. The most useful identities are
\begin{equation}
	\def\arraystretch{1.5}
	\begin{array}{c}
		u_{i \, a}w_{i \, b}-u_{i \, b}w_{i \, a}=\epsilon_{ab} \>, \hspace{0.5cm} \tilde{u}_{i \, \dot{a}}\tilde{w}_{i \, \dot{b}}-\tilde{u}_{i \, \dot{b}}\tilde{w}_{i \, \dot{a}}=\epsilon_{\dot{a}\dot{b}} \ , \\
		| u_i \cdot i \rangle = |u_j \cdot j \rangle \> , \hspace{0.5cm} | \tilde{u}_i \cdot i ]= | \tilde{u}_j \cdot j ] \hspace{0.5cm} \forall \> i,j = 1,2,k \ ,                                                \\
		|w_1 \cdot 1 \rangle + |w_2 \cdot 2 \rangle + |w_k \cdot k \rangle =0 \ ,                                                                                                                                  \\
		| \tilde{w}_1 \cdot 1 ] + | \tilde{w}_2 \cdot 2 ] + | \tilde{w}_k \cdot k ]=0 \ ,
	\end{array}
\end{equation}
where we used the shorthand notation $u_{i \, a} | i^a \rangle =|u_i \cdot i \rangle$ and $\tilde{u}_{i \, \dot{a}} | i^{\dot{a}}]= | \tilde{u}_i \cdot i ]$. These identities allow us to rewrite
\begin{equation}
	\def\arraystretch{1.3}
	\begin{array}{l}
		\Gamma_{abd}(\hat{1},2,\hat{k})\, \langle \hat{k}^d | = \langle\hat{1}_a| \,  u_{2 \, b} +\langle 2_b | \,  u_{\hat{1}\, a} \> , \\
		\widetilde{\Gamma}_{\dot{a}\dot{b}\dot{d}}(\hat{1},2,\hat{k})\, |\hat{k}^{\dot{d}} ] = |\hat{1}_{\dot{a}}] \,  \tilde{u}_{2 \, \dot{b}} +|2_{\dot{b}} ] \,  \tilde{u}_{\hat{1}\, \dot{a}} \> ,
	\end{array}
\end{equation}
which in turn leads to
\begin{equation}
	\def\arrastretch{1.5}
	\begin{array}{rl}
		(\text{A}) = \dfrac{i}{s_{12}}X^{a \dot{a}} & \left( \langle \hat{1}_a \, \hat{3}_{\dot{c}}] \langle \hat{3}_c \, \hat{1}_{\dot{a}}] \, u_{2 \, b} \tilde{u}_{2 \, \dot{b}} + \langle \hat{1}_a \, \hat{3}_{\dot{c}}] \langle \hat{3}_c \, 2_{\dot{b}}] \, u_{2 \, b} \tilde{u}_{1 \, \dot{a}}  \right. \\
		                                            & + \left. \langle 2_b \, \hat{3}_{\dot{c}}] \langle \hat{3}_c \, \hat{1}_{\dot{a}}] \, u_{\hat{1} \, a} \tilde{u}_{2 \, \dot{b}} +
		\langle 2_b \, \hat{3}_{\dot{c}}] \langle \hat{3}_c \, 2_{\dot{b}}] \, u_{\hat{1} \, a} \tilde{u}_{\hat{1} \, \dot{a}} \, \right) \>.
	\end{array}
\end{equation}

To further simplify the result, and to eliminate the residual SU(2) spinors, we make the following observations:
\begin{itemize}
	\item pairs of $u_i,\tilde{u}_j$ with $i \neq j$ can be immediately rewritten in terms of six-dimensional invariants as
	      \begin{equation}
		      \def\arraystretch{1.36}
		      \begin{array}{c}
			      u_{1 \, a}\tilde{u}_{2 \, \dot{b}}=\langle 1_a \, 2_{\dot{b}}] \>, \hspace{0.5cm} u_{2 \, b}\tilde{u}_{1 \, \dot{a}}=-\langle 2_b \, 1_{\dot{a}}] \>, \\
			      u_{2 \, b}\tilde{u}_{k \, \dot{c}}=\langle 2_b \, k_{\dot{c}}] \>, \hspace{0.5cm} u_{k \, c}\tilde{u}_{2 \, \dot{b}}=-\langle k_c \, 2_{\dot{b}}] \>.
		      \end{array}
	      \end{equation}
	\item pairs of $u_i,\tilde{u}_j$ with $i = j$ can be rewritten using the identity~\cite{Dennen:2009vk}
	      \begin{equation}
		      u_{i \, a}\tilde{u}_{i \, \dot{a}}= \frac{(-1)^{\mathcal{P}_{ij}}}{s_{iP}}\, \langle i_a |p_j P | i_{\dot{a}}] \>,
	      \end{equation}
	      where $p_j$ is any other momentum belonging to the same three-point amplitude as $p_i$, and $\mathcal{P}_{ij}=+1$ for clockwise ordering of the states $(i,j)$. Also $P$ is any given arbitrary momentum.
\end{itemize}

Repeating a similar reasoning on term (B) one gets
\begin{equation}
	\def\arrastretch{1.5}
	\begin{array}{rl}
		(\text{B}) = \dfrac{i}{s_{23}}X^{a \dot{a}} & \left( \langle \hat{1}_a \, \hat{3}_{\dot{c}}] \langle \hat{3}_c \, \hat{1}_{\dot{a}}] \, u_{2 \, b} \tilde{u}_{2 \, \dot{b}} + \langle \hat{1}_a \, \hat{3}_{\dot{c}}] \langle 2_b \, \hat{1}_{\dot{a}}] \, u_{3 \, c} \tilde{u}_{2 \, \dot{b}}  \right. \\
		                                            & + \left. \langle \hat{1}_a \, 2_{\dot{b}}] \langle \hat{3}_c \, \hat{1}_{\dot{a}}] \, u_{2 \, b} \tilde{u}_{\hat{3} \, \dot{c}} +
		\langle \hat{1}_a \, 2_{\dot{b}}] \langle 2_b \, \hat{1}_{\dot{a}}] \, u_{\hat{3} \, c} \tilde{u}_{\hat{3} \, \dot{c}} \, \right) \>.
	\end{array}
\end{equation}

The on-shell condition for the intermediate propagators in (A) and (B) defines two different BCFW shift parameters, which we label $z_A$ and $z_B$ respectively. By computing $z_A$ and $z_B$ one can see that they are related by
\begin{equation}
	z_B=-\frac{s_{23}}{s_{12}}\, z_A \>.
\end{equation}
Thanks to this relation multiple cancellations happen between terms in (A) and terms in (B). With some further algebra and removing the $X^{a \dot{a}}$ tensor, one  arrives at~\eqref{eq:TrF2nonminimal}.

The analytic expression of the six-dimensional form factor $F^{(0)}_{\mathcal{O}_2}(1_{a \dot{a}},2_{b \dot{b}},3_{c \dot{c}};q)$ could also be computed using Feynman diagrams, see for example~\cite{Davies:2011vt}. Due to the low multiplicity of this form factor, there is just a small number of contributing Feynman diagrams. The diagrammatic approach may thus be considered as equivalently viable as BCFW in this case, the latter method however leads to a far more compact expression with all the symmetries manifest.

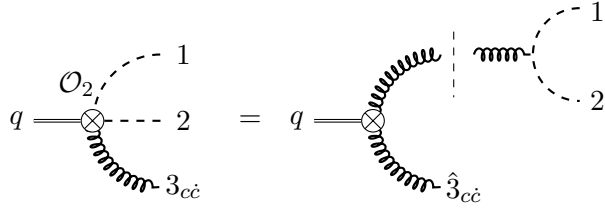
\begin{figure}
	\centering
	\begin{tikzpicture}[scale=25,auto,cross/.style={path picture={
							\draw[black]
							(path picture bounding box.south east) -- (path picture bounding box.north west) (path picture bounding box.south west) -- (path picture bounding box.north east);
						}}]

		%One loop the form factors

		\def\y{0}
		\def\x1{0}

		\node at (0.5pt+\x1,2pt+\y) (one){};
		\node at (-0.36pt+\x1,1pt+\y)(cross0){};
		\draw [thick,dashed](one.east) arc (90:180:1pt);
		\draw [thick,decorate, decoration={coil, amplitude=2.3pt, segment length=3pt}](cross0.center) arc (180:270:1pt);
		\node [draw,circle,cross, inner sep=3pt,fill=white] at (-0.36pt+\x1,1pt+\y)(cross){};
		\node at (-0.6pt+\x1,1.55pt+\y)(label){$\cO_2$};
		\node at (-1.5pt+\x1,1pt+\y)(Q){$q$};

		\draw [double](Q) -- (cross);

		\node at (1pt+\x1,2pt+\y) (oneB){$1$};
		\node at (1pt+\x1,1pt+\y) (two){$2$};
		\node at (1pt+\x1,0pt+\y) (three){$3_{c \dot{c}}$};

		\draw [thick,dashed](cross) -- (two);

		\node at (2pt+\x1,1pt+\y){$=$};

		\def\y{0}
		\def\x1{4.2pt}

		\node at (0.5pt+\x1,2pt+\y) (one){};
		\draw [thick,decorate, decoration={coil, amplitude=2.3pt, segment length=3pt}](one.east) arc (90:270:1pt);
		\node [draw,circle,cross, inner sep=3pt,fill=white] at (-0.36pt+\x1,1pt+\y)(cross){};
		\node at (-1.5pt+\x1,1pt+\y)(Q){$q$};

		\draw [double](Q) -- (cross);

		\node at (1pt+\x1,2pt+\y) (oneB){};
		\node at (1pt+\x1,0pt+\y) (three){$\hat{3}_{c \dot{c}}$};

		%the cut

		\node at (0.85pt+\x1,2.5pt+\y) (cut1){};
		\node at (0.85pt+\x1,1.2pt+\y) (cut2){};
		\draw [dashed] (cut1) -- (cut2);
		%\node at (0.95pt+\x1,2.4pt+\y)[rotate around={-90:(1pt,2.5pt)}]{\Cutright};

		%The tree-level three-point amplitude

		\node at (1pt+\x1,2pt+\y)(quattro){};
		\node at (2pt+\x1,2pt+\y)(center){};
		\node at (3pt+\x1,2.7pt+\y)(out1){$\hat{1}$};
		\node at (3pt+\x1,1.3pt+\y)(out2){$2$};

		\draw [thick,decorate, decoration={coil, amplitude=2.3pt, segment length=3pt}] (quattro) -- (center.center);
		%\draw [thick,decorate, decoration={coil, amplitude=2.3pt, segment length=3pt}] (out1) -- (center.center);
		%\draw [thick,decorate, decoration={coil, amplitude=2.3pt, segment length=3pt}] (out2) -- (center.center);

		\draw [thick,dashed](out1.west) arc (90:270:0.7pt);

	\end{tikzpicture}
	\caption{BCFW construction of the tree-level non-minimal $\Tr F^2$ form factor with two scalars.}\label{fig:TrF2nonminimal_scalars}
\end{figure}

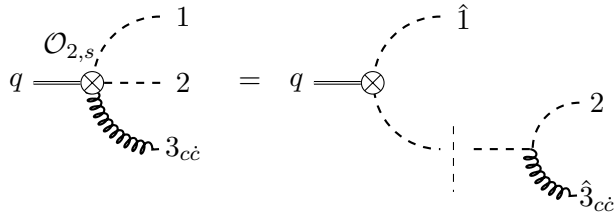
\begin{figure}[H]
	\centering
	\begin{tikzpicture}[scale=25,auto,cross/.style={path picture={
							\draw[black]
							(path picture bounding box.south east) -- (path picture bounding box.north west) (path picture bounding box.south west) -- (path picture bounding box.north east);
						}}]

		%One loop the form factors

		\def\y{0}
		\def\x1{0}

		\node at (0.5pt+\x1,2pt+\y) (one){};
		\node at (-0.36pt+\x1,1pt+\y)(cross0){};
		\draw [thick,dashed](one.east) arc (90:180:1pt);
		\draw [thick,decorate, decoration={coil, amplitude=2.3pt, segment length=3pt}](cross0.center) arc (180:270:1pt);
		\node [draw,circle,cross, inner sep=3pt,fill=white] at (-0.36pt+\x1,1pt+\y)(cross){};
		\node at (-0.7pt+\x1,1.53pt+\y)(label){$\cO_{2,s}$};
		\node at (-1.5pt+\x1,1pt+\y)(Q){$q$};

		\draw [double](Q) -- (cross);

		\node at (1pt+\x1,2pt+\y) (oneB){$1$};
		\node at (1pt+\x1,1pt+\y) (two){$2$};
		\node at (1pt+\x1,0pt+\y) (three){$3_{c \dot{c}}$};

		\draw [thick,dashed](cross) -- (two);

		\node at (2pt+\x1,1pt+\y){$=$};

		\def\y{0}
		\def\x1{4.2pt}

		\node at (0.5pt+\x1,2pt+\y) (one){};
		\draw [thick,dashed](one.east) arc (90:270:1pt);
		\node [draw,circle,cross, inner sep=3pt,fill=white] at (-0.36pt+\x1,1pt+\y)(cross){};
		\node at (-1.5pt+\x1,1pt+\y)(Q){$q$};

		\draw [double](Q) -- (cross);

		\node at (1pt+\x1,2pt+\y) (oneB){$\hat{1}$};
		\node at (1pt+\x1,0pt+\y) (three){};

		%the cut

		\node at (0.85pt+\x1,0.5pt+\y) (cut1){};
		\node at (0.85pt+\x1,-0.8pt+\y) (cut2){};
		\draw [dashed] (cut1) -- (cut2);
		%\node at (0.95pt+\x1,0.4pt+\y)[rotate around={-90:(1pt,2.5pt)}]{\Cutright};

		%The tree-level three-point amplitude

		\node at (1pt+\x1,0pt+\y)(quattro){};
		\node at (2pt+\x1,0pt+\y)(center){};
		\node at (3pt+\x1,0.7pt+\y)(out1){$2$};
		\node at (3pt+\x1,-0.7pt+\y)(out2){$\hat{3}_{c\dot{c}}$};

		\draw [thick,dashed] (quattro) -- (center.center);
		%\draw [thick,decorate, decoration={coil, amplitude=2.3pt, segment length=3pt}] (out1) -- (center.center);
		%\draw [thick,decorate, decoration={coil, amplitude=2.3pt, segment length=3pt}] (out2) -- (center.center);

		\draw [thick,dashed](out1.west) arc (90:180:0.7pt);
		\draw [thick,decorate, decoration={coil, amplitude=2.3pt, segment length=3pt}](center.center) arc (180:270:0.7pt);

	\end{tikzpicture}
	\caption{BCFW construction of the tree-level non-minimal $D\phi^2$ form factor.}\label{fig:Dphi2nonminimal}
\end{figure}

In a similar way but with much less involved calculation, we can find both the non-minimal form factors with two scalars and one gluon \eqref{eq::TrF2nonmininaml_scalars} and \eqref{eq::Dphi2nonminimal}. In Figure~\ref{fig:TrF2nonminimal_scalars} and Figure~\ref{fig:Dphi2nonminimal} we show the BCFW factorization channels for these calculations. The only missing ingredient is the three-point amplitude with two scalars and one gluon in six dimensions, which turns out to be very simple:

\begin{equation}
	\cA (1_{a \dot{a}},2,3) = i \, u_{1 a} \widetilde{u}_{1 \dot{a}}\ .
\end{equation}

\section{Integral Expressions}
\label{sec:integrals}

The integrals needed in this paper are:

\begin{equation}
	\begin{aligned}
		\begin{tikzpicture}[baseline={([yshift=-1mm]uno.base)},scale=15]
			\def\x{0}
			\def\y{0}

			\clip (-2.5pt,-2pt) rectangle (2.5pt,2pt);

			\def\x{0pt}
			\def\y{0}

			\node at (-0.75pt+\x,0pt+\y)(uno){};
			\node at (-1.5pt+\x,0pt+\y)(unoA){};
			\node at (0.75pt+\x,0pt+\y)(due){};
			\node at (1.5pt+\x,0.75pt+\y)(dueA){};
			\node at (1.5pt+\x,-0.75pt+\y)(dueB){};
			\node at (2pt+\x,0.75pt+\y){\footnotesize $p_1$};
			\node at (2pt+\x,-0.75pt+\y){\footnotesize $p_2$};
			\node at (0pt+\x,0+\y)[thick,scale=1.5]{};

			\draw [thick] (0+\x,0+\y) circle (0.75pt);
			\draw [thick,double] (uno.center) -- (unoA.center);
			\draw [thick] (due.center) -- (dueA.center);
			\draw [thick] (due.center) -- (dueB.center);
		\end{tikzpicture} \> & = \int \frac{\dd^{4-2\epsilon} l}{(2 \pi)^{4-2\epsilon}} \;\dfrac{1}{l^2\, (l+p_1+p_2)^2} = i \frac{c_{\Gamma}}{(4\pi)^{2-\epsilon}} \frac{(-s_{12})^{-\epsilon}}{\epsilon (1-2\epsilon)}\ ,       \\
		\begin{tikzpicture}[baseline={([yshift=-1mm]uno.base)},scale=15]
			\def\x{0}
			\def\y{0}

			\clip (-1.4pt,-2pt) rectangle (3.5pt,2pt);

			\node at (0+\x,0+\y)(uno){};
			\node at (1.5pt+\x,1pt+\y)(due){};
			\node at (1.5pt+\x,-1pt+\y)(tre){};
			%\node at (-0.75pt+\x,0.5pt+\y)(unoB){};
			%\node at (-0.75pt+\x,0.5pt+\y)(unoB){};
			\node at (-0.75pt+\x,0+\y)(unoS){};
			\node at (2.25pt+\x,1.5pt+\y)(dueB){};
			\node [right=0.02pt of dueB, yshift=0.4pt]{\footnotesize$p_1$};
			\node at (2.25pt+\x,-1.5pt+\y)(treB){};
			\node [right=-0.02pt of treB, yshift=-0.4pt]{\footnotesize$p_2$};
			\node at (0.75pt+\x,-1pt+\y) (end){};
			\node at (-4pt+\x,-2pt+\y)(start){};
			\node at (0.9pt+\x,0+\y)[thick,scale=1.5]{};

			\draw [thick] (uno.center) -- (due.center);
			\draw [thick] (due.center) -- (tre.center);
			\draw [thick] (uno.center) -- (tre.center);
			%\draw [thick] (uno.center) -- (unoB.center);
			%\draw [thick] (uno.center) -- (unoC.center);
			\draw [thick,double] (uno.center) -- (unoS.center);
			\draw [thick] (due.center) -- (dueB.center);
			\draw [thick] (tre.center) -- (treB.center);

			%\draw [dotted,thick](-0.6pt+\x,0.3pt+\y) arc (140:240:0.4pt);
		\end{tikzpicture}
		%\coloneqq I_{3}(p_1,p_2,q)
		                               & = \int \frac{\dd^{4-2\epsilon} l}{(2 \pi)^{4-2\epsilon}} \;\dfrac{1}{l^2\, (l+p_2)^2\, (l+p_1+p_2)^2} = - \frac{i\, c_{\Gamma}}{(4\pi)^{2-\epsilon}} \frac{(-s_{12})^{-1-\epsilon}}{\epsilon^2}\ , \\
		\begin{tikzpicture}[baseline={([yshift=-1mm]uno.base)},scale=15]
			\def\x{0}
			\def\y{0}

			\clip (-1.4pt,-2pt) rectangle (3.5pt,2pt);

			\node at (0+\x,0+\y)(uno){};
			\node at (1.5pt+\x,1pt+\y)(due){};
			\node at (1.5pt+\x,-1pt+\y)(tre){};
			%\node at (-0.75pt+\x,0.5pt+\y)(unoB){};
			%\node at (-0.75pt+\x,0.5pt+\y)(unoB){};
			\node at (-0.75pt+\x,0+\y)(unoS){};
			\node at (-1.2pt+\x,0pt+\y){};
			\node at (2.25pt+\x,1.5pt+\y)(dueB){};
			\node [right=0.02pt of dueB, yshift=0.4pt]{\footnotesize$p$};
			\node at (2.25pt+\x,0.5pt+\y)(dueC){};
			\node at (2.25pt+\x,-1.5pt+\y)(treB){};
			\node [right=-0.02pt of treB, yshift=-0.4pt]{\footnotesize$q$};
			\node at (0.75pt+\x,-1pt+\y) (end){};
			\node at (-4pt+\x,-2pt+\y)(start){};
			\node at (0.9pt+\x,0+\y)[thick,scale=1.5]{};

			\draw [thick] (uno.center) -- (due.center);
			\draw [thick] (due.center) -- (tre.center);
			\draw [thick] (uno.center) -- (tre.center);
			%\draw [thick] (uno.center) -- (unoB.center);
			%\draw [thick] (uno.center) -- (unoC.center);
			\draw [thick] (uno.center) -- (unoS.center);
			\draw [thick,double] (due.center) -- (dueB.center);
			\draw [thick,double] (tre.center) -- (treB.center);

			%\draw [dotted,thick](-0.6pt+\x,0.3pt+\y) arc (140:240:0.4pt);
		\end{tikzpicture}    & =  - \frac{i\, c_{\Gamma}}{(4\pi)^{2-\epsilon}} \frac{1}{\epsilon^2}\left[\frac{(-q^2)^{-\epsilon}-(-p^2)^{-\epsilon}}{q^2-p^2}\right]\ ,
	\end{aligned}
\end{equation}
and
\begin{equation}
	\begin{aligned}
		\begin{tikzpicture}[baseline={([yshift=-1mm]Base.base)},scale=15]

			\def\x{0}
			\def\y{0}

			\clip (-1.4pt,-1.4pt) rectangle (3.5pt,3.5pt);

			\node at (0pt+\x,1pt+\y) (Base) {};

			\node at (0+\x,0+\y) (A) {};
			\node at (2pt+\x,0+\y) (B) {};
			\node at (2pt+\x,2pt+\y) (C) {};
			\node at (0+\x,2pt+\y) (D) {};

			\node at (-1pt+\x, -1pt+\y) (A1) {\footnotesize$p_3$};
			\node at (3pt+\x,-1pt+\y) (B1) {\footnotesize$p_2$};
			\node at (3pt+\x,3pt+\y) (C1) {\footnotesize$p_1$};
			\node at (-1pt+\x,3pt+\y) (D1) {\footnotesize$q$};

			\draw [thick] (A.center) -- (B.center);
			\draw [thick] (B.center) -- (C.center);
			\draw [thick] (C.center) -- (D.center);
			\draw [thick] (D.center) -- (A.center);
			\draw [thick] (A1) -- (A.center);
			\draw [thick] (B1) -- (B.center);
			\draw [thick] (C1) -- (C.center);
			\draw [thick,double] (D1) -- (D.center);

		\end{tikzpicture}
		= & \int \frac{\dd^{4-2\epsilon} l}{(2 \pi)^{4-2\epsilon}} \;\dfrac{1}{l^2\,(l+p_1)^2\, (l+p_1+p_2)^2\,(l+p_1+p_2+p_3)^2}                                                                               \\
		= & -\frac{2\, i\, c_{\Gamma}}{(4\pi)^{2-\epsilon}} \frac{1}{s_{12} s_{23}} \Bigg\{ - \frac{1}{\epsilon^2}\left[(-s_{12})^{-\epsilon}+(-s_{23})^{-\epsilon}-(-q^2)^{-\epsilon}\right]+                  \\
		  & +{\rm Li}_{2}\left(1-\frac{s_{12}}{q^2}\right)+{\rm Li}_{2}\left(1-\frac{s_{23}}{q^2}\right)+\frac{1}{2}\log^2\left(\frac{s_{12}}{s_{23}}\right)+\frac{\pi^2}{6}\Bigg\}+\cO\left(\epsilon\right)\ ,
	\end{aligned}
\end{equation}
where
\begin{equation}
	c_{\Gamma}=\frac{\Gamma[1+\epsilon]\, \Gamma[1-\epsilon]^2}{\Gamma[1-2\epsilon]}\ .
\end{equation}

This results are exact to all orders in $\epsilon$, and the expression of the corresponding integral functions in a different number of dimensions can be obtained by simply replacing $\epsilon$ to the appropriate value, for instance $\epsilon \mapsto \epsilon - 1$ and $\epsilon \mapsto \epsilon - 2$ for $d=6-2\epsilon$ and $d=8-2\epsilon$, respectively. In particular it turns out that all the integrals which give the rational terms, \textit{i.e.} those with a non-trivial numerator written in \eqref{eq:muintegrals}, can always be expressed as integrals in higher dimensions \cite{Bern:1995db}. Indeed consider the general integral function
\begin{equation}
	\label{eq::rationalintegral}
	I^d_{n}[\mu^{2 p}] = \int \frac{\dd^{4-2\epsilon} l}{(2\pi)^{4-2\epsilon}} (\mu^2)^{p} f_n\left( \{p_i\}, l \right) = \int \frac{\dd^{4} l^{(4)}}{(2\pi)^{4}} \int \frac{\dd^{-2\epsilon} \mu}{(2\pi)^{-2\epsilon}} (\mu^2)^{p} f_n\left( \{ p_i \}, l \right)\ ,
\end{equation}
and the $\mu$-measure can be rewritten as
\begin{equation}
	\begin{split}
		\int \dd^{-2\epsilon} \mu\; (\mu^2)^p &= \frac{1}{2} \int \dd\Omega_{-1-2\epsilon} \int_{0}^{+\infty} \dd \mu^2\; (\mu^2)^{-1-\epsilon + p}\\ \vspace{5mm}
		&= \frac{\int \Omega_{-1-2\epsilon}}{\int \dd \Omega_{2p -1-2\epsilon}}\int \dd^{2 p - 2\epsilon} \mu \ .
	\end{split}
\end{equation}
Then \eqref{eq::rationalintegral} can be written as
\begin{equation}
	\begin{split}
		I^d_{n}[\mu^{2 p}] &= \frac{(2\pi)^{2p} \int \dd \Omega_{-1-2\epsilon}}{\int \dd \Omega_{2p -1-2\epsilon}} \int \frac{\dd^{4+2p-2\epsilon} l}{(2\pi)^{4+2p-2\epsilon}} f_n\left( \{p_i\}, l \right)\\ \vspace{5mm}
		&= -\epsilon (1-\epsilon)(2-\epsilon)\cdots (p-1-\epsilon) (4\pi)^p I^{d+2p}_{n}[1]\ ,
	\end{split}
\end{equation}
where
\begin{equation}
	\int \dd\Omega_x = \frac{2 \pi^{\frac{x+1}{2}}}{\Gamma [\frac{x+1}{2}]}\ .
\end{equation}

Then to compute this integrals becomes just simple algebra:

\makebox[0pt][l]{
	\begin{minipage}[h]{.5\linewidth}
		\begin{equation*}
			\begin{aligned}
				\begin{tikzpicture}[baseline={([yshift=-1mm]uno.base)},scale=15]
					\def\x{0}
					\def\y{0}

					\clip (-2.5pt,-2pt) rectangle (2.5pt,2pt);

					\def\x{0pt}
					\def\y{0}

					\node at (-0.75pt+\x,0pt+\y)(uno){};
					\node at (-1.5pt+\x,0pt+\y)(unoA){};
					\node at (0.75pt+\x,0pt+\y)(due){};
					\node at (1.5pt+\x,0.75pt+\y)(dueA){};
					\node at (1.5pt+\x,-0.75pt+\y)(dueB){};
					\node at (2pt+\x,0.75pt+\y){\footnotesize $p_1$};
					\node at (2pt+\x,-0.75pt+\y){\footnotesize $p_2$};
					\node at (0pt+\x,0+\y)[thick,scale=1.5]{};
					\node at (0+\x,0+\y){\footnotesize$\mu^2$};

					\draw [thick] (0+\x,0+\y) circle (0.75pt);
					\draw [thick,double] (uno.center) -- (unoA.center);
					\draw [thick] (due.center) -- (dueA.center);
					\draw [thick] (due.center) -- (dueB.center);
				\end{tikzpicture} \> & = \frac{-i}{(4\pi)^2}\cdot\frac{s_{12}}{6}+ \cO (\epsilon)\ ,       \\
				\begin{tikzpicture}[baseline={([yshift=-1mm]uno.base)},scale=15]
					\def\x{0}
					\def\y{0}

					\clip (-1.8pt,-2pt) rectangle (3.5pt,2pt);

					\node at (0+\x,0+\y)(uno){};
					\node at (1.5pt+\x,1pt+\y)(due){};
					\node at (1.5pt+\x,-1pt+\y)(tre){};
					\node at (-0.75pt+\x,0pt+\y)(unoB){};
					\node at (2.25pt+\x,1.5pt+\y)(dueB){};
					\node [right=0.02pt of dueB, yshift=0.4pt]{\footnotesize$p_1$};
					\node at (2.25pt+\x,-1.5pt+\y)(treB){};
					\node [right=-0.02pt of treB, yshift=-0.4pt]{\footnotesize$p_2$};
					\node at (0.75pt+\x,-1pt+\y) (end){};
					\node at (-4pt+\x,-2pt+\y)(start){};
					%		\node at (0.9pt+\x,0+\y)[thick,scale=1.5]{$\circlearrowright$};
					\node at (1pt+\x,0.1pt+\y){\footnotesize$\mu^2$};
					\node at (-0.75pt+\x,0+\y)(cinque){};
					%\node at (-1.25pt+\x,0+\y){\footnotesize $p_4$};

					\draw [thick] (uno.center) -- (due.center);
					\draw [thick] (due.center) -- (tre.center);
					\draw [thick] (uno.center) -- (tre.center);
					\draw [thick,double] (uno.center) -- (unoB.center);
					\draw [thick] (due.center) -- (dueB.center);
					\draw [thick] (tre.center) -- (treB.center);

				\end{tikzpicture}    & = \frac{i}{(4\pi)^{2}} \cdot \frac{1}{2} + \cO (\epsilon)\ ,        \\
				\begin{tikzpicture}[baseline={([yshift=-1mm]uno.base)},scale=15]
					\def\x{0}
					\def\y{0}

					\clip (-1.8pt,-2pt) rectangle (3.5pt,2pt);

					\node at (0+\x,0+\y)(uno){};
					\node at (1.5pt+\x,1pt+\y)(due){};
					\node at (1.5pt+\x,-1pt+\y)(tre){};
					\node at (-0.75pt+\x,0pt+\y)(unoB){};
					\node at (2.25pt+\x,1.5pt+\y)(dueB){};
					\node [right=0.02pt of dueB, yshift=0.4pt]{\footnotesize$p_1$};
					\node at (2.25pt+\x,-1.5pt+\y)(treB){};
					\node [right=-0.02pt of treB, yshift=-0.4pt]{\footnotesize$p_2$};
					\node at (0.75pt+\x,-1pt+\y) (end){};
					\node at (-4pt+\x,-2pt+\y)(start){};
					%		\node at (0.9pt+\x,0+\y)[thick,scale=1.5]{$\circlearrowright$};
					\node at (1pt+\x,0.1pt+\y){\footnotesize$\mu^4$};
					\node at (-0.75pt+\x,0+\y)(cinque){};
					%\node at (-1.25pt+\x,0+\y){\footnotesize $p_4$};

					\draw [thick] (uno.center) -- (due.center);
					\draw [thick] (due.center) -- (tre.center);
					\draw [thick] (uno.center) -- (tre.center);
					\draw [thick,double] (uno.center) -- (unoB.center);
					\draw [thick] (due.center) -- (dueB.center);
					\draw [thick] (tre.center) -- (treB.center);

				\end{tikzpicture}    & = \frac{i}{(4\pi)^{2}} \cdot \frac{s_{12}}{24}  + \cO (\epsilon)\ ,
			\end{aligned}
		\end{equation*}
	\end{minipage}
	\begin{minipage}[h]{.5\linewidth}
		\begin{equation*}
			\begin{aligned}
				\begin{tikzpicture}[baseline={([yshift=-1mm]Base.base)},scale=15]

					\def\x{0}
					\def\y{0}

					\clip (-1.55pt,-1.4pt) rectangle (3.5pt,3.5pt);

					\node at (0pt+\x,1pt+\y) (Base) {};

					\node at (0+\x,0+\y) (A) {};
					\node at (2pt+\x,0+\y) (B) {};
					\node at (2pt+\x,2pt+\y) (C) {};
					\node at (0+\x,2pt+\y) (D) {};

					\node at (-1pt+\x, -1pt+\y) (A1) {\footnotesize$p_3$};
					\node at (3pt+\x,-1pt+\y) (B1) {\footnotesize$p_2$};
					\node at (3pt+\x,3pt+\y) (C1) {\footnotesize$p_1$};
					\node at (-1pt+\x,3pt+\y) (D1) {\footnotesize$q$};
					\node at (1.1pt+\x, 1.1pt+\y) (Center) {\footnotesize$\mu^2$};

					\draw [thick] (A.center) -- (B.center);
					\draw [thick] (B.center) -- (C.center);
					\draw [thick] (C.center) -- (D.center);
					\draw [thick] (D.center) -- (A.center);
					\draw [thick] (A1) -- (A.center);
					\draw [thick] (B1) -- (B.center);
					\draw [thick] (C1) -- (C.center);
					\draw [thick,double] (D1) -- (D.center);

				\end{tikzpicture} & = \cO (\epsilon)\ ,                                      \\
				\begin{tikzpicture}[baseline={([yshift=-1mm]Base.base)},scale=15]

					\def\x{0}
					\def\y{0}

					\clip (-1.55pt,-1.4pt) rectangle (3.5pt,3.5pt);

					\node at (0pt+\x,1pt+\y) (Base) {};

					\node at (0+\x,0+\y) (A) {};
					\node at (2pt+\x,0+\y) (B) {};
					\node at (2pt+\x,2pt+\y) (C) {};
					\node at (0+\x,2pt+\y) (D) {};

					\node at (-1pt+\x, -1pt+\y) (A1) {\footnotesize$p_3$};
					\node at (3pt+\x,-1pt+\y) (B1) {\footnotesize$p_2$};
					\node at (3pt+\x,3pt+\y) (C1) {\footnotesize$p_1$};
					\node at (-1pt+\x,3pt+\y) (D1) {\footnotesize$q$};
					\node at (1.1pt+\x, 1.1pt+\y) (Center) {\footnotesize$\mu^4$};

					\draw [thick] (A.center) -- (B.center);
					\draw [thick] (B.center) -- (C.center);
					\draw [thick] (C.center) -- (D.center);
					\draw [thick] (D.center) -- (A.center);
					\draw [thick] (A1) -- (A.center);
					\draw [thick] (B1) -- (B.center);
					\draw [thick] (C1) -- (C.center);
					\draw [thick,double] (D1) -- (D.center);

				\end{tikzpicture} & = \frac{-i}{(4\pi)^2}\cdot\frac{1}{6}+ \cO (\epsilon)\ ,
			\end{aligned}
		\end{equation*}
	\end{minipage}
}

\section{The \texttt{SpinorHelicity6D} \texttt{MATHEMATICA} Package}
\label{Mathematica}

In this section we briefly describe the \texttt{Mathematica} package \texttt{SpinorHelicity6D} \cite{Accettulli2019git}, which we developed to facilitate our calculations involving six- (and four-)dimensional spinors and was initially inspired by the package \texttt{SpinorHelicity} \cite{Panerai2019git}.
In the following  we present  a subset  of all the available functions -- specifically, we focus on the routines needed to check our results. In the near future  a more complete documentation along with an updated version of the package will be released, including  tools needed for numerical evaluations in six-dimensional space. For the sake of concreteness, the functions will be presented applying them to an example calculation, namely $\Tr F^3$ in the all-plus helicity configuration\footnote{See also the Mathematica notebook in the Github repository.}.

\begin{figure}[H]
	\centering
	\makebox[\textwidth]{
		\fbox{\includegraphics[page=1,trim={1.5cm 10.4cm 4.5cm 7.6cm},clip]{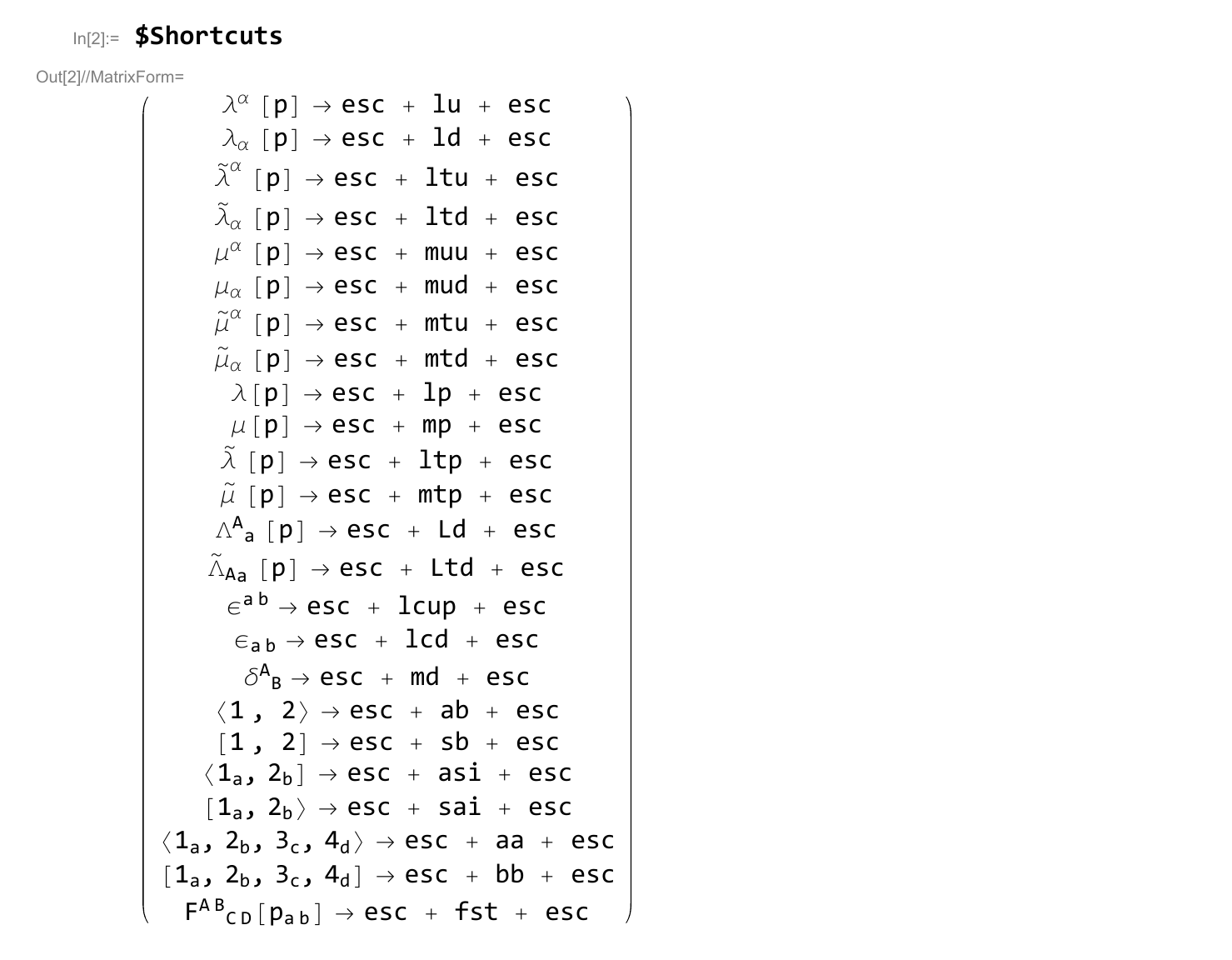}}}
	\caption{A complete list of all the shortcuts available in the package.}
	\label{fig::shortcutsss}
\end{figure}

\subsubsection*{The Building Blocks}

The basic input of most computations is given in terms of spinors and invariants built out of them. All of these can be input into \texttt{Mathematica} through a keyboard shortcut, a complete list of which is stored into the variable \texttt{\$Shortcuts}, see Figure~\ref{fig::shortcutsss}.
Alternatively one can use the more user-friendly palette which opens automatically upon loading the package or by typing \texttt{SpinorPalette}.

The basic objects are:
\begin{itemize}
	\item the standard four-dimensional spinors $\lambda$ and $\tilde{\lambda}$. These allow for upper and lower spinor-indices and have the usual contraction properties:
	      \begin{figure}[H]
		      \begin{center}
			      \makebox[\textwidth]{
				      \fbox{\includegraphics[page=1,trim={1.5cm 6.2cm 4.5cm 19cm},clip]{Example2}}}
		      \end{center}
	      \end{figure}
	      The indices of $\lambda$ and $\tilde{\lambda}$ do not mix, in other words the package is able to distinguish between dotted and undotted indices:
	      \begin{figure}[H]
		      \begin{center}
			      \makebox[\textwidth]{
				      \fbox{\includegraphics[page=1,trim={1.5cm 4.7cm 4.5cm 23.5cm},clip]{Example2}}}
		      \end{center}
	      \end{figure}
	      We adopted the convention that
	      \begin{equation}
		      \lambda(-p)=i \lambda(p)\ , \hspace{1cm} \tilde{\lambda}(-p)=i \tilde{\lambda}(p)\, ,
	      \end{equation}
	      and this is automatically applied both to the free as well as contracted spinors:
	      \begin{figure}[H]
		      \begin{center}
			      \makebox[\textwidth]{
				      \fbox{\includegraphics[page=2,trim={1.5cm 24.3cm 4.5cm 2.5cm},clip]{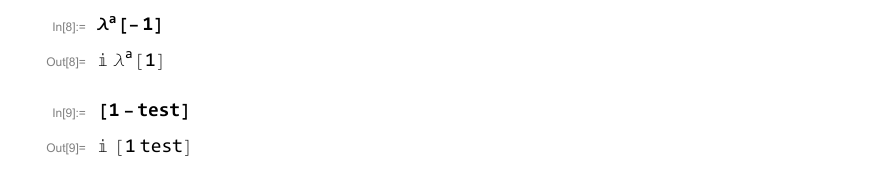}}}
		      \end{center}
	      \end{figure}
	      As can be seen from the above example one is free to choose whatever label for the momenta. However some caution is needed and we recommend avoiding labels containing minus signs\footnote{For example $n-1$ could be misinterpreted as the opposite of $1-n$ due to the automatic ordering applied by \texttt{Mathematica}.}, these are safe only if written in form of a string.

	\item $\mu$ and $\tilde{\mu}$ are the auxiliary spinors needed to write a massive momentum in terms of two massless ones as in~\eqref{eq:massivemom}. They share all the properties of $\lambda$ and $\tilde{\lambda}$ respectively.

	\item The set of spinors $\lambda$ and $\mu$ without indices are auxiliary objects used for example when defining properties or applying
	      transformations to both spinors with upper and lower Lorentz indices.

	\item \texttt{M} and $\tilde{\texttt{M}}$ are the ``masses'' arising from the fifth and sixth spacetime component of the momentum as in~\eqref{eq:extramass}.

	\item $\Lambda$ and $\tilde{\Lambda}$ are the six-dimensional spinors. These have been implemented only with lower little group indices, but in order to raise them one can simply use the two-dimensional Levi-Civita tensor.

	\item Clearly $\langle \cdot,\cdot \rangle$ and $[ \cdot , \cdot ]$ are the four-dimensional invariants and $\langle \cdot , \cdot ]$, $[\cdot , \cdot \rangle$, $\langle \cdot , \cdot ,\cdot , \cdot \rangle$ and $[ \cdot , \cdot ,\cdot , \cdot ]$ the six-dimensional ones.

\end{itemize}

It is important to point out that, despite the nice visualisation properties of the above presented objects, they are still interpreted as plain functions by \texttt{Mathematica}. This means that, for example, they can be safely copy-pasted and all the standard operations for functions can be applied. In order to access the explicit functional forms use \texttt{InputForm}.
\begin{figure}[H]
	\begin{center}
		\makebox[\textwidth]{
			\fbox{\includegraphics[page=2,trim={1.5cm 22.5cm 4.5cm 5.5cm},clip]{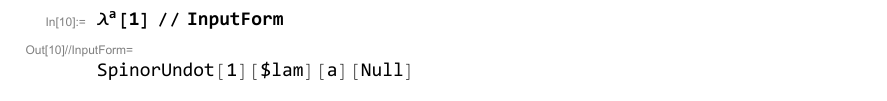}}}
	\end{center}
\end{figure}

\subsubsection*{Computing the Double-Cut}

The computation of $\Tr F^3$ (all-plus) begins by imposing four-dimensional kinematics on external particles, or equivalently declaring the corresponding momenta to be massless. This is done through the function \texttt{KillMasses}, and the list of massless momenta\footnote{By default all momenta are always considered massive.} is stored in \texttt{Momenta4D}.

\begin{figure}[H]
	\begin{center}
		\makebox[\textwidth]{
			\fbox{\includegraphics[page=2,trim={1.5cm 18.8cm 4.5cm 7.1cm},clip]{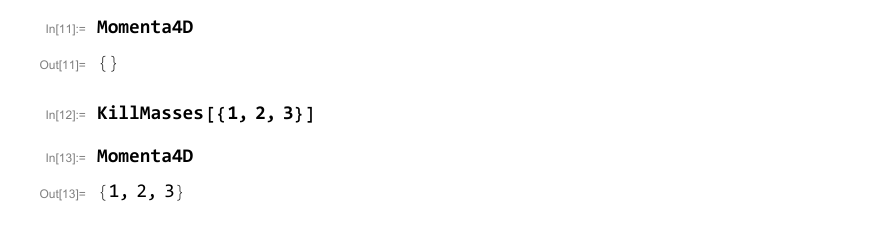}}}
	\end{center}
\end{figure}

In our specific case we need to break down the six-dimensional spinors and invariants in terms of the four-dimensional ones. This breakdown is achieved through the command \texttt{To4D} locally on a given expression, but it can also be turned on globally and applied automatically to all expressions through \texttt{To4DAlways[True]}. This setting can be reversed by \texttt{To4DAlways[False]}.

\begin{figure}[H]
	\begin{center}
		\makebox[\textwidth]{
			\fbox{\includegraphics[page=7,trim={1.5cm 12cm 4.4cm 11cm},clip]{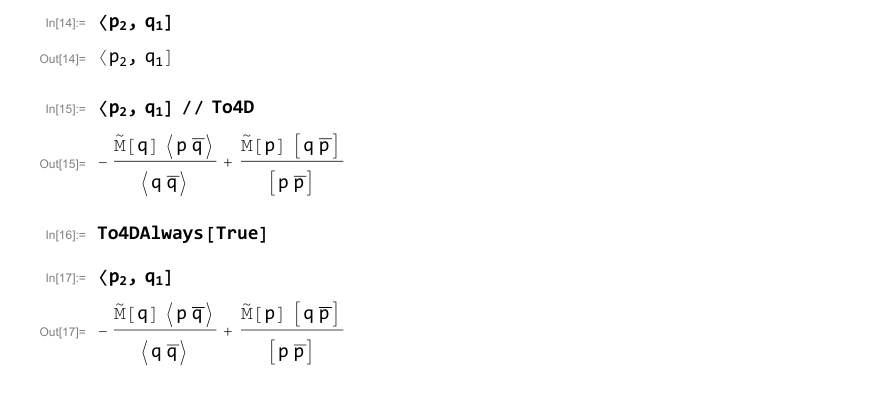}}}
	\end{center}
\end{figure}

One can then write down the following double-cut expression with gluons running through the loop:

\begin{figure}[H]
	\begin{center}
		\makebox[\textwidth]{
			\fbox{\includegraphics[page=2,trim={1.5cm 10cm 4.4cm 10.5cm},clip]{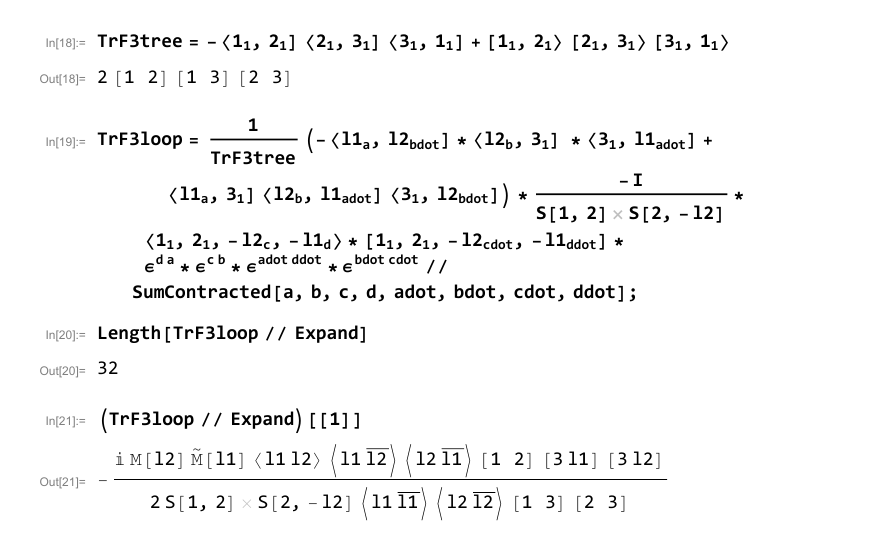}}}
	\end{center}
\end{figure}
\noindent
Here \texttt{TrF3tree} is the tree-level form factor used to normalize the one-loop expression, and \texttt{S} is the Mandelstam invariant $s_{ij}$ with momenta $i$ and $j$ in six-dimensions. We made use of the command \texttt{SumContracted}, which sums over the contracted little-group indices. As can be seen the result of the double cut is a 32-term expression, and we printed the first term of the sum to give an idea of their form. We can obtain a first simplification of the expression by applying momentum conservation in the form of~\eqref{eq:extramass}, which reduces the number of terms down to 24:

\begin{figure}[H]
	\begin{center}
		\makebox[\textwidth]{
			\fbox{\includegraphics[page=2,trim={1.5cm 7.3cm 4.4cm 19.5cm},clip]{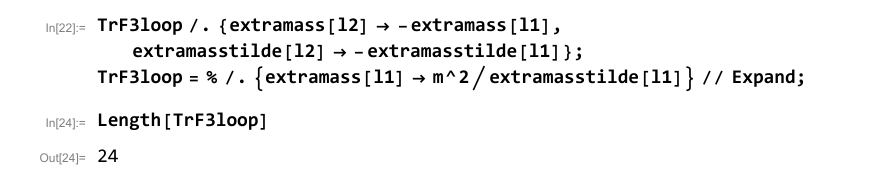}}}
	\end{center}
\end{figure}

\subsubsection*{Removing the Redundancy}

Now we can use the function \texttt{MuReplace} to eliminate the redundant degrees of freedom parame\-trised by the $\{\mu,\tilde{\mu}\}$ spinors.
These can be fixed to arbitrary values without affecting the final (little-group invariant) result. \texttt{MuReplace} goes through all
the $\mu$s ($\tilde{\mu}$) present in the given expression and replaces them with $\lambda$s ($\tilde{\lambda}$) in such a way that as many terms as possible vanish due to the antisymmetry of the angle and square brackets. This function allows for two options:
\begin{itemize}
	\item \texttt{DisplayReplacements}, default is False. If set to True the replacements chosen by\\
	\texttt{MuReplace} will be displayed along with the result.
	\item \texttt{GlobalReplacements}, default is False. If set to True the chosen replacements are stored and become globally defined\footnote{In other words the chosen replacements are applied from there on whenever the given spinors are encountered.}. The list of spinors defined to be equivalent
	      can be accessed through \texttt{FixedSpinors}. The global definitions stored in \texttt{FixedSpinors} can be cleared with \texttt{ClearSpinors}.
\end{itemize}
\begin{figure}[H]
	\begin{center}
		\makebox[\textwidth]{
			\fbox{\includegraphics[page=2,trim={1.5cm 3.8cm 4.4cm 22.5cm},clip]{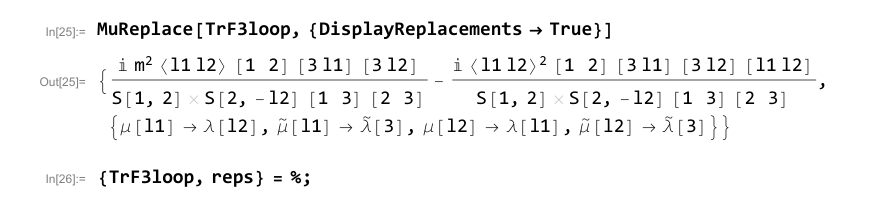}}}
	\end{center}
\end{figure}

%Input 16 just enforces momentum conservation in the form of
%\begin{equation}
%	m_{l_2}=-m_{l_1} \> , \hspace{0.5cm} \tilde{m}_{l_2}=-\tilde{m}_{l_1} \>.
%\end{equation}
The output of \texttt{MuReplace} upon setting \texttt{DisplayReplacements} to True is a list of two elements. The first is the expression of $\Tr F^3$ after the most suitable replacements have been applied. The second term is the list of replacements which have been found to be the most convenient.

It is also possible for the user to choose the replacements and apply them manually. This can be done through the function \texttt{SpinorReplace} or \texttt{SpinorReplaceSequential}, which allow to perform generic replacements of spinors inside given expressions. The difference between the two functions is that the first performs all the specified replacements simultaneously whereas the second performs them sequentially.

%Finally, we write the square bracket $\sqr{l_1}{l_2}$ as a function of $\agl{l_1}{l_2}$, $s_{12}$ and $m^2$ using the momentum conservation relation $s_{l_1 l_2}=s_{12}$. Upon replacing the $\mu$ spinors with the previously chosen ones we get
%\begin{equation}
%	\sqr{l_1}{l_2}=-\frac{s_{12}-2m^2}{\agl{l_1}{l_2}} \>,
%\end{equation}
%and thus:
Finally, before performing any other manipulations, we restore the momenta $l_1$ and $l_2$ to massive four-dimensional momenta, as discussed in more detail in Section~\ref{sec:TrF2nonmin}. This is done by the function \texttt{CompleteToMassive}, which takes as input an expression and the list of replacements used to remove the $\mu$s. One gets
\begin{figure}[H]
	\begin{center}
		\makebox[\textwidth]{
			\fbox{\includegraphics[page=3,trim={1.5cm 23.7cm 4.4cm 2.5cm},clip]{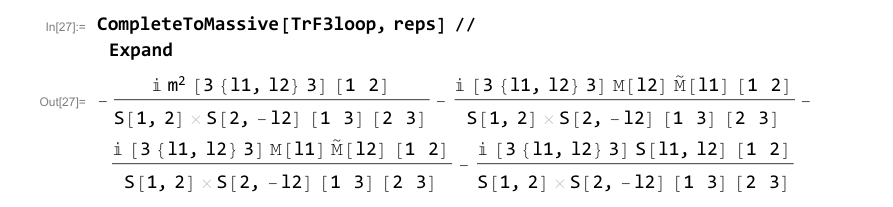}}}
	\end{center}
\end{figure}
where the spinor invariants $[ 3 \, l_1 ] \langle l_1 \, l_2 \rangle [l_2 \, 3]$ has been closed to $[3 \, \slashed{l}_1 \, \slashed{l}_2 \, 3]$ and new dependencies on the masses $m$ and $\tilde{m}$ appeared. Using once again momentum conservation to get rid of $m_{l_2}$ and $\tilde{m}_{l_2}$, and to replace $s_{l_1l_2}$ with $s_{12}$ we get
\begin{figure}[H]
	\begin{center}
		\makebox[\textwidth]{
			\fbox{\includegraphics[page=3,trim={1.5cm 20.3cm 4.4cm 6cm},clip]{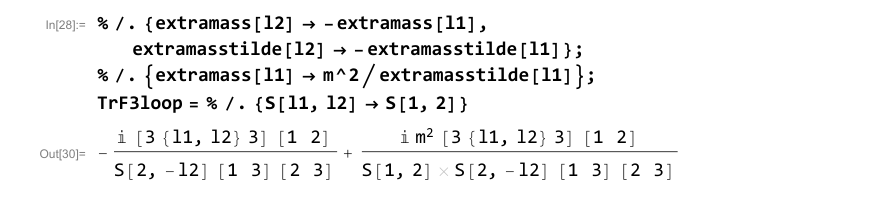}}}
	\end{center}
\end{figure}

\subsubsection*{The Scalar Subtraction}

Once the computation with the gluons in the loop has been completed we can move on to the scalars. One has to go through exactly the same steps as before. First write down the double-cut expression, then evaluate it and remove the redundant $\mu$ spinors to simplify the result, and finally rewrite the expression in terms of four-dimensional massive momenta.
%This time, for demonstration purposes, we use the function \texttt{SpinorReplaceSequential} which requires the user to choose the replacements\footnote{The difference between \texttt{SpinorReplace} and \texttt{SpinorReplaceSequential} is that the first performs all the replacements at the same time, whereas the second performs them sequentially one at the time in the given order.}.

\begin{figure}[H]
	\begin{center}
		\makebox[\textwidth]{
			\fbox{\includegraphics[page=3,trim={1.5cm 11.5cm 4.4cm 9.3cm},clip]{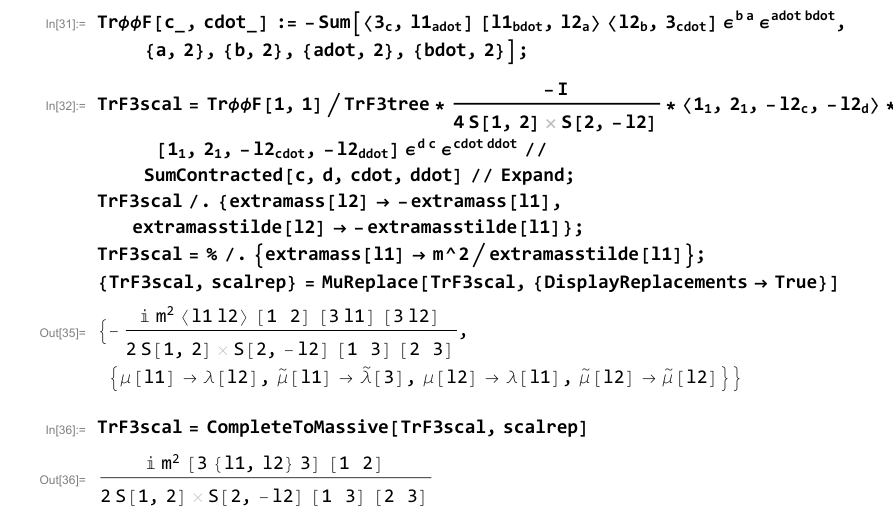}}}
	\end{center}
\end{figure}

Performing the scalar subtraction leads then to the final result:
\begin{figure}[H]
	\begin{center}
		\makebox[\textwidth]{
			\fbox{\includegraphics[page=3,trim={1.5cm 9.5cm 4.4cm 18cm},clip]{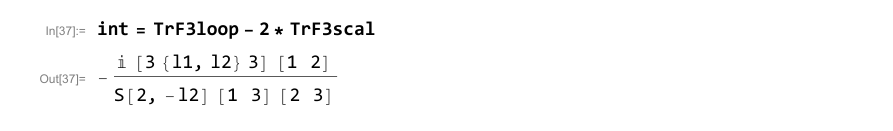}}}
	\end{center}
\end{figure}

From here on one can take two possible routes: either manipulate the final expression by hand and obtain~\eqref{eq::TrF3comp}, where the IR divergences are clearly visible in the form of a one-mass triangle. If one is interested in a completely reduced expression, one can proceed as follows:
\begin{itemize}
	\item complete the denominator to Mandelstam invariants and further contract all possible expressions of the form $\langle i \, j \rangle [j \, i]$ to $s_{ij}$. This is accomplished with \texttt{CompleteDenomina\-tors} and \texttt{CompleteMandelstam} respectively. The variables \texttt{S4} is the Mandelstam invariant $s_{ij}$ with $p_i$ and $p_j$ in four dimensions, where \texttt{S4} can be related to \texttt{S} through \texttt{CompleteToMassive}.
	      \begin{figure}[H]
		      \begin{center}
			      \makebox[\textwidth]{
				      \fbox{\includegraphics[page=3,trim={1.5cm 8cm 4.4cm 20cm},clip]{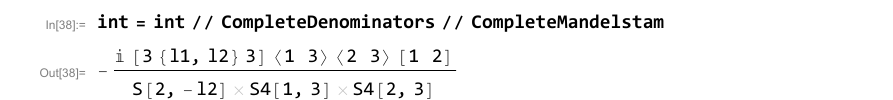}}}
		      \end{center}
	      \end{figure}

	\item Since we removed the helicity structure, \texttt{int} must be expressible in terms of Mandelstam invariants only. In other words we expect the numerator to be  of the form $\langle i \, \slashed{p}_j \cdots \slashed{p}_k \, i ]$, which is a trace involving a helicity projector. To contract the numerator into the above form we use \texttt{ToChain}:

	      \begin{figure}[H]
		      \begin{center}
			      \makebox[\textwidth]{
				      \fbox{\includegraphics[page=3,trim={1.5cm 6cm 4.4cm 22cm},clip]{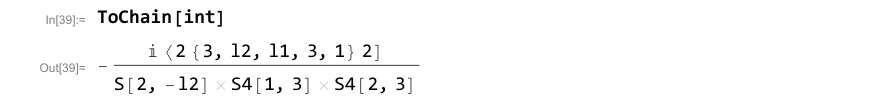}}}
		      \end{center}
	      \end{figure}
	      and then we use \texttt{ToTrace} to evaluate the resulting trace. The option \texttt{KillEpsilon} sets the contributions proportional to the Levi-Civita tensor to zero. We can do this since the kinematics of the problem does not allow for more than three independent momenta in the final  answer and thus similar terms would be vanishing anyway. The scalar products appearing after the trace are four dimensional.
	      \begin{figure}[H]
		      \begin{center}
			      \makebox[\textwidth]{
				      \fbox{\includegraphics[page=4,trim={1.5cm 23.8cm 4.4cm 2.5cm},clip]{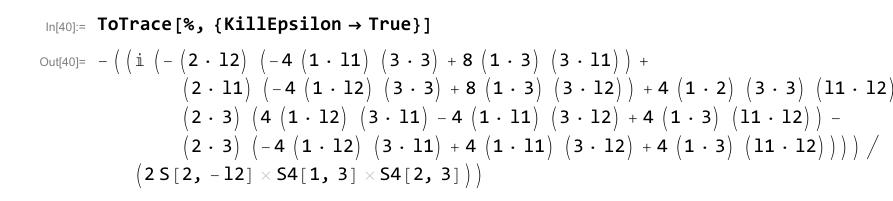}}}
		      \end{center}
	      \end{figure}
	\item rewrite the four-dimensional scalar products as six-dimensional Mandelstam invariants and masses, using \texttt{ScalProdToS}
	      \begin{figure}[H]
		      \begin{center}
			      \makebox[\textwidth]{
				      \fbox{\includegraphics[page=4,trim={1.5cm 17.5cm 4.4cm 6.5cm},clip]{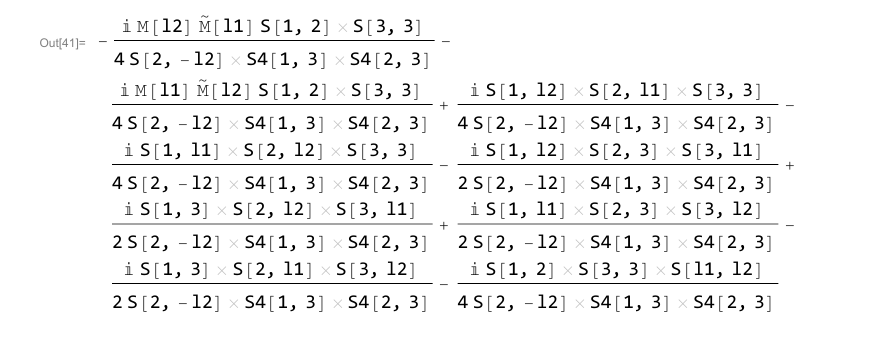}}}
		      \end{center}
	      \end{figure}
	\item then we use momentum conservation once again, and since all the \texttt{S4} appearing in the expression involve only external four-dimensional momenta, we uplift it to \texttt{S}.
	      \begin{figure}[H]
		      \begin{center}
			      \makebox[\textwidth]{
				      \fbox{\includegraphics[page=4,trim={1.5cm 11.2cm 4.4cm 12cm},clip]{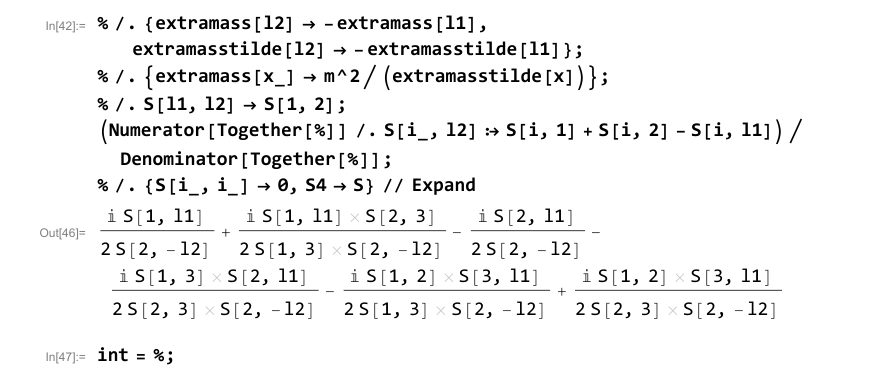}}}
		      \end{center}
	      \end{figure}
	\item Finally we can perform the reduction of the above expression using IBP identities, for example with \texttt{LiteRed} \cite{Lee:2012cn,Lee:2013mka}.
	      We export our result to the \texttt{LiteRed} notation using the function \texttt{Relabel}, which allows to relabel the momenta inside the scalar products and Mandelstam invariants, as well as reassign a new name to the scalar product. We also uplift the cut and then perform the reduction:
	      \begin{figure}[H]
		      \begin{center}
			      \makebox[\textwidth]{
				      \fbox{\includegraphics[page=6,trim={1.5cm 6.5cm 4.4cm 20cm},clip]{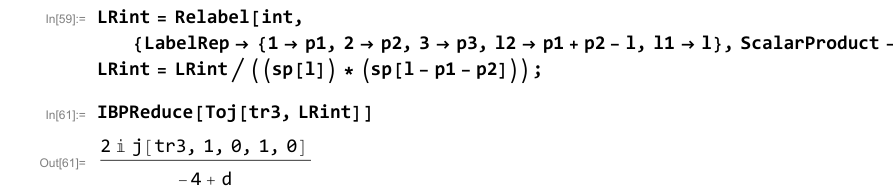}}}
		      \end{center}
	      \end{figure}
	      which of course matches with the reduction of expression~\eqref{eq::TrF3comp}:
	      \begin{figure}[H]
		      \begin{center}
			      \makebox[\textwidth]{
				      \fbox{\includegraphics[page=6,trim={1.5cm 4.5cm 4.4cm 23.5cm},clip]{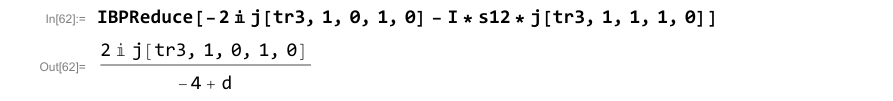}}}
		      \end{center}
	      \end{figure}

	      where
	      \begin{figure}[H]
		      \begin{center}
			      \makebox[\textwidth]{
				      \fbox{\includegraphics[page=7,trim={1.5cm 20.5cm 4.4cm 5cm},clip]{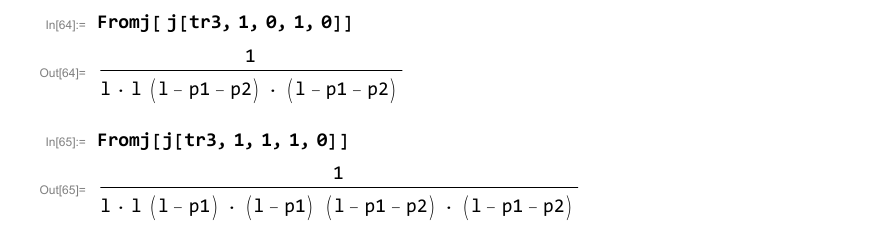}}}
		      \end{center}
	      \end{figure}

\end{itemize}

\newpage

\bibliographystyle{utphys}
\bibliography{biblio}

\end{document}